\documentclass[a4paper,11pt,fleqn]{article}

\usepackage{graphicx}
\usepackage{dcolumn}
\usepackage{bm}
\usepackage{amsmath}
\usepackage{amssymb}
\usepackage{amsfonts}
\usepackage{subfigure}
\usepackage{esvect}
\usepackage[dvipsnames]{xcolor}
\usepackage{fix-cm} 
\usepackage{cite}
\usepackage{soul}
\usepackage{mdframed}
\usepackage{lipsum}
\usepackage{multicol}
\usepackage{float}
\usepackage{longtable}
\usepackage{setspace}
\usepackage{tabularx}
\usepackage{ltablex}
\usepackage{newfloat}
\DeclareFloatingEnvironment[fileext=frm,placement={!ht},name=Box]{boxfloat}
\usepackage{capt-of}
\usepackage[htt]{hyphenat} 
\usepackage{hyperref}
\usepackage{url}

\usepackage{tikz}
\usetikzlibrary{decorations.pathreplacing}

\newcommand{\codeline}[2][]{\textcolor{red}{$#2$}}

\usepackage{listings}
\usepackage[frozencache]{minted}
\newminted{shell-session}{
    frame=lines, breaklines=true}
\definecolor{mintbackground}{RGB}{246, 246, 246}

\newcommand{\insertcppcode}[4]{\medskip
\noindent \texttt{#1}
\inputminted[framesep=2mm,
baselinestretch=1.2,
bgcolor=mintbackground,
fontsize=\small,
linenos, firstline=#2, lastline=#3, breaklines=true]{C++}{#4}}

\newcommand{\inserttxtcode}[4]{\vspace{0.5cm}
\noindent \texttt{#1}
\inputminted[frame=single,linenos, firstline=#2, lastline=#3, breaklines=true]{text}{#4}
}

\newminted{C++}{framesep=2mm,
baselinestretch=1.2,
bgcolor=mintbackground,
fontsize=\small,breaklines=true}

\usepackage{empheq} 

\usepackage[symbol]{footmisc}

\newcommand{\CL}{{\tt ${\mathcal C}$osmo${\mathcal L}$attice}~}
\newcommand{\CLns}{{\tt ${\mathcal C}$osmo${\mathcal L}$attice}}
\newcommand{\dx}{\ensuremath{\delta x}}

\newcommand{\deta}{\delta\eta}
\newcommand{\bn}{{\bf n}}
\newcommand{\dd}{\text{d}}

\newcommand{\be}{\begin{equation}}
\newcommand{\ee}{\end{equation}}
\newcommand{\bea}{\begin{eqnarray}}
\newcommand{\eea}{\end{eqnarray}}

\newcommand{\piSc}{\tilde\pi_\phi}
\newcommand{\piSingl}{\tilde\pi_\varphi}

\newcommand{\piA}{\tilde\pi_A}
\newcommand{\piApar}{\left(\piA\right)}
\newcommand{\piDoubl}{\widetilde\pi_\Phi}

\newcommand{\piB}{\tilde\pi_B}
\newcommand{\piBpar}{\left(\piB\right)}

\newcommand{\kersutwoComp}{\mathcal{K}_{B_i^a}}

\newcommand{\addressIFIC}{\it Instituto de F\'isica Corpuscular (IFIC), Consejo Superior de Investigaciones \\Cient\'ificas (CSIC) and Universitat de Valencia (UV), Valencia, Spain.}
\newcommand{\addressEPFL}{\it Institute of Physics, Laboratory of Particle Physics and Cosmology (LPPC), \'Ecole \\ Polytechnique F\'ed\'erale de Lausanne (EPFL), CH-1015 Lausanne, Switzerland.}
\newcommand{\addressUNIBAS}{\it Department of Physics, University of Basel, \\ Klingelbergstr. 82, CH-4056 Basel, Switzerland.}

\newcommand{\addressSUNY}{Center for Nuclear Theory, Department of Physics and Astronomy, Stony Brook University \\ Stony Brook, New York 11794, USA.}

\begin{document}
\hypersetup{
    pdfborder={0 0 0},
}

\vspace*{-0.25cm}
\begin{figure}
\hspace{0.9cm}
\includegraphics[width = 0.9\textwidth]{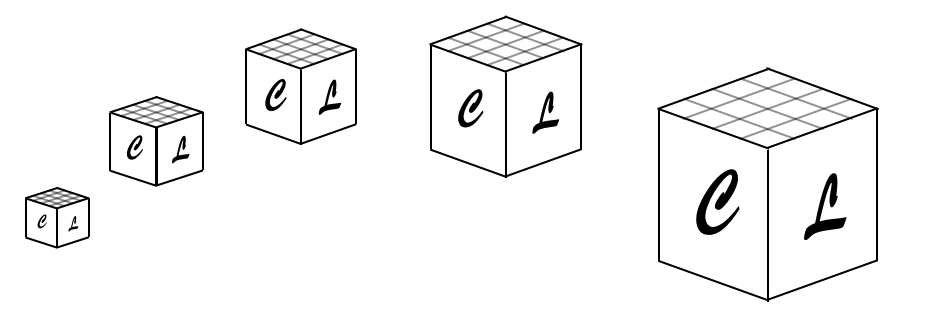}
\end{figure}

\begin{center}

\vspace*{-0.25cm}

{\fontsize{72.0}{0} {\bf ${\mathcal C}$osmo${\mathcal L}$attice}}
\\[0.30cm]
{\fontsize{21.3}{0} \it 
A modern code for lattice simulations of scalar\\and gauge field dynamics in an expanding universe}
\vspace*{0.5cm}\\
{\bf \Large -- User Manual --}\\[0.15cm]
 \CL {\tt v1.1}, {\it \today}
\\{\footnotesize ({\it Previous Versions: {\tt v1.0}, February 2, 2021})}
\\[2.15cm]

{\Large  Daniel~G.~Figueroa\footnote[1]{daniel.figueroa@ific.uv.es}}\\
\addressIFIC
\\[0.5cm]
{\Large {\rm Adrien Florio}\footnote[2]{adrien.florio@stonybrook.edu}}\\
\addressSUNY
\\[0.5cm]
{\Large {\rm Francisco Torrenti}\,\footnote[3]{f.torrenti@unibas.ch}}\\ \addressUNIBAS
\\[0.5cm]
{\Large {\rm Wessel Valkenburg}\,\footnote[4]{wessel.valkenburg@epfl.ch}}\\
\addressEPFL

\vspace{.3cm}
\end{center}

\newpage

\title{{\bf${\mathcal C}$osmo${\mathcal L}$attice}: {\it User Manual}
\\ \vspace*{-2mm} {\large {\it Version 1.1, \today}} \\
\vspace*{-2mm}{\footnotesize ({\it Previous Versions: {\tt v1.0}, February 2, 2021})}}

\author{Daniel G. Figueroa\,$^1$, Adrien Florio\,$^2$, Francisco Torrent\'i\,$^3$ and Wessel Valkenburg\,$^4$\vspace*{0.35cm}\\
$^1${\it
Instituto de F\'isica Corpuscular (IFIC), CSIC-Universitat de Valencia, Spain.}\\
$^{2}${\it Center for Nuclear Theory, Department of Physics and Astronomy}\\{\it Stony Brook University, New York 11794, USA.} \\
$^{2,4}${\it Institute of Physics, Laboratory of Particle Physics and Cosmology (LPPC),} \vspace*{-0.1cm}\\{\it \'Ecole Polytechnique F\'ed\'erale de Lausanne (EPFL), CH-1015 Lausanne, Switzerland.}\\
$^3${\it Department of Physics, University of Basel, Klingelbergstr. 82, CH-4056 Basel, Switzerland.}
}

\date{}
\maketitle

\begin{abstract}
This is the user manual for \CLns, a modern package for lattice simulations of the dynamics of interacting scalar and gauge fields in an expanding universe. \CL incorporates a series of features that makes it very versatile and powerful: $i)$ it is written in {\tt C++} fully exploiting the object oriented programming paradigm, with a modular structure and a clear separation between the physics and the technical details, $ii)$ it is {\tt MPI}-based and uses a discrete Fourier transform parallelized in multiple spatial dimensions, which makes it specially appropriate for probing scenarios with well-separated scales, running very high resolution simulations, or simply very long ones, $iii)$ it introduces its own symbolic language, defining field variables and operations over them, so that one can introduce differential equations and operators in a manner as close as possible to the continuum, $iv)$ it includes a library of numerical algorithms, ranging from $\mathcal{O}(\delta t^2)$ to $\mathcal{O}(\delta t^{10})$ methods, suitable for simulating global and gauge theories in an expanding grid, including the case of `self-consistent' expansion sourced by the fields themselves. Relevant observables are provided for each algorithm (e.g.~energy densities, field spectra, lattice snapshots) and we note that remarkably all our algorithms for gauge theories always respect the Gauss constraint to machine precision. In this manual we explain how to obtain and run \CL in a computer (let it be your laptop, desktop or a cluster). We introduce the general structure of the code and describe in detail the basic files that any user needs to handle. We explain how to implement any model characterized by a scalar potential and a set of scalar fields, either singlets or interacting with $U(1)$ and/or $SU(2)$ gauge fields. \CL is publicly available at \href{http://www.cosmolattice.net}{\color{blue} http://www.cosmolattice.net}.
\end{abstract}

\tableofcontents

\newpage

\section*{About This Manual}
\label{sec:ThisManual}
\addcontentsline{toc}{section}{About this manual}

~~~~~This document is the manual for the code \CLns, a modern package for lattice simulations of the dynamics of interacting scalar and gauge fields in an expanding universe. The present document complements Ref.~\cite{Figueroa:2020rrl} -- {\it The art of simulating the early Universe} -- , a dissertation meant as a primer on lattice techniques for the simulation of scalar-gauge field theories. The theoretical basis for the equations implemented in \CL is described in full detail in Ref.~\cite{Figueroa:2020rrl}. The focus of this manual is, instead, on explaining \textit{how} to use \CLns. Therefore, we will just quote results from \cite{Figueroa:2020rrl} whenever needed, without explaining meticulously their origin and/or derivation. In that regard, we recommend to any user without previous experience on lattice techniques to read first sections 2, 3 and 4 of Ref.~\cite{Figueroa:2020rrl}, in order to have a better understanding of the basic lattice concepts used in \CLns. In particular, Section 2 of~\cite{Figueroa:2020rrl} reviews the formulation of scalar and gauge field interactions in the continuum, both in flat and {\it Friedmann-Lema\^itre-Robertson-Walker} (FLRW) backgrounds, whereas Section 3 of~\cite{Figueroa:2020rrl} introduces the basic tools for discretizing any bosonic field theory in an expanding background, including a discussion on {\it lattice gauge-invariant} techniques for both {\it Abelian} and {\it non-Abelian} gauge theories. Section 4 of Ref.~\cite{Figueroa:2020rrl} describes the formulation and properties of the evolution algorithms used in \CL to simulate the dynamics of interacting singlet scalar fields. Furthermore, if the user is inexperienced in the simulation of the dynamics of Abelian and non-Abelian gauge theories, we recommend them to also read sections 5 and 6 of Ref.~\cite{Figueroa:2020rrl}, where we discuss in detail different evolution algorithms implemented in \CL specialized for gauge theories. This manual is self-contained, so it is not really mandatory to read all the mentioned sections in Ref.~\cite{Figueroa:2020rrl} in order to follow it. However, it will definitely help users to grasp better the motivations sustaining the equations and techniques used in \CLns, particularly to those users with less experience in lattice simulations.

\begin{center}
-----------------
\end{center}

The present manual is structured as follows. In Section \ref{sec:Overview} we provide a short overview on \CLns. In Section \ref{sec:Introduction} we briefly review physically-motivated scenarios suitable for lattice simulations, we introduce the file structure of the code, and we present the basic field equations that \CL is ready to solve. In Section \ref{sec:Basics} we review basic but necessary concepts on lattice techniques, including lattice gauge invariant field theory formulations. In Section \ref{sec:MyFirstModelScalars} we present all necessary steps to run an example model with interacting scalar (singlet) fields. We introduce the important concept of \textit{program variables}, corresponding to appropriate re-scaling of the scalar field amplitudes and space-time variables, so that a given scenario can be simulated in a computer. We also explain there the basic commands to compile and run \CLns, how to define a new model, how to introduce the different parameters of the simulation, and how to interpret the output produced by the code. In Section \ref{sec:MyFirstModelGauge} we expand over the previous section, explaining how to use \CL to simulate  models with scalar fields interacting among themselves and charged under U(1) or SU(2) gauge symmetries, and hence coupled to Abelian and non-Abelian gauge fields. In Section \ref{sec:graybox} we elaborate on the physics captured by \CLns, including details on how fields are initialized, how the equations of motion are solved, and what are the relevant observables that can be measured in a run. In Section \ref{sec:UsefulFeatures} we describe some of the technical features implemented in \CLns, including its parallel support and back-up options. The manual is also complemented with few appendices: Appendix \ref{app:Installation} describes the installation process in detail, of both \CL itself and of different tools and libraries that it uses (some of them compulsory, others optional). Appendices \ref{App:TableParameters}-\ref{app:ImplemFunc} contain respectively a list of the most relevant parameters, variables, functions, and CMake flags, used in \CLns.

\newpage
\section*{Quick installation and execution}
\label{sec:QuickInstallation}
\addcontentsline{toc}{section}{Quick installation and execution}

\medskip
\medskip

~~~~ Here we provide a brief description of the steps to download, compile and run a job with \CLns. As an example we load a model called {\tt lphi4}, with input parameters specified in the file {\tt lphi4.in} (details of this model and of its parameters can be found in Section~\ref{sec:MyFirstModelScalars}, here we just use these files as a demonstration example). An appropriate version of FFTW needs to be installed in order to handle the code's Fourier transforms, either for serial jobs run on a single processor, or for parallelized jobs run on multiple processors. For serial jobs, that is all it is needed. For parallelized jobs, in addition, the appropriate MPI compiler needs also to be installed. Optionally, if one wants to print two- or three-dimensional snapshots of a simulation, or simply to stop/restart a run at a certain time, the HDF5 library also needs to be installed. Also optionally, the simulations can be parallelized in more than one spatial dimension, to speed up jobs of long duration or to run massive simulations. This requires the library PFFT (parallel Fast Fourier transforms) \cite{Pi13} to be installed. For further details on the installation and use of the code and the libraries, please see Appendix~\ref{app:Installation}.\\

\medskip
\medskip

\noindent\textbf{Requirements:} \texttt{CMake} v3.0 (or above), \texttt{fftw3}, \texttt{g++} v$5.0$ (or above) or \texttt{clang++} v$3.4$ (or above). For parallel use: \texttt{MPI}. {\it Optional}: \texttt{HDF5} and \texttt{PFFT}.

\medskip
\medskip
\medskip
\medskip

\noindent\textbf{Download:} You can download \CL from \href{http://www.cosmolattice.net}{\color{blue} http://www.cosmolattice.net}\,, or alternatively use {\color{gray} \tt git clone} as indicated below.
\vspace*{0.8cm}

\noindent\textbf{Personal computer: }\vspace*{-0.4cm}
\begin{shell-sessioncode}
git clone https://github.com/cosmolattice/cosmolattice
cd cosmolattice                  # Enter into main code folder
mkdir build                      # Create a new directory
cd build                         # and go inside it.
cmake -DMODEL=lphi4 ../          # Selects model phi^4 for serial runs
                                 # OR
cmake -DMODEL=lphi4 -DMPI=ON ../ # Selects model phi^4 and activates parallelization
make cosmolattice                # Compiling
./lphi4 input=../src/models/parameter-files/lphi4.in # Executes serial run (input parameter file 'lphi4.in')
                                 # OR
mpirun -n 8 lphi4 input=../src/models/parameter-files/lphi4.in # Parallelized run on 8 cores (input file 'lphi4.in')
\end{shell-sessioncode}
\vspace*{0.4cm}

\noindent\textbf{High-Performance cluster:} Assuming it uses Environment Modules, otherwise one needs to see how to load the required libraries.\vspace*{-0.4cm}
\begin{shell-sessioncode}
git clone https://github.com/cosmolattice/cosmolattice
cd cosmolattice   # Enter into main code folder
mkdir build       # Create a new directory
cd build          # and go inside it.
module list       # Displays down the names of the libraries you need.
module load CMAKE # Here we will call C++, FFTW3, CMAKE and MPI.
module load C++   # The order is important, C++ before MPI before FFTW3
module load MPI   # Needed to run parallelized simulations
module load FFTW3 # Needed for Fourier transforms in parallelized simulations

cmake -DMODEL=lphi4 ../          # Selects model phi^4 for serial runs
                                 # OR
cmake -DMODEL=lphi4 -DMPI=ON ../ # Selects model phi^4 and activates parallelization
make cosmolattice                # Compiling
# Now you can run lphi4. How you do it depends on the cluster.
\end{shell-sessioncode}
\medskip
\medskip

\noindent \textit{Optional:} Install \texttt{PFFT} and simply add the flag \texttt{-DPFFT=ON} when you call \texttt{CMake}:\vspace*{-0.4cm}

\begin{shell-sessioncode}
cmake -DMODEL=lphi4 -DMPI=ON -DPFFT=ON ../ # For parallel use
\end{shell-sessioncode}
\medskip

\noindent \textit{Optional:} Install \texttt{PFFT} and simply add the flag \texttt{-DPFFT=ON} when you call \texttt{CMake}:
\vspace*{-0.4cm}

\begin{shell-sessioncode}
cmake -DMODEL=lphi4 -DHDF5=ON ../          # For serial use
cmake -DMODEL=lphi4 -DMPI=ON -DHDF5=ON ../ # For parallel use (also works with PFFT)
\end{shell-sessioncode}
\medskip
\medskip

\noindent\textbf{Installing \texttt{fftw3}, \texttt{PFFT} and \texttt{HDF5}}:
\vspace*{-0.5cm}

\begin{shell-sessioncode}
cd .../dependencies/
bash fetchall.sh MyLibs # Install everything. You are done if you do that.
bash fftw3.sh MyFFTW3 --parallel # Install only fftw3 in MyFFTW3. Can remove --parallel.
bash pfft.sh MyPFFT MyFFTW3 # Install only PFFT in MyPFFT. MyFFTW3 is the path to fftw3.
bash hdf5.sh MyHDF5 --parallel   # Install only hdf5 in MyHDF5. Can remove --parallel.
\end{shell-sessioncode}

\medskip
\medskip

\noindent \textbf{Note:} Some of the above commands are only indicative, as they can change from machine to machine. For more detailed explanations about the installation of \CL and the libraries it uses (or can use), see Appendix~\ref{app:Installation}.

\newpage

\section*{Conventions and Notation}
\label{sec:Conventions}
\addcontentsline{toc}{section}{Conventions}

Unless otherwise specified, we use the following conventions throughout the document:

\begin{itemize}

\item We use natural units $c=\hbar=1$ and metric signature $(-1,+1,+1,+1)$.

\item We use interchangeably the Newton constant $G$, the full Planck mass $M_p \simeq 1.22\cdot 10^{19}$ GeV, and the reduced Planck mass $m_p \simeq 2.44\cdot 10^{18}$ GeV, all related through $M_p^2 = 8\pi m_p^2 = 1/G$.

\item Latin indices $i, j, k, ... = 1,2,3$ are reserved for spatial dimensions, and Greek indices $\alpha, \beta, \mu, \nu,... = 0,1,2,3$ for space-time dimensions. We use the {\it Einstein convention} of summing over repeated indices {\it only in the continuum}. However, {\tt in the lattice, unless stated otherwise, repeated indices do not represent summation}.

\item We consider a flat FLRW metric $ds^2 = -a^{2\alpha}(\eta)d\eta^2 + a^2(\eta) \, \delta_{ij} \, dx^i dx^j$ with $\alpha \in \mathcal{R}e$ a constant chosen conveniently in each scenario. For $\alpha = 0$, $\eta$ denotes the {\it coordinate time} $t$, whereas for $\alpha = 1$, $\eta$ denotes the {\it conformal time} $\tau = \int {dt'\over a(t')}$. For arbitrary $\alpha$, we will refer to the time variable as the {\it $\alpha$-time}.

\item We reserve the notation $()^{\cdot}$ for derivatives with respect to cosmic time with $\alpha = 0$, and $()'$ for derivatives with respect to $\alpha$-time with arbitrary $\alpha$.

\item Physical momenta are represented by ${\bf p}$, comoving momenta by ${\bf k}$, the $\alpha$-time Hubble rate is given by $\mathcal{H} = a'/a$, whereas the physical Hubble rate is denoted by $H = \mathcal{H}|_{\alpha = 0}$.

\item Cosmological parameters are fixed to the CMB values given in \cite{Aghanim:2018eyx,Akrami:2018odb}.

\item Our Fourier transform convention in the continuum is given by
\begin{eqnarray}\label{eq:FTcont}
f({\bf x}) = \frac{1}{(2 \pi)^3} \int d^3 {\bf k} \, f({\bf k}) \, e^{+i {\bf k} {\bf x}}\, ~~~ \Longleftrightarrow ~~~  f({\bf k}) = \int d^3 {\bf x} \, f ( {\bf x}) \, e^{-i {\bf k} {\bf x}}\,.\nonumber
\end{eqnarray}

\item Our discrete Fourier transform (DFT) is defined by
\begin{eqnarray}\label{eq:FTdiscreteAux}
f({\bf n}) \equiv {1\over N^3}\sum_{\tilde n} e^{+i{2\pi\over N} {\bf \tilde n n}} f({\bf \tilde n}) ~~~~ \Leftrightarrow ~~~~  f({\bf \tilde n}) \equiv \sum_{n} e^{-i{2\pi\over N} {\bf n \tilde n} }f({\bf n})\,.\nonumber
\end{eqnarray}

\item A scalar field living in a generic lattice site $n = (n_o,\bn) = (n_o,n_1,n_2,n_3)$, i.e.~$\phi_n = \phi(n)$, will be simply denoted as $\phi$. If the point is displaced in the $\mu-$direction by one unit lattice spacing/time step, $n + \hat\mu$, we will then use the notation $n+\mu$ or simply by $+\mu$ to indicate this, so that the field amplitude in the new point is expressed as $\phi_{+\mu} \equiv \phi(n+\hat\mu)$.

\item When representing explicitly gauge fields in the lattice, we will automatically understand that they live in the middle of lattice points, i.e.~$A_{\mu} \equiv A_{\mu}(n+{1\over2}\hat\mu)$. It follows then that e.g.~$A_{\mu,+\nu} \equiv A_{\mu}\big(n + {1\over2}\hat\mu +  \hat\nu\big)$. In the case of links, we will use the notation $U_\mu \equiv U_{\mu,n} \equiv U_\mu(n+{1\over2}\hat\mu)$, and hence $U_{\mu,\pm\nu} = U_{\mu,n\pm\nu} \equiv U_\mu(n + {1\over2}\hat\mu \pm \hat\nu)$.

\item Even though the {\it lattice spacing} $\dx$ and the {\it time step} $\delta t$ do not need to be equal, we will often speak loosely of corrections of order $\mathcal{O}(\dx)$, independently of whether we are referring to the lattice spacing or the time step (the latter is actually always forced to be smaller than the former).

\end{itemize}

\newpage

\section{Overview}
\label{sec:Overview}

~~~~ \CL is a program designed to simulate the evolution of interacting 
fields in an expanding universe. It can simulate (so far) the dynamics of $i)$ {\it global theories}, $ii)$ {\it Abelian $U(1)$ gauge theories}, and $iii)$ {\it non-Abelian $SU(2)$ gauge theories}, i.e.~it can handle scenarios 
including singlet scalar fields, scalar fields charged under a $U(1)$ and/or $SU(2)$ gauge symmetry, and the corresponding Abelian and/or non-Abelian gauge vector fields. \CL can simulate the dynamics of such fields either in a flat space-time background, or in a homogeneous and isotropic (spatially flat) expanding background. In the latter case the fields can evolve either over a fixed background (e.g.~with a power-law scale factor), or {\it self-consistently} with the fields determining themselves the expansion rate of the universe. In all cases \CL provides symplectic integrators, with accuracy ranging from $\mathcal{O}(\delta t^2)$ up to $\mathcal{O}(\delta t^{10})$. Appropriate observables are also provided for each algorithm, like the energy density components of each field, their relevant spectra, or dynamical constraints. Our algorithms conserve energy up to the accuracy set by the order of the evolution algorithm, reaching even machine precision in the case of the highest order integrators. Notably, our algorithms for gauge theories, either Abelian or non-Abelian, always respect the Gauss constraint to machine precision, independently of the integrator and even in the case of self-consistent expansion.

\CL is written in C++, and fully exploits the {\it object oriented  programming} nature of this language, with a modular structure that separates well all the ingredients involved. This allows \CL to have a clear separation between the physics and the technical implementation details. The code is designed so that we can simulate a given scenario with different parameters, without requiring to re-compile each time. More importantly, the code allows for an easy implementation of new models with either singlet or gauge interactions. \CL is fully parallelized using {\it Message Passing Interface} (MPI), and uses a discrete Fourier Transform parallelized in multiple spatial dimensions. This makes it ideal for probing physical problems with well-separated mass/length scales, running very high resolution simulations, or simply shortening the running time of long simulations. \CL is actually a general package that defines field variables and their operations, by introducing its own symbolic language. Once you become familiar with the basic ‘vocabulary’ of \CL language, editing the code or implementing your own model (resembling how you would write it in the continuum), should become a simple task.

\CL can be used at multiple levels of complexity. For instance, a {\it basic level} user, say someone with no experience at all in parallelization techniques and with little to no experience programming in general, will be able to run fully parallelized simulations of their favourite models (say using hundreds of processors in a cluster), while being completely oblivious to the technical details of the algorithm implementation or parallelization. \CL automatically prints a collection of relevant observables, such as volume averages, field spectra, or dynamical constraints, which can be used to monitor the evolution of the system. However, an {\it intermediate level} user, say with certain programming experience, may want however to modify the type of output \CL generates, change the initial condition routine (e.g.~turn it into a Monte-Carlo generator for thermal configurations), or even design their own evolution algorithms and add these to the family of integrators available in \CLns. Finally, {\it advanced level} users may want to edit the hard core inner parts of \CLns, in an attempt to understand or to improve the most advanced technical aspects of the code. This could include e.g.~the handling of the parallelization, which is something that typically will remain as a black box for the majority of users. What type of user you want to be is entirely up to you: it simply depends on your programming expertise (or lack of it), and on your will to learn how the code works internally.

We have developed \CL with the intention of providing a new up-to-date, relevant numerical tool for the scientific community working in the physics of the early universe. Presently \CL is able to simulate canonical scalar-gauge field theories in an expanding universe, and as such, it is already an extremely useful tool for many physics scenarios. However, we conceive \CL as an evolving package that we plan to upgrade constantly, for instance by incorporating new evolution algorithms or new modules dedicated to specialized tasks. In that regard, we plan to further develop new modules for \CLns, that will be made publicly available in due time. For example, to mention just a few, we plan to add the computation of gravitational waves, an initializer for the creation a cosmic defects, the handing of theories with non-canonical kinetic terms, and the inclusion of axion-couplings to $F\tilde F$ of a gauge sector. The interested reader can check Section 9 of Ref.~\cite{Figueroa:2020rrl} for further details on the aspects we plan to implement in the future.


\CL is freely available to anyone who wants to use or modify it, as long as you give us credit for its creation. If you have any questions or comments about \CLns, please email us (you can find our emails in the front page of this document). We would love to hear how the program performs for you, and we will be happy to assist you with any question you might have, bug reports, suggestions for future improvements, etc. We welcome everyone to use \CL for their own projects. Whenever using \CL in your research, no matter how much (or little) you modify the code, please cite this manual together with our dissertation {\it The art of simulating the early Universe}~\cite{Figueroa:2020rrl}, where the basic algorithms and techniques implemented in \CL are explained.

\begin{center}
You can download \CL at any time from:\\
\href{http://www.cosmolattice.net}{\color{blue} http://www.cosmolattice.net}
\end{center}

\begin{mdframed}
{\bf Note -.} If you would like to help developing some aspect of \CLns, or even implement your own modules with some new functionality we have not envisaged, please contact us and let us know about your idea(s). \CL introduces a natural language describing fields and operations between them, so it is a natural platform to implement new libraries (related or not to cosmology).
\end{mdframed}


\section{Introduction to \CL} \label{sec:Introduction}

~~~~~Here we review first briefly, in Section~\ref{subsec:EU}, physically-motivated early universe scenarios, suitable for lattice simulations. We introduce the reader to \CL in Section~\ref{subsec:Purpose}, where we discuss its purpose and capabilities, and introduce its file structure. In Section~\ref{subsec:BasicEOM} we present the basic field equations that \CL is ready to solve. A reader familiar that has read Section~1 of Ref.~\cite{Figueroa:2020rrl}, can skip Section~\ref{subsec:EU} and jump directly into Sections~\ref{subsec:Purpose} or \ref{subsec:BasicEOM}.

\subsection{The {\it Numerical} Early Universe}
\label{subsec:EU}

~~~~~The phenomenology of high energy physics in the early universe is vast and very rich, and it is often characterized by non-linear dynamics. The {\it numerical early universe}, i.e.~the study with numerical techniques of high energy non-linear field theory phenomena in the early universe, is an emerging field increasingly gaining relevance, especially as a methodology to assess our experimental capabilities to constrain the physics of this epoch. The details of nonlinear phenomena are often too difficult, when not impossible, to be described by analytic means. In order to fully understand the non-linearities developed in the dynamics of a given scenario, the use of numerical techniques becomes mandatory. The outcome from non-linear early universe phenomena represents, more and more, an important perspective in determining the best observational strategies to probe the unknown physics from this era. It is therefore crucial to develop numerical techniques, as efficient and robust as possible, to simulate these phenomena. Numerical algorithms developed for this purpose must satisfy a number of physical constraints (e.g.~energy conservation), and keep the numerical integration errors under control. It is actually useful to develop as many techniques as possible, to validate and double check results from simulations. Only in this way, we will achieve a certain robustness in the predictions of the potentially probeable implications from these phenomena.

It is precisely because we recognize the importance of the above circumstances, that we have created \CLns, a modern code for lattice simulations of scalar-gauge field theories in an expanding universe. \CL allows for the simulation of the evolution of interacting (singlet) scalar fields, charged scalar fields under U(1) and/or SU(2) gauge groups, and the corresponding associated Abelian and/or non-Abelian gauge fields. \CL is capable of solving the evolution of such field $dof$'s whenever their dynamics develop large occupation numbers $n_k \gg 1$, so that the fields' quantum nature can be neglected. When such circumstance is met, classical field theory can be used as a powerful tool to solve complicated field dynamics, including the case when non-linear interactions characterize the field evolution, non-perturbative particle production effects are present, or out-of-equilibrium field distributions are developed.

\CL is particularly suitable for solving the non-linear dynamics of post-inflationary scenarios like e.g.~(p)reheating. Preheating scenarios are actually characterized by non-perturbative particle production mechanisms, which cannot be described with standard perturbative quantum field theory techniques: particle species are typically created far away from thermal equilibrium with exponentially growing occupation numbers, so that when they eventually ‘backreact’ onto the system, the dynamics become non-linear from that moment onward. This is the
case of parametric resonance of scalar fields\footnote{If the species created are fermions, non-perturbative particle production can also take place, see e.g.~\cite{Greene:1998nh,Greene:2000ew,Peloso:2000hy,Berges:2010zv}, but no resonance can be developed due to Pauli blocking.} either during preheating~\cite{Traschen:1990sw,Kofman:1994rk,Shtanov:1994ce,Kaiser:1995fb,Khlebnikov:1996mc,Prokopec:1996rr,Kofman:1997yn,Greene:1997fu,Kaiser:1997mp,Kaiser:1997hg} or in other circumstances like the non-perturbative decay of the curvaton~\cite{Enqvist:2008be,Enqvist:2012tc, Enqvist:2013qba, Enqvist:2013gwf} or of the Higgs field of the Standard Model~\cite{Enqvist:2013kaa,Enqvist:2014tta,Figueroa:2014aya,Kusenko:2014lra,Figueroa:2015rqa,Enqvist:2015sua,Figueroa:2016dsc}. Similarly, an explosive decay of a field condensate can take place in theories with flat-directions~\cite{Olive:2006uw,Basboll:2007vt,Gumrukcuoglu:2008fk}. In theories with spontaneous symmetry breaking, tachyonic effects can also lead to non-perturbative and out-of-equilibrium particle production, like in tachyonic preheating scenarios after hybrid inflation ~\cite{Felder:2000hj,Felder:2001kt,Copeland:2002ku,GarciaBellido:2002aj}, or in preheating after hilltop inflation~\cite{Antusch:2015nla,Antusch:2015vna,Antusch:2015ziz}. Preheating effects have also been studied in models with gravitationally non-minimal coupled fields~\cite{Bassett:1997az,Tsujikawa:1999jh,Tsujikawa:1999iv,Tsujikawa:1999me,Ema:2016dny,Crespo:2019src,Crespo:2019mmh}, and in particular, recently, in multi-field inflation scenarios~\cite{DeCross:2015uza,DeCross:2016fdz,DeCross:2016cbs,Nguyen:2019kbm,vandeVis:2020qcp}.

The presence of gauge fields has also been considered in multiple scenarios. For instance, if a field enjoys a shift-symmetry, then a coupling $\phi F\tilde F$ between such field and some gauge sector is allowed. Particle production in axion-inflation scenarios, where such topological interaction is present, provide an extremely efficient mechanism to reheat the universe, leading to potentially observable phenomenology~\cite{Adshead:2015pva,Adshead:2016iae,Figueroa:2017qmv,Adshead:2018doq,Cuissa:2018oiw,Adshead:2019lbr,Figueroa:2019jsi,Adshead:2019igv}. Interactions between a singlet inflaton and an Abelian gauge sector, via $f(\phi)F^2$, or a non-Abelian $SU(2)$ gauge sector, via $f(\phi)\rm{Tr \,G^2}$, have also been explored in the context of preheating~\cite{Deskins:2013lfx,Adshead:2017xll}. In Hybrid preheating scenarios, the excitation of gauge fields have also been addressed extensively, both for Abelian and non-Abelian scenarios, obtaining a very rich phenomenology~\cite{Rajantie:2000nj,Copeland:2001qw,Smit:2002yg,GarciaBellido:2003wd,Tranberg:2003gi,Skullerud:2003ki,vanderMeulen:2005sp,DiazGil:2007dy,DiazGil:2008tf,Dufaux:2010cf,Tranberg:2017lrx}. Preheating via parametric resonance with an inflaton charged under a gauge symmetry, has been also studied in detail in~\cite{Lozanov:2016pac,Figueroa:2020rrl}, both for Abelian $U(1)$ and non-Abelian $SU(2)$ gauge groups. A natural realization of an inflationary set-up where the inflaton is charged under a gauge group is Higgs-Inflation~\cite{Bezrukov:2007ep,Bezrukov:2010jz}, where the SM Higgs is the inflaton. There the electroweak gauge bosons can experience parametric excitation effects during the oscillations of the Higgs after inflation~\cite{Bezrukov:2008ut,GarciaBellido:2008ab,Figueroa:2009jw,Figueroa:2014aya,Repond:2016sol,Ema:2016dny,Sfakianakis:2018lzf}. If the SM Higgs is rather a spectator field during inflation, the post-inflationary explosive non-perturbative decay of the Higgs into SM fields has also been considered~\cite{Figueroa:2015rqa,Enqvist:2015sua,Kohri:2016wof,Figueroa:2017slm,Ema:2017loe}.

The techniques developed for studying nonlinear dynamics of classical fields are actually common to many other non-linear problems in the early universe, like the production of stochastic gravitational wave backgrounds by parametric effects~\cite{Khlebnikov:1997di,Easther:2006gt,Easther:2006vd,GarciaBellido:2007af,Dufaux:2007pt,Dufaux:2008dn,Dufaux:2010cf,Zhou:2013tsa,Bethke:2013aba,Bethke:2013vca,Antusch:2016con,Antusch:2017flz,Antusch:2017vga,Liu:2018rrt,Figueroa:2017vfa,Fu:2017ero,Lozanov:2019ylm,Adshead:2019lbr,Adshead:2019igv,Armendariz-Picon:2019csc} (for a review on stochastic backgrounds see~\cite{Caprini:2018mtu}), the dynamics of phase transitions~\cite{Hindmarsh:2001vp,Rajantie:2000fd,Hindmarsh:2001vp,Copeland:2002ku,GarciaBellido:2002aj,Figueroa:2017hun,Brandenburg:2017neh,Brandenburg:2017rnt,Figueroa:2019jsi} and their emission of gravitational waves~\cite{Hindmarsh:2013xza,Hindmarsh:2015qta,Hindmarsh:2017gnf,Cutting:2018tjt,Cutting:2019zws,Pol:2019yex,Cutting:2020nla,Di:2020ivg} (for a review on early universe phase transitions see~\cite{Hindmarsh:2020hop}), cosmic defect formation~\cite{Hindmarsh:2000kd,Rajantie:2001ps,Rajantie:2002dw,Donaire:2004gp,Dufaux:2010cf,Hiramatsu:2012sc,Kawasaki:2014sqa,Fleury:2016xrz,Moore:2017ond,Lozanov:2019jff}, their later evolution~\cite{Vincent:1997cx,Bevis:2006mj,Hindmarsh:2014rka,Daverio:2015nva,Lizarraga:2016onn,Hindmarsh:2018wkp,Eggemeier:2019khm,Hindmarsh:2019csc,Gorghetto:2018myk,Gorghetto:2020qws,Hindmarsh:2021mnl} and gravitational wave emission~\cite{Dufaux:2010cf,Figueroa:2012kw,Hiramatsu:2013qaa,Figueroa:2020lvo,Gorghetto:2021fsn}, axion-like field dynamics~\cite{Kolb:1993hw,Kitajima:2018zco,Amin:2019ums,Buschmann:2019icd,Hindmarsh:2019csc,Gorghetto:2018myk,Gorghetto:2020qws,Hindmarsh:2021vih,Buschmann:2021sdq}, oscillon dynamics \cite{Amin:2011hj,Zhou:2013tsa,Antusch:2016con,Antusch:2017flz,Lozanov:2017hjm,Amin:2018xfe,Liu:2018rrt,Kitajima:2018zco,Lozanov:2019ylm,Antusch:2019qrr,Kasuya:2020szy}, the post-inflationary evolution of the equation of state \cite{Podolsky:2005bw,Lozanov:2016hid,Figueroa:2016wxr,Lozanov:2017hjm,Krajewski:2018moi,Antusch:2020iyq}, moduli dynamics~\cite{Giblin:2017wlo,Amin:2019qrx}, etc. These techniques can also be used in applications of interest not only to cosmology, but also to other high energy physics areas. For example, classical-statistical simulations have been used to compute quantities such as the sphaleron-rate \cite{Philipsen:1995sg,Ambjorn:1995xm,Arnold:1995bh,Arnold:1996dy,Arnold:1997yb,Moore:1997sn,Bodeker:1998hm,Moore:1998zk,Moore:1999fs,Bodeker:1999gx,Arnold:1999uy, Tang:1996qx,Ambjorn:1997jz,Moore:2000mx,DOnofrio:2012phz,DOnofrio:2015gop}, and to study the Abelian~\cite{Buividovich:2015jfa,Buividovich:2016ulp,Figueroa:2017hun,Figueroa:2019jsi,Mace:2019cqo,Mace:2020dkp} and non-Abelian~\cite{Akamatsu:2015kau} dynamics associated to the chiral anomaly. They have also been used to study spectral quantities~\cite{Boguslavski:2018beu,Schlichting:2019tbr}, and some properties of the quark-gluon plasma~\cite{Laine:2009dd,Laine:2013lia,Panero:2013pla,Boguslavski:2020tqz}.

In summary the study of non-linear dynamics of early universe high-energy phenomena, represents an important emerging and phenomenologically rich field, which will help to determine best our observational strategies to probe the unknown physics from this era. Its study requires the development of appropriate numerical techniques, as efficient and robust as possible, to simulate such phenomena.

\subsection{Purpose, capabilities and structure of \CL}
\label{subsec:Purpose}

~~~~A number of public packages for lattice simulations have appeared over the years, mostly dedicated to the simulation of interacting scalar fields, like \texttt{LatticeEasy}~\cite{Felder:2000hq}, \texttt{ClusterEasy}~\cite{Felder:2007nz},  \texttt{Defrost}~\cite{Frolov:2008hy}, {\tt CUDAEasy}~\cite{Sainio:2009hm}, \texttt{HLattice}~\cite{Huang:2011gf},  \texttt{PyCOOL}~\cite{Sainio:2012mw} and \texttt{GABE}~\cite{Child:2013ria}, which use finite difference techniques and a FLRW background metric. Other packages are suitable for full general relativistic evolution, like {\tt GABERel}~\cite{Giblin:2019nuv} or the recent {\tt GRChombo}~\cite{Andrade:2021rbd}. Others use pseudo-spectral techniques, like {\tt PSpectRe}~\cite{Easther:2010qz} and {\tt Stella}~\cite{Amin:2018xfe}. \texttt{Latfield2}~\cite{Daverio:2015ryl}, on the other hand, is a library in C++ designed to simplify writing parallel codes for solving partial differential equations, and hence can be used for field dynamics as long as the users implement their own lattice equations of motion. Finally, \texttt{GFiRe}~\cite{Lozanov:2019jff} is a package dedicated to Abelian gauge theories, and even though the code itself has not been made publicly available yet, their algorithm is clearly spelled out in their publication.

\CL differs from the above codes in a number of aspects. To begin with, \CL is ready to simulate not only the evolution of global scalar and Abelian $U(1)$ gauge theories, but also non-Abelian $SU(2)$ gauge theories. More importantly, \CL has been designed as a `platform' to implement any system of dynamical equations suitable for discretization on a lattice, i.e.~\CL is not just meant as a code for one type of simulation, but it is rather a more evolved concept. It is a package that introduces its own \textit{symbolic language}, by defining field variables and operations over them. Once the user becomes familiar with the basic `vocabulary' of the new language, they can write their own code: let it be for the time evolution of the relevant field variables in a given model of interest, or for some other operation, like e.g.~a Monte-Carlo generator for thermal configurations. One of the main advantages of \CL is that it clearly separates the $physics$ (e.g.~definition of the field content, operations between fields, evolution equations, etc) from the $implementation~details$ (e.g.~parallelization aspects, Fourier transforms, etc). For example, let us consider a beginner user with little experience in programming, and with no experience at all in parallelization techniques. With \CLns, they will be able to run a fully parallelized simulation of their favourite model (say using hundreds of processors in a cluster), while being completely oblivious to the technical details. They will just need to write a basic \textit{model file} in the language of \CLns, containing the details of the model being simulated. If, on the contrary, the user is rather an experienced programmer and wants to look inside the core routines of \CLns, and modify, say the MPI-implementation, they can always do so, and perhaps even contribute to their improvement.

\CL comes with symbolic scalar, complex and $SU(2)$ algebras, which allows to use vectorial and matrix notations without sacrificing performances. The code includes also a {\it library} of basic field theory equations, as well as routines and field-theoretical operations. At the time of writing (Jan~2021), \CL is ready to simulate scenarios including singlet scalar fields, scalar fields charged under a $U(1)$ and/or $SU(2)$ gauge symmetry, and the corresponding Abelian and/or non-Abelian gauge vector fields. Simulations can be done either in a flat space-time background, or in a homogeneous and isotropic (spatially flat) expanding FLRW background. In the latter case the fields can evolve either over a fixed background (e.g.~with a power-law scale factor), or {\it self-consistently}, i.e.~`dictating' themselves the expansion of the universe as sourced by their volume averaged energy and pressure densities. \CL provides symplectic integrators, with accuracy ranging from $\mathcal{O}(\delta t^2)$ up to $\mathcal{O}(\delta t^{10})$, to simulate the
non-linear dynamics of the appropriate fields in comoving two- or three-dimensional lattices. Appropriate observables are also provided for each algorithm, like the energy density components of each field, their relevant spectra, or dynamical constraints. Our algorithms conserve energy up to the accuracy set by the order of the evolution algorithm, reaching even down to machine precision in the case of the highest order integrators. Our algorithms for gauge theories, either Abelian or non-Abelian, respect always (independently of the integrator) the Gauss constraint to machine precision, even in the case of self-consistent expansion.  Furthermore, \CL can use a discrete Fourier Transform parallelized in multiple spatial dimensions~\cite{Pi13}, which makes it a very powerful code for probing physical problems with well-separated scales, running very high resolution simulations, or simply very long ones. All the above aspects constitute clear advantages for using \CL as a platform to implement any scenario desired, over writing your own code from scratch.

\CL is structured in such a way that all the technicalities, such as memory handling or parallelization tasks, remain mostly hidden to a typical user. These are integrated in a set of libraries called {\tt TempLat}, which in principle, a standard user will never need to edit. In {\tt TempLat} we have implemented a new language that can be used to define new fields and operations between them in a natural way. For example, let us imagine that we have two fields {\tt f} and {\tt g} in a lattice and we want to sum them. Without {\tt TempLat}, we would need to explicitly write a loop that sums the amplitudes of both fields at each node of the lattice. Instead, with {\tt TempLat} we can just write {\tt f + g}, and the hidden structure handles the whole operation of summing their values everywhere in the lattice. At the same time, we have developed another collection of libraries called {\tt CosmoInterface}, where all relevant aspects of the physics of scalar-gauge theories are handled, such as the initialization, evolution equations, or relevant field observables. This makes the physics part of the code easy to understand and well separated from technical details. This separation significantly simplifies the process of writing new operations for your own purposes.

\noindent The basic folder tree structure of \CL is the following:
\begin{eqnarray}
{\tt cosmolattice}:
\left\lbrace
\begin{array}{l}
{\it CMakeLists.txt}~ {\rm[+~other~files]}\vspace*{0.2cm}\\
{\tt dependencies} \vspace*{0.2cm}\\
{\tt docs} \vspace*{0.2cm}\\
{\tt src}:\left\lbrace
     \begin{array}{l}
                {\it cosmolattice.cpp}~ {\rm[+~other~files]}\vspace*{0.2cm}\\
                {\tt cmake} \vspace*{0.2cm}\\
                {\tt models} \vspace*{0.2cm}\\
                {\tt tests} \vspace*{0.1cm}\\
                {\tt include}:\left\lbrace
                \begin{array}{l}
                {\tt TempLat}:\left\lbrace
                \begin{array}{l}
                {\it cosmolattice.h}\\
                {\tt fft}\\
                {\tt lattice}\\
                {\tt parallel}\\
                {\tt parameters}\\
                {\tt session}\\
                {\tt util}
                \end{array} \right.\\
                {\tt CosmoInterface}:\left\lbrace
                \begin{array}{l}
                {\it CosmoInterface.h}~ {\rm[+~other~files]}\\
                {\tt definitions}\\
                {\tt evolvers}\\
                {\tt initializers}\\
                {\tt measurements}
                \end{array} \right.
                \end{array} \right.
     \end{array} \right.
\end{array} \right.\nonumber
\end{eqnarray}\vspace*{0.5cm}

\noindent with the content of each folder summarized as:\\

\begin{longtable}{|r|l|}
\hline
{\bf folder name(s)} & {\bf brief description of each folder}\\\hline
\vspace*{-0.45cm}\, & \, \\\hline\vspace*{-0.3cm}\, & \, \\
{\tt dependencies} & scripts to install external libraries \\
{\tt docs} & documentation files\\
{\tt src} & source code (contains {\tt cmake, models, tests, include})\vspace*{-0.3cm}\\ \, & \, \\\hline
\, & \, \vspace*{-0.3cm} \\
{\tt cmake} & files for compilation\\
{\tt models} & model files\\
{\tt tests} & files for testing purposes\\
{\tt include} & libraries for lattice operations ({\tt TempLat}) and field dynamics ({\tt CosmoInterface})\vspace*{-0.3cm}\\
\, & \, \\\hline
\, & \, \vspace*{-0.3cm}\\

{\tt TempLat} & library for lattice operations \tiny{(contains~{\tt fft, lattice, parallel, parameters, session, util})}\\
{\tt CosmoInterface} & library for field dynamics \tiny{(contains {\tt definitions, evolvers, initializers, measurements})}
\vspace*{-0.3cm}\\
\, & \, \\\hline
\, & \, \vspace*{-0.3cm}\\
{\tt fft} & library to handle Fourier transformations\\
{\tt lattice} & library for basic lattice definitions and field operations\\
{\tt parallel} & library for parallelization routines\\
{\tt parameters} & library for parsing parameters from the command-line/files.\\
{\tt session} & library for taking care of initialization and destruction of external libraries \\
{\tt util} & library for basic useful operations \vspace*{-0.3cm}\\
\, & \, \\\hline
\, & \, \vspace*{-0.3cm}\\
{\tt definitions} & library for basic field definitions (EOM terms, energy terms, etc)\\
{\tt evolvers} & library for evolution algorithms\\
{\tt initializers} & library for initialization algorithms\\
{\tt measurements} & library for observables (energy densities, field spectra, etc)\vspace*{-0.3cm}\\\, & \, \\\hline
\end{longtable}

\begin{mdframed}
{\bf Note -.} A remarkable feature of \CL is that the operations and parallelization of \texttt{TempLat} can actually work in an arbitrary number of spatial dimensions $d$, including $d < 3$ and $d>3$. This feature is not exploited in {\tt version 1.0} of \CLns, but the \texttt{TempLat} library has this capability, which makes it a perfect basis for developing future interfaces dealing with field dynamics on lower- or higher-dimensional lattices. Visit \href{https://cosmolattice.net/technicalnotes/}{\color{blue} https://cosmolattice.net/technicalnotes/} to check for additional modules incorporated in successive updated versions of \CL to run in $d \neq 3$ spatial dimensions.
\end{mdframed}

\subsection{Basic Field Equations implemented (so far) in \CL}
\label{subsec:BasicEOM}

Let us consider scalar fields of the type
\begin{eqnarray} \label{eq:ChargedScalars}
\begin{array}{c|c|c}
{\rm Singlet} & U(1){\rm-charged} & SU(2){\rm-charged~Doublet}
\\\hline\vspace{-0.3cm} \, & \, &\\
\phi \in \mathcal{R}e &  \varphi \equiv {1\over\sqrt{2}}(\varphi_0 +i\varphi_1) & \Phi = \left(
\begin{array}{c}
\varphi^{(0)} \\ \varphi^{(1)}
\end{array}
\right) =
{1\over\sqrt{2}}
\left(
\begin{array}{c}
\varphi_0 +i\varphi_1 \vspace*{0.1cm}\\ \varphi_2 +i\varphi_3
\end{array}
\right)
\end{array}\,~~~~,
\eea
and define standard {\it gauge covariant derivatives} $D_{\mu}^{\rm A} \equiv \partial _{\mu} - i Q_A g_{_A}A_\mu $, $D_{\mu} \equiv  \mathcal{I}D^{\rm A}_\mu
- i g_B Q_B B_{\mu}^a \,T_a$ (here $Q_{A}$ and $Q_B$ denote the Abelian and non-Abelian charges), and {\it field strength} tensors $F_{\mu \nu} \equiv  \partial_{\mu}  A_{\nu} - \partial_{\nu} A_{\mu}$, and $G_{\mu \nu} \equiv \partial_{\mu} B_{\nu} - \partial_{\nu} B_{\mu} - i[B_\mu,B_\nu]$, where $A_\mu$ and $B_\mu = B_\mu^aT_a$ are Abelian and non-Abelian gauge fields, $\mathcal{I}$ is the $2\times 2$ identity matrix, and $\lbrace T_a \equiv \sigma_a / 2 \rbrace$ ($a=1,2,3$) are the $SU(2)$ group generators, with $\sigma_a$ the {\it Pauli matrices}. \CL is then ready to solve the following type of equations (here written in cosmic time, with $a(t)$ the scale factor):
\begin{eqnarray}
\ddot{\phi} - a^{-2} {\vv\nabla}^{\,2} \hspace{-1mm}\phi + 3 \frac{\dot{a}}{a} \dot{\phi} &=& - V_{,\phi} \ , \label{eq:singletEOM} \\
\ddot{\varphi} - a^{-2} {\vv D}_{\hspace{-0.5mm}A}^{\,2}\varphi + 3\frac{\dot{a}}{a}  {\dot \varphi} &=& - {1\over2}{\varphi \over |\varphi|}V_{,|\varphi|} \ , \label{eq:higgsU1EOM}\\
\ddot{\Phi} - a^{-2} {\vv D}^{\,2}\Phi + 3 \frac{\dot a}{a}  {\dot{\Phi}} &=& - {1\over2}{\Phi \over |\Phi |}V_{,|\Phi|} \ , \label{eq:higgsSU2EOM}
\\
\partial_0 F_{0i} - a^{-2} \partial_j F_{ji} +  \frac{\dot a}{a} F_{0i} &=&
 J^A_i \ , \label{eq:U1EOM}
\\
(\mathcal{D}_0 )_{a b} (G_{0i})^b -  a^{-2} ( \mathcal{D}_j )_{a b} (G_{ji} )^b + \frac{\dot a}{a} (G_{0i} )^b &=& (J_i)_a \ , \label{eq:SU2EOM}
\end{eqnarray}
with as many copies as desired of each type of field\footnote{\CL version 1.0 allows to simulate field theories with one gauge field of each kind, i.e.~one Abelian field $A_{\mu}$ and one non-Abelian field $B_{\mu}^a$, and only one $SU(2)$ doublet when it couples to an $SU(2)$ gauge field. We are currently testing the possibility of simulations with multiple gauge fields in \CLns, so we will make available this option as soon as possible in a future update of the code.}, and where 
$V \equiv V(\phi,|\varphi|, |\Phi|)$ is the potential describing the interactions among the scalar fields. The Abelian and non-Abelian currents in the $rhs$ of the gauge field EOM (\ref{eq:U1EOM})-(\ref{eq:SU2EOM}), correspond to $J_A^\mu \equiv 2 g_AQ_A^{(\varphi)} \mathcal{I}m [ \varphi^{*} ( D_A^{\mu} \varphi )] + 2 g_AQ_A^{(\Phi)} \mathcal{I}m [ \Phi^\dag (D^{\mu} \Phi  )]$ and $J_a^\mu \equiv 2g_BQ_B\mathcal{I}m [ \Phi^{\dag} T_a( D^{\mu} \Phi )]$. \CL guarantees that the constraint equations
\begin{eqnarray}\label{eq:GaussU1}
\partial_i F_{0i} &=& a^2J^A_0\,, \\ \label{eq:GaussSU2}
(\mathcal{D}_i )_{a b} (G_{0i})^b &=& a^2(J_0)_a\,,
\end{eqnarray}
which represent the $U(1)$ and $SU(2)$ Gauss constraints in an expanding background, are preserved all throughout the evolution.

In the case of self-consistent expansion, \CL obtains numerically the scale factor $a(t)$ by solving the {\it Friedmann} equation (here written in cosmic time)
\be
\hspace{0.6cm} {\ddot a\over a} = - \frac{1}{6 m_p^2}[ \bar\rho + 3 \bar p ]\,,\label{eq:Friedmann-full}
\ee
while checking that the other Friedmann equation -- the {\it Hubble constraint} -- (also written in cosmic time),
\be\label{eq:HC}
H^2 \equiv \left({\dot a\over a}\right)^2 =  \frac{\bar\rho}{3 m_p^2} \ ,
\ee
is verified throughout the evolution. Here $\bar\rho \equiv \langle \rho \rangle$ and $\bar p \equiv \langle p \rangle $ are the background energy and pressure densities, obtained from a volume average of the local expressions contributed by the matter fields (both scalar and gauge fields),
\begin{eqnarray}
\rho &=& {K}_{\phi} + {K}_{\varphi} + {K}_{\Phi} + {G}_{\phi} + {G}_{\varphi} + {G}_{\Phi} + {K}_{U(1)} + {G}_{U(1)} + {K}_{SU(2)} + {G}_{SU(2)} + {V} \ ,  \label{eq:rhoLocal}\\
p &=& {K}_{\phi} + {K}_{\varphi} + {K}_{\Phi} -{1\over3}({G}_{\phi} + {G}_{\varphi} + {G}_{\Phi}) + {1\over3}({K}_{U(1)} + {G}_{U(1)}) + {1\over3}({K}_{SU(2)} + {G}_{SU(2)}) - {V} \ ,  \label{eq:pLocal}
\end{eqnarray}
with $V$ the interacting scalar potential, and $K_x$ and $G_x$ the kinetic and gradient energy densities of each field species [for their exact expression see Eq.~(51) in Ref.~\cite{Figueroa:2020rrl}, or  e.g.~Eq.~(\ref{eq:energy-contrib}) in this manual]. This procedure determines the evolution of the background metric of the universe within a given volume $L^3$, with $L$ the length scale of the simulation box. As long as $L$ is sufficiently large compared to the typical wavelengths excited in the fields, this procedure should lead to a well-defined notion of a 'homogeneous and isotropic' expanding background, within the given volume $L^3$ of the box.

\CL can also allow for a fixed expansion rate of the universe, with the scale factor given by a power-law function (again written in cosmic time),
\be\label{eq:aFixed}
a(t) = a (t_* ) \left(1 + \frac{3 (1 + w)}{2} H (t_*) (t-t_*) \right)^{\frac{2}{3(1 + \omega) } \,, } \ ,  \ee
where $a(t_*)$ and $H (t_*)$ are the scale factor and Hubble parameter evaluated at the initial time of the simulation $t = t_*$, and $w$ is the constant equation of state of an external fluid sourcing the expansion (assumed to be energetically dominant with respect to the fields actually being simulated). Eq.~(\ref{eq:aFixed}) acts as an input for the field Eqs.~(\ref{eq:singletEOM})-(\ref{eq:SU2EOM}).

We note that the dynamical equations presented before were expressed in cosmic time just for simplicity. In reality, \CL can solve them in any time variable of the user's preference, such as conformal time. More importantly,
we note that, of course, \CL does not really solve exactly the continuum differential equations as formulated in Eqs.~(\ref{eq:singletEOM})-(\ref{eq:SU2EOM}) or Eq.~(\ref{eq:Friedmann-full}), nor it really checks the differential constraint Eqs.~(\ref{eq:GaussU1})-(\ref{eq:GaussSU2}) or Eq.~(\ref{eq:HC}). \CL rather solves and/or checks a set of finite difference equations -- the {\it lattice equations} -- that approximate the above equations in the continuum. The lattice equations, and hence their numerical solutions, can reproduce the continuum results with higher or lower accuracy, depending on the integrator algorithm on which the lattice equations are based on. Some algorithms can solve lattice equations with numerical solutions that satisfy the (lattice) constraint equations down to machine precision. For a detailed description of different numerical integration algorithms, we refer the interested reader to Sections 3.3-3.5 of Ref.~\cite{Figueroa:2020rrl}. The detailed implementation of such algorithms in \CLns, as specialized for the dynamics of singlet scalar fields, Abelian-gauge theories and non-Abelian gauge theories, can be found in sections 4, 5, and 6 of Ref.~\cite{Figueroa:2020rrl}, respectively. 

\section{Brief review on lattice techniques} \label{sec:Basics}

\begin{mdframed}
{\bf Note -.} If the reader is already familiar with scalar and gauge field lattice simulations, or they have already read Section 3 of Ref.~\cite{Figueroa:2020rrl}, they can skip this section and jump right ahead into Sections~\ref{sec:MyFirstModelScalars} or \ref{sec:MyFirstModelGauge} of the present document, in order to set up their first scalar or scalar-gauge field lattice simulations, respectively. If the reader is familiar with scalar field lattice simulations but not with gauge lattice field theories, we recommend them to read at least Section~\ref{subsec:LGT}. The present section~\ref{sec:Basics} represents, in any case, a summary of the more extended discussion about basic lattice concepts presented in Section 3 of Ref.~\cite{Figueroa:2020rrl}.
\end{mdframed}

\subsection{Basic lattice definitions}

~~~~\CL simulates the dynamics of interacting fields in a regular cubic lattice of $N^{d}$ points in total, with $N$ the number of lattice sites per dimension, and $d$ the number of spatial dimensions. In in this document we set $d = 3$ unless otherwise specified, as \CL works by default in 3-spatial dimensions. The complete set of points in a lattice can then be labeled as
\begin{eqnarray}
{\bf n} = (n_1,n_2,n_3),~~~~ {\rm with}~~ n_i = 0,1,...,N-1 \,,~~~i = 1,2,3\,.
\end{eqnarray}
We note that it is actually such set of points that is collectively referred to as {\it the lattice}. Alternatively,  we will also refer to them as {\it the grid}, or even more colloquially, as {\it the box}. In the case of scalar field theories, \CL also allows for simulations in $d = 2$ and $d=1$ spatial dimensions\footnote{Let us note that the usability of the dynamics in lower spatial dimensions has not been robustly tested in the same manner as $d=3$, and hence any results obtained with this must be taken with great care. We plan to review this feature and guarantee its robustness in the near future.}, so that the lattice sites would be labelled in such cases as ${\bf n} = (n_1,n_2)$, or ${\bf n} = n_1$, respectively. For convenience, we stick to $d = 3$ in the following discussion. Apart from $N$, the user must also choose a {\it length side} $L$ for the box. After fixing $N$ and $L$, the smallest possible distance between two sites, the so called {\it lattice spacing}, is given by
\be \delta x \equiv \frac{L}{N} \ . \ee

\begin{mdframed}
{\it\bf Important to know -.} For serial runs (i.e.~jobs running in a single processor) the number of points per dimension $N$ can be arbitrary. For parallelized runs using MPI (i.e.~jobs running simultaneously in multiple processors), $N$ must be divisible by the number of processors $n_p$ if the parallelization is in one spatial dimension, whereas $N$ must be divisible by both $n_1$ and $n_2$ when the paralellization is done in two spatial dimensions in a number $n_p = n_1\cdot n_2$ of processors. See Section \ref{sec:UsefulFeatures} for further clarifications.
\end{mdframed}

A continuum function ${\tt f}(\bf x)$ in space is represented in the lattice by a function $f({\bf n})$, which has the same value as ${\tt f}(\bf x)$ at ${\bf x}={\bf n}\,\dx$. We note that whereas in a flat background, positions $\lbrace \bf x \rbrace$ and their corresponding lattice sites $\lbrace \bf n \rbrace$ represent physical spatial coordinates, in an expanding background they will represent {\it comoving} spatial coordinates. Unless specified otherwise, we consider {\it periodic boundary conditions}, so that $f({\bf n} + \hat \imath N) = f({\bf n})$, $i  = 1,2$ or $3$, with $\hat 1 \equiv (1,0,0)$, $\hat 2 \equiv (0,1,0)$ and $\hat 3 \equiv (0,0,1)$, unit vectors corresponding to positive displacements of one lattice spacing in each independent direction $x$, $y$ and $z$ in the lattice. The periodic boundary conditions in coordinate space imply that the momenta must be discretized, whereas the discretization of the spatial coordinates implies that any definition of a discrete Fourier transform must be periodic. For each lattice we can always consider a {\it reciprocal lattice} representing $Fourier$ modes, with its sites labeled as
\begin{eqnarray}
\tilde{\bf n} = (\tilde n_1, \tilde n_2, \tilde n_3), ~~~~{\rm with}~~
\tilde n_i = -\frac{N}{2}+1, -\frac{N}{2}+2, ... ,-1,0,1, ... , \frac{N}{2} - 1, \frac{N}{2}  \,,~~~ i  = 1,2,3\,.
\end{eqnarray}
A discrete Fourier transform (DFT) is then defined by
\begin{eqnarray}\label{eq:FTdiscrete}
f({\bf n}) \equiv {1\over N^3}\sum_{\tilde n} e^{-i{2\pi\over N} {\bf \tilde n n}} f({\bf \tilde n}) ~~~~ \Leftrightarrow ~~~~  f({\bf \tilde n}) \equiv \sum_{n} e^{+i{2\pi\over N} {\bf n \tilde n} }f({\bf n})\,,
\end{eqnarray}
from where it follows that Fourier-transformed functions are periodic in the reciprocal lattice, with periodic boundary conditions as $ f({\bf\tilde{n}} + {\hat \imath} N) =  f({\bf\tilde{n}})$, with ${ \hat \imath}$ analogous unit vectors as before but defined now in the reciprocal lattice.

From the above discussion, it follows that we can only represent momenta down to a minimum infrared (IR) cut-off
\begin{eqnarray}
k_{\rm IR} = \frac{2\pi}{L} = \frac{2\pi}{N\dx}\,,
\end{eqnarray}
and hence $\tilde{\bf n}$ labels the continuum momentum values ${\bf k} = (\tilde n_1, \tilde n_2, \tilde n_3)\, k_{\rm IR}$. There is also a maximum ultraviolet (UV) momentum
that we can capture within each spatial dimension,
\begin{eqnarray}
k_{i,\rm UV} = {N\over2}k_{\rm IR} = {\pi\over \dx}\,.
\end{eqnarray}
The maximum momentum we can capture in a 3-dimensional reciprocal lattice is therefore the diagonal of the box,
\begin{eqnarray}
k_{\rm max} = \sqrt{k_{1,\rm UV}^2+k_{2,\rm UV}^2+k_{3,\rm UV}^2} = {\sqrt{3}\over2}Nk_{\rm IR} = \sqrt{3}{\pi\over \dx}\,\,.
\end{eqnarray}
We note that for given $N$, fixing $k_{\rm IR}$ automatically determines $L$. Fixing $k_{\rm IR}$ can be very useful if one has an {\it a priori} understanding  of the typical momenta scales expected to be excited in the scenario to be simulated.

Finally, we also note that a time-step $\delta \eta$ must be chosen in order to run any simulation. As a rule of thumb, stability of the solution typically requires $\delta \eta / \delta x < 1 / \sqrt{d}$. Continuum derivatives, either spatial or temporal, need to be replaced in the lattice with different finite expressions that have a correct continuum limit, i.e.~approximations to the continuum derivative to some order in the lattice spacing/time step. Simple definitions of a lattice derivative are the {\it neutral} derivative
\begin{equation}
\label{eq:neutrald}
[\nabla^{(0)}_\mu f] = \frac{f({n}+\hat\mu) - f({n}-\hat\mu)}{2\dx ^\mu} ~~\longrightarrow ~~ \partial_i{\tt f}({x})\big|_{{x}\,\equiv\, {\bf n}\dx+n_0\deta} + \mathcal{O}(\dx_\mu^2)\,,
\end{equation}
with $\dx^\mu$ referring to the lattice spacing $\dx$ in the case of spatial derivatives, and to the time step $\delta\eta$  in the case of temporal derivatives. The expression to the right-hand side of the arrow indicates where and to what order in the lattice spacing/time step the continuum limit is recovered. Also standard are the {\it forward} and {\it backward} derivatives
\begin{eqnarray}
\label{eq:forwardbackwardd}
[\nabla^\pm_\mu f] = \frac{\pm f({n}\pm \hat\mu) \mp f({n})}{\dx^\mu} ~~\longrightarrow ~~ \left\lbrace\begin{array}{l}
\partial_i{\tt f}({x})\big|_{{x}\,\equiv\, {\bf n}\dx+n_0\deta} + \mathcal{O}(\dx_\mu)  \vspace*{0.2cm}\\
\partial_i{\tt f}({x})\big|_{{x}\,\equiv\, ({n} \pm \hat\mu/2)\dx^\mu} + \mathcal{O}(\dx_\mu^2)
\end{array}\right.\,,
\end{eqnarray}
which recover the continuum limit to linear or to quadratic order in the lattice spacing/time step, depending on whether we interpret that they live in ${n}$, or in between the two lattice sites involved ${n} \pm \hat\mu/2$. This shows that in order to recover a continuum differential operation in the lattice, not only it is important to use a suitable discrete operator, but also to determine where it `lives'. To improve accuracy, one can also consider lattice derivatives which involve more points, typically leading to definitions that have a symmetry either around a lattice site or around half-way between lattice sites, see e.g.~\cite{Frolov:2008hy}.

Depending on the choice of lattice operator $\nabla_{i}$ for the spatial derivatives, the discrete Fourier transform leads to different {\it lattice momenta}. The Fourier transform of a derivative $[\nabla_if]$ can be written as~\cite{Figueroa:2020rrl}
\begin{eqnarray}
{\nabla_i f}(\tilde{\bf n}) \equiv -i{\bf k}_{\rm L}(\tilde{\bf n}) f(\tilde{\bf n})\,,
\end{eqnarray}
which leads, for the neutral derivative (\ref{eq:neutrald}), to
\begin{eqnarray}
\label{eq:neutralMomentum}
k_{{\rm Lat},i}^0=\frac{\sin(2\pi \tilde{n}_i/N)}{\dx}\,,
\end{eqnarray}
and for forward/backward derivatives~(\ref{eq:forwardbackwardd}) to
\begin{eqnarray}
\label{eq:ForwardBackwardMomentum}
&& k_{{\rm Lat},i}^+ = k_{{\rm Lat},i}^- = 2\frac{\sin(\pi \tilde{n}_i/N)}{\dx}\,,~~~{\rm if} ~{\bf l} = {\bf n} \pm {\hat{\imath}\over2}\,,\\
\label{eq:ForwardBackwardMomentumII}
&& k_{{\rm Lat},i}^\pm = \frac{\sin(2\pi \tilde{n}_i/N)}{\dx} \pm i
\frac{1-\cos(2\pi \tilde{n}_i/N)}{\dx}\,,~~~{\rm if} ~{\bf l} = {\bf n}\,.
\end{eqnarray}

Finally, in order to mimic the {\it power spectrum} of a continuum function ${\tt f}({\bf x})$, which by definition characterizes its ensemble average $\langle {\tt f}^2 \rangle$ as
\bea
		&&\langle {\tt f}^2 \rangle = \int d\log k~\Delta_{\tt f}(k)~~,\label{eq:continuumPSII}\\
		&&\Delta_{\tt f}(k) \equiv {k^3\over 2\pi^2}\mathcal{P}_{\tt f}(k)~~,~~~ \langle {\tt f}_{\bf k} {\tt f}_{{\bf k}^{\prime}} \rangle = (2\pi)^3 \mathcal{P}_{\tt f}(k) \delta (\mathbf{k}-\mathbf{k^{\prime}})~, \label{eq:continuumPS}
\eea
we define in the lattice the expression of a discrete power spectrum as
	\be
		 \Delta_f(k) \equiv \frac{k^3(\tilde {\bf n})}{2\pi^2}\left(\frac{\delta x}{N}\right)^3 \big\langle \big|f(\tilde{\bf n})\big|^2\big\rangle_{R(\tilde{\bf n})}\,, \label{eq:discretePS}
	\ee
where $\langle ( ... ) \rangle_{R(\tilde{\bf n})} \equiv \frac{1}{4\pi|\tilde{\bf n}|^2}\sum_{\tilde{\bf n}^{\prime}\in R(\tilde{\bf n})}( ... )$ is an angular average over a spherical shell of radius $\tilde{\bf n}^{\prime}\in \big[|\tilde{\bf n}|,|\tilde{\bf n}+ \Delta\tilde{\bf
n}|\big)$, with $\Delta \tilde{\bf n}$ a given radial binning. In this way, we obtain
\bea
\left<f^2\right>_V \equiv {1\over N^3}\sum_{\bf n}f^2({\bf n}) = \sum_{\tilde{\bf n} \neq 0}{k_{\rm IR}\over k(\tilde {\bf n})} \Delta_f(k(\tilde {\bf n}))\,,\label{eq:discretePSII}
\eea
with $\langle ( ... ) \rangle_{V}$ representing now a volume average over the lattice, equivalent to the $lhs$ of the continuum expression Eq.~(\ref{eq:continuumPSII}), whilst the $rhs$ of Eq.~(\ref{eq:discretePSII}) mimics exactly the $rhs$ of  Eq.~(\ref{eq:continuumPSII}).

\begin{mdframed}
{\bf Important Note -.} The definition of the scalar power spectrum ${\Delta}_{\phi} (k)$ in the lattice has evolved since \CL {\tt version 1.0} was first released. From {\tt version 1.1} and above, different versions of a scalar field power spectrum can be output, see  \href{https://cosmolattice.net/technicalnotes/}{\color{blue} {\tt Technical Note I}} for an explanation of the different choices. Visit regularly \href{https://cosmolattice.net/technicalnotes/}{\color{blue} https://cosmolattice.net/technicalnotes/} to check for new features (ranging from new definitions, options, algorithms, interactions, etc) incorporated in successive updated versions of \CLns.
\end{mdframed}

\subsection{Lattice gauge invariant techniques}
\label{subsec:LGT}

~~~~~Discretizing a gauge theory requires a special care in order to preserve gauge invariance at the lattice level. It is not enough to recover gauge invariance in the continuum, i.e.~in the limit of zero lattice spacing/time step, as gauge invariance is meant to remove spurious transverse degrees of freedom.  If we were to discretize a gauge theory by substituting all ordinary derivatives in the continuum EOM by finite differences like those in Eqs.~(\ref{eq:neutrald}), (\ref{eq:forwardbackwardd}), the gauge symmetry would not be preserved and the spurious degrees of freedom would propagate in the lattice. Lattice gauge invariant techniques are meant to avoid this type of trouble.

In order to build an action or EOM for any gauge theory that preserves a discretized version of the gauge symmetry, it is customary to define \textit{link} variables as\footnote{$Pexp\lbrace...\rbrace$ means \textit{path-ordered} along the integration trajectory, as the construction of links is based on the definition of a {\it parallel transporter}, connecting two points in space-time as $U(x,y) = Pexp\left\lbrace-ie\int_{x}^{y}dx^{\mu}A_\mu  \right\rbrace$.}
\begin{eqnarray}
U_{0,n} \equiv Pexp\left\lbrace-ie\int_{x(n)}^{x(n+\hat0)}dt'A_0 \right\rbrace \approx  e^{-ie\delta t A_0}\,, ~~~~ U_{i,n} \equiv Pexp\left\lbrace-ie\int_{x(n)}^{x(n+\hat\imath)} dxA_i \right\rbrace \approx e^{-ie\dx A_i}\,,
\end{eqnarray}
where the gauge field $A_\mu$, and hence the link $U_\mu$, is considered to live in the point $n + {\hat\mu\over2}$. We also define $U_{-\mu,n} = U_{\mu,n-\mu}^\dag \equiv U_\mu^\dag(n-{1\over 2}\hat\mu)$. In the continuum limit, the gauge fields can be recovered simply from $-i(\mathcal{I}- U_{\mu,n})/(e\delta x^\mu) \longrightarrow  A_\mu\big(n+{1\over2}\hat\mu\big) + \mathcal{O}(\dx^\mu)$.\\

\begin{mdframed}
{\it\bf Important to know -.} To simplify the notation on the lattice, a scalar field living in a generic lattice site $n = (n_o,\bn) = (n_o,n_1,n_2,n_3)$, i.e.~$\phi_n = \phi(n)$, will be simply denoted as $\phi$. If the point is displaced in the $\mu-$direction by one unit lattice spacing/time step, $n + \hat\mu$, we use the notation $n+\mu$ or simply by $+\mu$ to indicate this, so that the field amplitude in the new point is expressed as $\phi_{+\mu} \equiv \phi(n+\hat\mu)$. In the case of gauge fields, whenever represented explicitly in the lattice, we will automatically understand that they live in the middle of lattice points, i.e.~$A_{\mu} \equiv A_{\mu}(n+{1\over2}\hat\mu)$. It follows then that e.g.~$A_{\mu,+\nu} \equiv A_{\mu}\big(n + {1\over2}\hat\mu +  \hat\nu\big)$. In the case of links, we will use the notation $U_\mu \equiv U_{\mu,n} \equiv U_\mu(n+{1\over2}\hat\mu)$, and hence $U_{\mu,\pm\nu} = U_{\mu,n\pm\nu} \equiv U_\mu(n + {1\over2}\hat\mu \pm \hat\nu)$.
\end{mdframed}
\vspace*{0.5cm}

One can actually build an action or EOM for any gauge theory, preserving a discretized version of the gauge symmetry, using only link variables and no gauge fields. That is known as the {\it compact formulation} of lattice gauge theories, and this can be applied to both Abelian and non-Abelian gauge theories. In the case of non-Abelian theories, compact formulations are actually the only way to discretize them while respecting gauge invariance in the lattice. For Abelian gauge theories, however, it is still possible to make use of an explicit representation of the  gauge fields, in the so called {\it non-compact formulation}. Below we provide both. We introduce standard definitions for $links$, $plaquettes$ and {\it lattice covariant derivatives}, specialized to both Abelian and non-Abelian gauge groups, setting back $e=g_A Q_A$. We provide also basic definitions, together with useful approximations and expressions (in the case of Abelian theories for both compact and non-compact formulations). All these ingredients, summarized in the $U(1)$ and $SU(2)$ toolkits below, represent all one needs to know in order to discretize gauge theories while preserving the gauge invariance at the lattice level.\\

\begin{mdframed}
\begingroup
\allowdisplaybreaks
\begin{center}
----- U(1) toolkit -----
\end{center}
\begin{eqnarray}
&&{\rm Links:}~ V_{\mu} \equiv e^{-i g_AQ_A \dx_{\mu} A_{\mu}} = \cos(g_AQ_A\dx_{\mu} A_{\mu}) - i \sin (g_AQ_A\dx_{\mu} A_{\mu}) ;~~~ V_{- \mu} \equiv V_{\mu,-\mu}^* ;~~~ V_{\mu}^* V_{\mu} = 1\,;\vspace*{0.4cm}\nonumber\\
&&{\rm Plaquettes}:~ V_{\mu \nu} \equiv V_{\mu} V_{\mu,+\mu} V_{\mu, +\nu}^* V_\nu^* \simeq e^{-i g_AQ_A\dx_{\mu} \dx_{\nu} [ F_{\mu \nu} + \mathcal{O}(\dx)] };~~~ V_{\mu\nu}^* = V_{\nu\mu}\,;
\vspace*{0.4cm}\nonumber\\
&&{\rm Covariant~Derivs.}:  (D_{\mu}^\pm\varphi)({\bf l}) = \pm{1\over \dx^\mu}(V_{\pm\mu}\varphi_{\pm\mu} - \varphi)\,,~~{\bf l} = {\bf n} \pm {1\over2}{\hat\mu}\,
\vspace*{0.6cm}\nonumber\\
&&{\rm Expansions}:
\left\lbrace
\begin{array}{rcl}
(D_{\mu}^\pm\varphi)({\bf l})  & \longrightarrow & (D_{\mu}\varphi)({\bf l}) + \mathcal{O}(\dx^2)\,~~{\bf l} = {\bf n} \pm {1\over2}{\hat\mu}\vspace*{0.2cm}\\
\mathcal{R}e\lbrace V_{\mu \nu} \rbrace  & \longrightarrow & 1 - \frac{1}{2} \dx_{\mu}^2 \dx_{\nu}^2g_A^2Q_A^2 F_{\mu \nu}^2 + \mathcal{O}(\delta x^5)\,,~~{\bf l} = {\bf n} + {1\over2}{\hat\mu} + {1\over2}{\hat\nu}\vspace*{0.2cm}\\ \mathcal{I}m\lbrace V_{\mu \nu} \rbrace & \longrightarrow & - \dx_{\mu} \dx_{\nu} g_AQ_AF_{\mu \nu} + \mathcal{O}(\delta x^3)\,,~~{\bf l} = {\bf n} + {1\over2}{\hat\mu} + {1\over2}{\hat\nu}
\end{array}\right.
\vspace*{0.6cm}\\
&&{\rm Expressions}:
\left\lbrace
\begin{array}{l}
\left.
\begin{array}{l}
\sum_n {1\over 4}F_{\mu \nu}^2 \cong -{1\over 2}\sum_n{\mathcal{R}e\lbrace V_{\mu \nu} \rbrace \over \dx_{\mu}^2 \dx_{\nu}^2g_A^2Q_A^2} = -{1\over4}\sum_n {(V_{\mu \nu}+V_{\mu \nu}^*)\over \dx_{\mu}^2 \dx_{\nu}^2g_A^2Q_A^2} + \mathcal{O}(\dx^2)\vspace*{0.2cm}\\
\sum_n {1\over4}F_{\mu \nu}^2 \simeq \sum_n {1\over4}{\mathcal{I}m^2\lbrace V_{\mu \nu} \rbrace \over \dx_{\mu}^2 \dx_{\nu}^2g_A^2Q_A^2} = -\sum_n {1\over4}{(V_{\mu \nu}-V_{\mu \nu}^*)^2\over \dx_{\mu}^2 \dx_{\nu}^2g_A^2Q_A^2} + \mathcal{O}(\dx^2)
\end{array}\right]~~({\tt Compact})\vspace*{0.5cm}\\
\left.
\begin{array}{l}
\sum_n {1\over4}F_{\mu \nu}^2 \simeq {1\over4}\sum_n (\Delta^+_\mu A_\nu - \Delta^+_\nu A_\mu)^2 + \mathcal{O}(\dx^2)
\end{array}\right]~~({\tt Non-Compact})
\end{array}\right.
\vspace*{0.6cm}\nonumber\\
&&{\rm Gauge~Trans}
\left.
\left\lbrace
\begin{array}{cll}
\phi  & \longrightarrow & e^{+ig_AQ_A\alpha}\phi\vspace*{0.2cm}\\
A_\mu & \longrightarrow & A_\mu - \Delta_\mu^+\alpha\vspace*{0.2cm}\\
V_{\pm \mu}  & \longrightarrow & V_{\pm \mu}e^{ig_AQ_A(\alpha_{\pm\mu}-\alpha)}
\end{array}\right.
\right] ~~\Longrightarrow ~~ \left\lbrace
\begin{array}{cll}
D_\mu^\pm\phi & \longrightarrow & e^{ig_AQ_A\alpha}(D_\mu^\pm\phi)\vspace*{0.2cm}\\
V_{\mu\nu}  & \longrightarrow & V_{\mu\nu} ~{\rm (gauge~inv.~!)}
\end{array}\right.
\nonumber
\label{eq:U1toolkit}
\end{eqnarray}
\endgroup
\\ \end{mdframed}
\vspace*{4cm}

\begin{mdframed}
\begin{center}
----- SU(N) toolkit -----
\end{center}
\begin{eqnarray}
&&{\rm Links}:~ U_{\mu} \equiv e^{-i g_B Q_B \dx B_\mu} = e^{-i g_B Q_B \dx B_{\mu}^a T_a} ;~~~ U_{- \mu} \equiv U_{\mu,-\mu}^{\dagger} ;~~~ U_{\mu}^{\dagger} U_{\mu} = \mathcal{I} \vspace*{0.4cm}\nonumber\\
&&{\rm Plaquettes}:~ U_{\mu \nu} \equiv U_{\mu} U_{\nu,+ \mu} U_{\mu, +\nu}^{\dagger} U_{\nu}^{\dagger} \simeq e^{-ig_B Q_B \dx_{\mu} \dx_{\nu} [ G_{\mu \nu}^a T_a + \mathcal{O} (\dx_{\mu} ) ] }\,; ~~~ U_{\mu \nu}^\dag = U_{\nu\mu} \vspace*{0.4cm}\nonumber\\
&&{\rm Covariant~Derivs.}:  (D_{\mu}^\pm\Phi)({\bf l}) = \pm{1\over \dx^\mu}(U_{\pm\mu}\Phi_{\pm\mu} - \Phi) ~\longrightarrow~ (D_{\mu}\Phi)({\bf l}) + \mathcal{O}(\dx^2),~~{\bf l} = {\bf n} \pm {1\over2}{\hat\mu}
\nonumber\vspace*{0.4cm}\\
&&{\rm Expansions}:
\left\lbrace
\begin{array}{ccl}
(D_{\mu}^\pm\Phi)({\bf l}) & \longrightarrow & (D_{\mu}\Phi)({\bf l}) + \mathcal{O}(\dx^2)\,,~~{\bf l} = {\bf n} \pm {1\over2}{\hat\mu}\vspace*{0.2cm}\\
(U_{\mu \nu} - U_{\mu \nu}^\dag ) & \longrightarrow & -2ig_B Q_B\delta x_{\mu} \delta x_{\nu}G_{\mu \nu} + \mathcal{O} (\dx_{\mu}^3)\,,~~{\bf l} = {\bf n} + {1\over2}{\hat\mu} + {1\over2}{\hat\nu} \vspace*{0.2cm}\\ {\rm Tr} [ U_{\mu \nu} ] & \longrightarrow & 2 - \frac{\dx_{\mu}^2 \dx_{\nu}^2g_B^2 Q_B^2}{4}\sum_a (G_{\mu \nu}^a)^2 + \mathcal{O} (\dx_{\mu}^5)\,,~~{\bf l} = {\bf n} + {1\over2}{\hat\mu} + {1\over2}{\hat\nu}
\end{array}\label{eq:SU2toolkit}
\right.
\vspace*{0.6cm}\\
&&{\rm Expressions}:
\left\lbrace
\begin{array}{l}
{1\over2}{\rm Tr}[G_{\mu\nu}G^{\mu\nu}] = {1\over4}\sum_a (G_{\mu\nu}^a)^2 \cong -{{\rm Tr} [ U_{\mu \nu} ]\over \dx_{\mu}^2 \dx_{\nu}^2 g_B^2 Q_B^2} + \mathcal{O}(\dx^2)\,, \vspace*{0.2cm}\\
G_{\mu \nu} = G_{\mu \nu}^aT_a \simeq \frac{i}{2\dx_{\mu} \dx_{\nu}g_B Q_B} (U_{\mu \nu} - U_{\mu \nu}^\dag) + \mathcal{O}(\dx^2) \,,\vspace*{0.2cm}\\
G_{\mu \nu}^a \simeq \frac{1}{\dx_{\mu} \dx_{\nu}g_B Q_B} {\rm Tr} [ (i T_a)  (U_{\mu \nu} - U_{\mu \nu}^\dag ) ] + \mathcal{O}(\dx^2)
\end{array}\right.
\vspace*{0.4cm}\nonumber\\
&&{\rm Gauge~Trans.}
\left.
\left\lbrace
\begin{array}{cll}
\Phi  & \longrightarrow & \Omega\,\Phi\,,~~~ \Omega \equiv e^{+ig_B Q_B\alpha_aT_a}\vspace*{0.2cm}\\
U_{\pm \mu}  & \longrightarrow & \Omega \,U_{\pm \mu}\,\Omega^\dag_{\pm \mu}
\end{array}\right.
\right] ~~\Longrightarrow ~~ \left\lbrace
\begin{array}{cll}
D_\mu^\pm\Phi & \longrightarrow & \Omega\,(D_\mu^\pm\Phi)\vspace*{0.2cm}\\
U_{\mu\nu}  & \longrightarrow & \Omega \,U_{\mu\nu}\,\Omega^\dag
\vspace*{0.2cm}\\
{\rm Tr}\lbrace U_{\mu\nu} \rbrace  & \longrightarrow & {\rm Tr}\lbrace U_{\mu\nu} \rbrace
\end{array}\right.
\label{eq:SUNtoolkit}
\nonumber
\end{eqnarray}
\\ \end{mdframed}
\medskip

\section{My first model of (singlet) scalar fields} \label{sec:MyFirstModelScalars}

~~~~~We now introduce the user to the basic functionalities of \CLns. As an example, we explain  step by step how to implement a specific model of interacting scalar fields in \CLns, for the simulation of a simple preheating scenario. This section is structured as follows. In
Section \ref{subsec:LatticeScalars} we first introduce the concept of \textit{program variables} for scalar fields, which are a new set of re-scaled dimensionless variables suitable for their introduction in a computer. In Section \ref{sec:ScTheModel} we present an example model and define its corresponding program variables and potential specific to it. We then explain in Section \ref{sec:MyFirstRun} how to compile and run the code. After that, in Section \ref{sec:TheModelFile} we walk the user through the \textit{model file}, where the model details are actually implemented. Finally, in Section \ref{sec:WhatHappensAuto} we provide a summarized picture of what happens 'under the hood', giving details on how the fields are initialized in the simulation, how their dynamical evolution is solved, and how different measurements are obtained. By the end of this chapter, the reader should have enough information to implement any model involving interacting scalar fields. Users interested in including gauge fields in their models should proceed to read Section \ref{sec:MyFirstModelGauge}.

\subsection{Program variables}
\label{subsec:LatticeScalars}
\vspace*{0.3cm}
\begin{mdframed}
{\bf Important -.} In the lattice, we operate with a particular set of dimensionless field and spacetime variables, $\{\tilde{\phi},\tilde{\eta},\tilde{x}^i\}$, which we call \textit{program variables}. The transformation from `physical' to program variables, $t \rightarrow \tilde{\eta}$, $x^i \rightarrow \tilde{x}^i$, and $\phi \rightarrow \tilde{\phi}$, is given by the following relations
\begin{eqnarray}
\label{eq:FieldSpaceTimeNaturalVariables}
\tilde\phi \equiv {\frac{\phi}{f_*}}\,,~~~~ d\tilde\eta \equiv a^{- \alpha}  \omega_* dt\,,~~~~ d\tilde x^i \equiv \omega_* dx^i\,,
\end{eqnarray}
where $f_*$ and $\omega_*$ are two constants with dimensions of energy. Program variables will be tagged with the diacritic $\sim$, as well as all quantities defined in terms of them.
\end{mdframed}

The reader familiar with {\tt LatticeEasy}, might have notticed that the above transformations are similar to the ones carried out in that code to define their program variables, if we set $A = 1/f_*$, $B= \omega_*$, $r=0$, and $s=-\alpha$ in their notation~\cite{Felder:2000hq}. As we will see, the main difference is that the evolution algorithms implemented in \CL \textit{do not require a conformal rescaling of the fields}, so we do not need to introduce a parameter analogous to $r$.

Before simulating a particular model, the user must choose a certain set of values for $\{f_*,\omega_*,\alpha\}$, which will define the program variables used in the lattice via Eq.~(\ref{eq:FieldSpaceTimeNaturalVariables}). The choice of $f_*$ and $\omega_*$ can be made arbitrary, as they only re-scale all numbers by constant factors. However, if we simulate a scenario in which, e.g.~a resonance is triggered by an oscillatory field, it can be convenient to set $f_*$ and $\omega_*$ to the initial amplitude and oscillation frequency of that field respectively. This way, the numbers produced by the code will be close to unity, which will help the interpretation of results. In general, in every scenario there is always a natural choice (at least of the order of magnitude) of $f_*$ and $\omega_*$, related to the typical field amplitudes and time scales of the the problem. Choosing those natural values will help us interpret more easily (in a more intuitive physical manner), the numbers that the code outputs.

A correct choice of $\alpha$ is perhaps more relevant, as a wrong choice could spoil the stability of the numerical solution at late times. For example, let us go back to the case of an oscillating homogeneous scalar field dominating the energy budget of the Universe. Our evolution algorithms operate with a constant time step, so it would be a good idea would be to choose $\alpha$ so that the oscillation frequency of the program field variable is approximately constant when expressed in the corresponding $\alpha$-time. In this way, we will be able to resolve each physical oscillation with a similar accuracy. For example, let us consider the common case of an oscillatory field with monomial potential $V(\phi) \propto |\phi|^p$ sourcing the expansion of the Universe. As described extensively in Ref.~\cite{Figueroa:2020rrl}, the oscillation frequency will be initially constant if we choose
\be \alpha = 3 \left( \frac{p-2}{p+2} \right) \ . \label{eq:Alpha-PowLaw}\ee
We recommend to use Eq.~(\ref{eq:Alpha-PowLaw}) for any scenario where there is an energetically dominant scalar field with potential $V(\phi) \propto \phi^p$. The choice of $\alpha$ for more complex scenarios must be done in a case by case basis.

\subsection{The model} \label{sec:ScTheModel}

~~~~ We will consider a simple preheating scenario for illustrative purposes, consisting of an inflaton $\phi$ with quartic potential $V(\phi) \propto \phi^4$, coupled to a secondary massless scalar field $\chi$ through a quadratic interaction. Denote the total number of scalar fields in a theory as $N_s$, then $N_s=2$ in our case. To describe the expansion of the universe, we consider a flat {\it Friedmann-Lema\^itre-Robertson-Walker} (FLRW) metric with line element
\be
\dd s^2 = g_{\mu\nu}\dd x^\mu\dd x^\nu = - a(\eta)^{2 \alpha} \dd \eta^2 + a(\eta)^2 \delta_{ij} \dd x^i \dd x^j \ , \label{eq:FLRWmetric}
\ee
where $a(\eta)$ is the scale factor, $\delta_{ij}$ is the Euclidean metric, and $\alpha$ is a constant parameter that will we choose conveniently in a moment. The choice $\alpha = 0$ would identify $\eta$ with {\it cosmic time} $t$, whereas $\alpha = 1$ would identify it with {\it conformal time} $\tau \equiv \int {dt'  a^{-1}(t')}$. For now, we will consider $\alpha$ as an unspecified constant, and we will refer to $\eta$ as the {\it $\alpha$-time variable}.

\begin{mdframed}
{\bf Note -.} We remind the reader that we reserve the symbol $\dot f \equiv {df/dt}$ for derivatives with respect to cosmic time, and $f' \equiv {df/ d\eta}$ for derivatives with respect to $\alpha$-time.
\end{mdframed}

In this metric, the action of the field theory we want so simulate is the following,
\begin{align} \label{eq:ScalarActionCont}
  S_{\rm S} &= - \int d\eta d^3 x a(\eta)^{3+\alpha} \left\{{1\over2} \partial^{\mu} \phi \partial_{\mu}\phi +{1\over2} \partial^{\mu} \chi \partial_{\mu}\chi + V(\phi,\chi)\right\} \ , \\
  V(\phi,\chi) &\equiv \sum_{m=0}^{N_p-1} V^{(m)} (\phi, \chi) =  \frac{\lambda}{4}\phi^4 +\frac{1}{2}g^2 \phi^2\chi^2\,, \label{eq:potentialExampleI}
  \end{align}
where $V(\phi, \chi)$ is the scalar potential, and $\lambda$ and $g$ are dimensionless parameters. The potential contains two different terms: the quartic potential of the inflaton and the interaction between both fields, which we denote as $V^{(m)}$ with $m=0,1$ respectively. The total number of terms is defined as $N_p$ ($= 2$). Indices are raised/lowered using the FLRW metric defined in Eq.~\eqref{eq:FLRWmetric}, e.g. $\partial^{\mu} \phi \partial_{\mu}\phi=g^{\mu\nu}\partial_\mu\phi\partial_\nu\phi$. The field equations of motion in $\alpha$-time read~\cite{Figueroa:2020rrl}
 \begin{align}
   \phi'' - a^{-2(1 - \alpha)} {\vv\nabla}^{\,2} \hspace{-1mm}\phi + (3 - \alpha)\frac{{a'}}{a} {\phi'} &= - a^{2 \alpha} V_{,\phi} \ , \\
   &=- a^{2 \alpha} (\lambda \phi^3 +g^2\phi\chi^2) \label{eq:scEOM}\\
 \chi'' - a^{-2(1 - \alpha)} {\vv\nabla}^{\,2} \hspace{-1mm}\chi + (3 - \alpha)\frac{{a'}}{a} {\chi'} &= - a^{2 \alpha} V_{,\chi} \\
 &=- a^{2 \alpha} g^2\phi^2\chi \ ,
\end{align}
with the evolution of the scale factor $a(\eta)$ given by the Friedmann equations. If these two fields constitute the only energy sources in the Universe (or at least the dominant ones), either of the Friedmann equations
\begin{eqnarray}
\mathcal{H}^2 \equiv  \left({a'\over a}\right)^2 &=&  \frac{a^{2 \alpha}}{3 m_p^2}\left\langle  {K}  + {G} +  {V} \right\rangle \,,
\\
{a''\over a} &=& \frac{a^{2 \alpha}}{3 m_p^2}\left\langle (\alpha-2)  {K} + \alpha  {G} + (\alpha + 1)  {V}  \right\rangle \,,  \label{eq:sfEOM}\nonumber
\end{eqnarray}
can be solved self-consistently, together with the fields' equations of motion. Here $\langle \dots \rangle$ indicates a volume average, and $K$ and $G$ are the total kinetic and gradient energies. All scalar fields contribute to these quantities as $K \equiv \sum_{n=0}^{N_s-1} {K}^{(n)}$ and $G \equiv \sum_{n=0}^{N_s-1} {G}^{(n)} $, with [$\phi_0 \equiv \phi$, $\phi_1 \equiv \chi$],
\be
{K}^{(n)} = \frac{1}{2 a^{2\alpha} } \phi_n^{'2} \ , \hspace{0.4cm} {G}^{(n)} = \frac{1}{2 a^2} \sum_i (\nabla_i \phi_n)^2 \ .
\ee
In other scenarios, one could have the expansion of the Universe to be fixed by an external, energetically-dominant fluid with (constant) equation of state $w$. In this case, the evolution of the scale factor and the Hubble parameter in program variables is given by the following functions,
\be\label{eq:ScaleFactorPowerLaw}
a(\tilde \eta) = a (\tilde \eta_* ) \left(1 + \frac{1}{p}\mathcal{H}_* (\tilde \eta- \tilde\eta_*) \right)^p \,,\hspace{0.3cm} \mathcal{H}(\eta) = {\mathcal{H}_*\over \left(1 + \frac{1}{p}\mathcal{H}_* (\tilde \eta- \tilde \eta_*) \right)}  \,,\hspace{0.5cm} p \equiv \frac{2}{3(1 + \omega) - 2 \alpha } \ . \ee

As mentioned before, numerical simulations are carried out in the dimensionless program variables defined in Eq.~(\ref{eq:FieldSpaceTimeNaturalVariables}). Therefore, we need to appropriately choose values for $\{ f_*, \omega_*, \alpha\}$ in this model. We take them as follows,
\begin{align} \label{eq:lphi4-ProgVar}
  f_*=\overline{\phi}_{*}\,,~~~~ \omega_*=\lambda^{1/2}  \overline{\phi}_*,~~~~ \alpha=1
\end{align}
where $\overline{\phi}_{*}$ and $\lambda^{1/2} \overline{\phi}_*$ are the amplitude and oscillation frequency of the inflaton at the end of inflation, see e.g.~Ref.~\cite{Greene:1997fu}. The constant $\alpha$ was chosen according to Eq.~(\ref{eq:Alpha-PowLaw}), which guarantees that the oscillation frequency remains approximately constant as long as the oscillatory inflaton field $\phi$ dominates the energy budget.

In \CLns, any field theory is implemented by means of the \textit{program potential}, which is a dimensionless quantity defined in terms of the program variables as follows,
\bea\label{eq:PotNat}
\widetilde V( \tilde\phi, \tilde\chi ) \equiv \frac{1}{f_*^2 \omega_*^2}V(f_*\tilde \phi,f_*\tilde\chi) = \frac{1}{4}\tilde\phi^4 +\frac{1}{2}\frac{g^2}{\lambda}\tilde\phi^2\tilde\chi^2 \ .
\eea
Similarly, we define each of the individual contributions of this quantity as $\widetilde{V}^{(m)} \equiv {V}^{(m)} /(f_*^2 \omega_*^2)$. We also define the following (dimensionless) \textit{program energy/pressure densities} as
\bea
\tilde\rho \equiv \frac{\rho}{f_*^2 \omega_*^2} = \widetilde{K} + \widetilde{G}  + \widetilde V \,,~~~~ ; ~~~~ \tilde p \equiv \frac{p}{f_*^2 \omega_*^2} = \widetilde{K} - \frac{1}{3} \widetilde{\rm G} - \widetilde V \ . \eea
with $\tilde V$ given by Eq.~(\ref{eq:PotNat}), and where we have introduced the following \textit{program kinetic and gradient energies} as $\widetilde{K} \equiv \sum_{n=0}^{N_s-1} \widetilde{K}^{(n)}$ and $\widetilde{G} \equiv \sum_{n=0}^{N_s-1} \widetilde{G}^{(n)} $, with
\bea\label{eq:KandGprogramUnits}
\widetilde{K}^{(n)} = \frac{1}{2 a^{2\alpha}} ({\tilde \phi_n}')^2  \ , \hspace{0.4cm}
\tilde{G}^{(n)} = \frac{1}{2 a^2 }  \sum_{i} (\widetilde\nabla_i \tilde \phi_{n})^2 \ .
\eea
We denote the corresponding volume-averaged energy density components as
\bea\label{eq:EK_EG_EV}
{\widetilde E}_K \equiv \left\langle \tilde{K} \right\rangle\,,~~~ {\widetilde E}_G \equiv \left\langle \tilde{G} \right\rangle\,, ~~~{\widetilde E}_V \equiv \left\langle \tilde{V} \right\rangle \ ,
\eea
and their partial contributions as ${\widetilde E}_K^{(n)}$, ${\widetilde E}_G^{(n)}$ and ${\widetilde E}_V^{(m)}$ respectively.  Using these notations, the equations of motion, still in the continuum but already expressed in program variables, read

\begin{mdframed}
\begin{align}
\tilde\phi'' - a^{-2 (1  - \alpha )} \tilde\nabla^2 \tilde\phi + (3 - \alpha)\frac{a'}{a} \tilde\phi'_a  &= -  a^{2 \alpha} \widetilde V_{,\tilde\phi}\,\label{eq:EOMscalarContinuumNat}\\
&= -  a^{2 \alpha}\left( \tilde \phi^3 + \frac{g^2}{\lambda}\tilde\phi\tilde\chi^2\right) \\
\tilde\chi'' - a^{-2 (1  - \alpha )} \tilde\nabla^2 \tilde\chi + (3 - \alpha)\frac{a'}{a} \tilde\chi'_a  &= -  a^{2 \alpha} \widetilde V_{,\tilde\chi}\,\\
&=- a^{2 \alpha}  \frac{g^2}{\lambda}\tilde\phi^2\tilde\chi
\\
\label{eq:NewFriedmannEQsII}
{a''\over a} &=  \frac{a^{2\alpha}}{3} \left( \frac{ f_*}{m_p} \right)^2 \Big[ (\alpha - 2){\widetilde E}_{K}  + \alpha {{\widetilde E}_{G}} + (\alpha + 1 ) {{\widetilde E}_V} \Big] \,,\\
\label{eq:NewFriedmannEQsI}
a'^{\,2} &= \frac{a^{2\alpha + 2}}{3} \left( \frac{ f_*}{m_p} \right)^2 \Big[ {\widetilde E}_{K}  + {{\widetilde E}_{G}} + {{\widetilde E}_V} \Big] \,.
\end{align}
\end{mdframed}

We note that the numerical schemes implemented in \CL use exclusively the second-order differential equation (\ref{eq:NewFriedmannEQsII}) to solve for the scale factor, whereas Eq.~\eqref{eq:NewFriedmannEQsI} is used simply as a constraint to monitor the accuracy of the obtained solution.

\subsection{My first run } \label{sec:MyFirstRun}

\CL comes with a set of ready-to-run models, which are available in the folder \texttt{src/models/}. In particular, the file \texttt{src/models/lphi4.h} contains the implementation of the model presented in the previous section, characterized by the potential given in Eq.~(\ref{eq:potentialExampleI}). We now show how to run the code and pass different parameters to the simulation. We also show how to modify/create model files in order to implement other scalar theories.

\subsubsection{Compilation}

First, we need to choose the location where the code will be compiled. This can be anywhere on your machine, \textbf{except in \texttt{src/} or any of its sub-folders}. As an example, let us create a \texttt{build/} directory and move inside it,
\begin{shell-sessioncode}
  cd cosmolattice
  mkdir build
  cd build
\end{shell-sessioncode}
\CL uses CMake for compilation (see Section \ref{app:CMakeArgs} for more details). The model \texttt{lphi4.h} is compiled by typing the following commands,
\begin{shell-sessioncode}
  cmake -DMODEL=lphi4 ../
  make cosmolattice
\end{shell-sessioncode}

Some explanations are of order.  The last argument of the \textcolor{gray}{\texttt{cmake}} command is the path to the CMake configuration file, which is located at the root of the \CL folders. In our case, its relative path with respect to the \texttt{build/} folder is \texttt{../}. The first argument \textcolor{gray}{\texttt{-DMODEL=lphi4}} is passed to CMake, and tells it to compile the model \texttt{lphi4.h}. Changing this argument to any other model present inside the \texttt{src/models/} will determine which model is compiled. Note that this is not a \CLns-specific CMake argument, see Appendix \ref{app:CMakeArgs} for an exhaustive list.

\begin{mdframed}
{\bf Important Note -.} Every time you call CMake, it is a good practice to first remove the \texttt{CMakeCache.txt} file that was previously generated.
\end{mdframed}

At this point, if everything went smoothly, you should have generated an executable named \texttt{lphi4}. If this is the case, move on to the next section. If not, continue reading.

\subsubsection*{Troubleshooting}

A common problem that will happen to some users at this stage is that CMake does not find your FFTW installation, typically because it is not installed in a standard path. If that is the case, you can indicate the location of FFTW by calling CMake as follows,

\begin{shell-sessioncode}
  cmake -DMODEL=lphi4 ../   #Does not work because your fftw3 is not found.
  rm CMakeCache.txt         #We want to clear the CMake before running it again.
  cmake  -DMYFFTW3_PATH="/path/to/fftw3/" -DMODEL=lphi4 ../   #And now this works!
\end{shell-sessioncode}
with \texttt{/path/to/fftw3/} the path where fftw3 is located. You can also call \textcolor{gray}{\texttt{make clean-cmake}} to remove the \texttt{CMakeCache.txt} file.

If this solves your problem, you can avoid having to specify the FFTW path each time you compile by modifying line \codeline[V2]{52} of the \texttt{CMakeLists.txt} file (located at the root of the \CL files), as highlighted below:

\vspace{0.5cm}\noindent\texttt{CMakeLists.txt:}~\inputminted[ firstline=50, lastline=53,linenos, frame=single]{CMake}{code_files/CMakeLists.txt}
where, again \texttt{/path/to/fftw3/} is the path where fftw3 is located.

\subsubsection{Running the program with an input parameter file}\label{subsec:Input-Scalars}

Now that we have generated the executable \texttt{lphi4}, we are ready to run our first simulation as follows:

\begin{shell-sessioncode}
  ./lphi4 input=../src/models/parameter-files/lphi4.in
\end{shell-sessioncode}

This will launch the model \texttt{lphi4} with the parameters specified in the input file located in \path{src/models/parameter-files/lphi4.in}. Let us have a look at it.

\vspace{0.5cm}\noindent\texttt{src/models/parameter-files/lphi4.in:}\inputminted[frame=single,linenos]{text}{code_files/lphi4.in}

One of the perks of using \CL is its very flexible way of handling parameters. The standard way of passing parameters to the program is to bundle them in an input file such as \texttt{lphi4.in}, and indicate its path when calling the program with the \texttt{input=...} argument. The structure of the input file is rather straightforward. First, if we want to pass a single parameter, we just write down its name followed by an equal sign, and then define its value. Second, if we are passing parameters that admit multiple values, these must be separated with white spaces, as e.g.~line \codeline[V2]{25} of \texttt{lphi4.in} above. And third, the character \texttt{\#} is use for comments, so everything following such character in a given line will be ignored. Note that the order in which the parameters are specified does not matter.  To sum it up, the way of defining parameters in an input file is
\begin{minted}[frame=single]{text}
  singleParameterName = value
  multipleParametersName = value1 value2 value3 ...
\end{minted}
A convenient feature of \CL is that we can also pass arguments directly through the console. Moreover, this feature can be used together with an input file: even if the argument is already specified in the file, it will always be overwritten by the one passed through the command-line. For instance,
\begin{shell-sessioncode}
  ./lphi4 input=../src/models/parameter-files/lphi4.in N=64
\end{shell-sessioncode}
will launch \texttt{lphi4} with the parameters specified in \texttt{src/models/parameter-files/lphi4.in} except for \texttt{N}, which is the size of the lattice and was specified through the command line to be $N=64$. Note that {\bf when passing arguments through the command-line, you should not use spaces around the equal sign}, so e.g.~\texttt{N=64} is correct, but \texttt{N = 64} is not. If you want to pass arguments that take multiple values, you should protect the values by double quotes as in the following example,
\begin{shell-sessioncode}
  ./lphi4 input=../src/models/parameter-files/lphi4.in initial_momenta="0 0" N=64
\end{shell-sessioncode}
In this case, both fields are initialized with zero velocity, and the lattice size is set to $N=64$.

A table of the most important parameters is the following:
\begin{center} \small
\begin{tabular}{ | m{3.7cm} | m{12.8cm}| } \hline
{\bf Parameters} & {\bf Explanation} \\ \hline
{\tt N} & Number of lattice points per dimension.\\ \hline
\tt kIR  & Infrared cutoff of the lattice \textbf{in program units}, i.e.~$\tilde{k}_{\rm IR} \equiv k_{\rm IR} / \omega_*$.  \\ \hline
\tt lSide  & Length of the box \textbf{in program units}, i.e.~$\tilde{L} \equiv L \omega_*$. \\ \hline
\tt dt & Time step of the evolution algorithm \textbf{in program units}, i.e.~$\delta \tilde{\eta}$.  \\ \hline
\tt expansion & Expanding universe or not.  If {\tt false}, the scale factor is fixed to unity and field dynamics occur in Minkowski. If {\tt true} (default value), the scale factor evolves self-consistently according to the Friedmann equations. A fixed background expansion rate can be further specified by the parameter \texttt{fixedBackground}.\\ \hline
\tt evolver & Type of evolution algorithm. Options `VV2', `VV4', `VV6', `VV8', and `VV10' solve the field equations with the velocity-verlet algorithm of the corresponding order, while `LF' solves them with the staggered-leapfrog method. Check \href{https://www.cosmolattice.net/technicalnotes}{\color{blue} https://www.cosmolattice.net/technicalnotes} for addition of new evolvers. \\ \hline
\tt t0 & Initial time of the simulation \textbf{in program units} (set to 0 by default).  \\ \hline
\tt tMax & Final time of the simulation \textbf{in program units}.  \\ \hline
\tt tOutputFreq & Time interval between the printing of \textit{frequent output} in program units.  \\ \hline
\tt tOutputInfreq & Time interval between the printing of \textit{infrequent output} in program units.  \\ \hline
\tt tOutputRareFreq & Time interval between the printing of \textit{very infrequent (rare) output} in program units.  \\ \hline
\tt kCutOff & If specified, the given cutoff (\textbf{in program units}) is imposed in the spectrum of initial fluctuations for all scalar fields: the amplitude of the field modes at larger momenta is set to zero up to machine precision. Not specifying {\tt kCutOff} implies not having an initial cut-off, whereas {\tt kCutOff = 0} implies initially vanishing fluctuations. \\ \hline
\tt fixedBackground & If set to {\tt true}, turns off the self consistent expansion and replace it by a fixed background expansion. \\ \hline
\tt omegaEoS & Barotropic equation of state parameter $\omega\equiv p/\rho$ required for a fixed background expansion. Note that {\bf fractions are not allowed}, so one must write e.g.~for a RD universe, `{\tt omegaEoS=0.333}' instead of `{\tt omegaEoS=1/3}'. \\ \hline
\tt H0 & Initial Hubble rate ({\bf in GeV}) used for the fixed background expansion. \\ \hline

\end{tabular}
\end{center}

An exhaustive list of all available parameters is given in Appendix \ref{App:TableParameters}. Let us remark that many of these parameters have default values, so they do not need to be specified unless needed otherwise. However, there are certain parameters that do not have a default value, which we call \textit{mandatory} parameters (such as $N$). These must be always specified.

\subsubsection{Outputs}

The code generates three different kinds of output files, classified according to the information they contain:

\begin{itemize}
    \item \textbf{Averages:} Volume-averages of field quantities (e.g.~mean amplitude, variance), or other quantities that are independent of the lattice site (e.g.~scale factor). Their printing frequency is controlled by the parameter {\tt tOutpufFreq}.
    \item \textbf{Spectra:} Binned spectra of fields and other quantities in momentum space. Their printing frequency is controlled by the parameter {\tt tOutpufInfreq}. Their computation is generally more time-consuming than averages, as they imply Fourier transforming the whole lattice forth and back.
    \item \textbf{Snapshots:} Values of a certain quantity (such as energy components) at all points of the lattice. These files are printed in HDF5 format, and their printing frequency is controlled by the parameter {\tt tOutputRareFreq}. Their computation is also typically  time-consuming, and the produced files are significantly heavier than other files.
\end{itemize}

When the simulated model contains only scalar singlets, the files generated by the simulation and the information they contain are the following:

\begin{itemize}
    \item {\tt average\_energies.txt}: Energy density volume-averaged components in the following order:

    $\tilde{\eta}$, $\tilde{E}_K^{(0)}$, $\tilde{E}_G^{(0)}$, ... , $\tilde{E}_K^{(N_s-1)}$, $\tilde{E}_G^{(N_s-1)}$, $\tilde{E}_V^{(0)}$, ... , $\tilde{E}_V^{(N_p-1)}$, $\langle \tilde{\rho} \rangle$.

    \item {\tt average\_energy\_conservation.txt}:

    \begin{itemize}
    \item[$\star$] If there is no expansion, it prints the relative degree of energy conservation as follows:

    $\tilde{\eta}$, $1 - \frac{\langle \tilde{\rho} (\tilde{\eta} ) \rangle}{\langle \tilde{\rho} (\tilde{\eta}_*  ) \rangle}$.

    \item[$\star$]  In the case of self-consistent expansion, it prints the degree of relative conservation of the Hubble constraint as follows [here LHS and RHS are the left and hand sides of Eq.~(\ref{eq:NewFriedmannEQsI})]:

    $\tilde{\eta}$, $\frac{\langle\text{LHS} - \text{RHS}\rangle}{\langle \text{LHS} + \text{RHS}\rangle}$, $\langle  \text{LHS} \rangle$, $\langle \text{RHS} \rangle$.
    \end{itemize}

    \item {\tt average\_scalar\_[n].txt}: One file is produced for each individual scalar field, containing the following averages: $\tilde{ \eta}$, $\langle \tilde{\phi}_n \rangle$, $\langle \tilde{\phi}'_n \rangle$, $\langle \tilde{\phi}_n^2 \rangle$, $\langle \tilde{\phi}^{'2}_n \rangle$, $\text{rms} (\tilde{\phi}_n)$, $\text{rms} (\tilde{\phi}'_n)$.

    \item {\tt average\_scale\_factor.txt}: Scale factor and their derivatives: $\tilde{\eta}$, $a$, $a'$, $a' \over a$.

    \item {\tt spectra\_scalar\_[nfld].txt}: One file is produced for each individual scalar field, in which the following data is printed: $\tilde{k}$,  $\widetilde{\Delta}_{\tilde \phi} (\tilde k)$, $\widetilde{\Delta}_{\tilde \phi'} (\tilde k)$, ${\tilde n}_{\tilde k}$, and $\Delta n_{bin}$ (multiplicity = lattice sites/bin), where $\tilde k \equiv k/\omega_*$, the dimensionless power spectra are related to their dimensionful counterparts [see Eq.~(\ref{eq:discretePS})] by $\Delta_{\phi} \equiv \widetilde \Delta_{\tilde\phi}f_*^2$, and $\Delta_{\phi'} \equiv \widetilde\Delta_{\tilde\phi'}f_*^2\omega_*^2$, and we have defined a (dimensionless) {\it lattice occupation number} as
    \be
    \label{eq:OccuppationNum}
    {\tilde n}_{\tilde k}(\tilde{\bf n}) = \frac{a^2\tilde{L}^3}{2N^6} \frac{f_*^2}{\omega_*^2} \left( \tilde{\omega}_{\tilde k(\tilde{\bf n})} \left\langle\Big| \tilde \phi_{\tilde k(\tilde{\bf n})}\Big|^2\right\rangle_{R(\tilde {\bf n})} + \frac{a^{2(1 - \alpha)}}{\tilde{\omega}_{\tilde k(\tilde{\bf n})}} \left\langle\Big| \tilde{\phi}_{\tilde k(\tilde{\bf n})}' + \frac{a'}{a} \tilde{\phi}_{\tilde k(\tilde{\bf n})}  \Big|^2\right\rangle_{R(\tilde {\bf n})} \right)  ~\,, \ee
    where $\left\langle ... \right\rangle_{R(\tilde {\bf n})}$ denotes an angular average within the spherical shells of each bin, and we have defined $\tilde{\omega}_{\tilde k(\tilde{\bf n})}^2 \simeq {\tilde k^2(\tilde{\bf n})} + a^{2} \left\langle \frac{\partial^2 \tilde{V} }{\partial \tilde{\phi}^2}\right\rangle_{L^3}$, with $\left\langle ... \right\rangle_{L^3}$ a volume average. We note that the occupation number ${\tilde n}_{\tilde k}(\tilde{\bf n})$ is independent of either $N$ or $\delta \tilde x$ at a given $\tilde n$. However, the total number density of particles $n_\phi \equiv \int {d^3k\over(2\pi)^3} n_k = {\tilde n}_\phi\omega_*^3$ with ${\tilde n}_\phi \simeq {1\over {\tilde L}^3}\sum_{\tilde {\bf n}}{\tilde n}_{\tilde k(\tilde{\bf n})}$, may depend of the choice of $N$ and $\delta \tilde x$, as these determine the infrared and ultraviolet extremes of momenta in the reciprocal lattice.

    \item {\tt average\_spectra\_times.txt}: List of times at which the above spectra are outputted.

    \item {\tt [energy\_term]\_scalar.h5}: If indicated in the parameters file, these files contain the entire distribution throughout the lattice of a given energy component (e.g.~kinetic, gradient, potential).

    \item {\tt [model\_name].infos}: Information about the run, such as parameter values, time of onset and end of the simulation, etc.

\end{itemize}

\begin{mdframed}
{\bf Important Note -.} Definitions of the output variables just defined may vary in successive updates of \CLns. For instance, whereas the form to calculate the scalar power spectrum $\widetilde{\Delta}_{\tilde \phi} (\tilde k)$ was unique in \CL {\tt v1.0}, in \CL {\tt v1.1} one can choose between multiple options to output different versions of a scalar field power spectrum, see \href{https://cosmolattice.net/technicalnotes/}{\color{blue} {\tt Technical Note I}} for further details. In general, we invite the user to visit regularly \href{https://www.cosmolattice.net/technicalnotes}{\color{blue} https://www.cosmolattice.net/technicalnotes} to check for new features (ranging from new definitions, options, algorithms, interactions, etc) incorporated in successive updated versions of \CLns.
\end{mdframed}

\subsection{The model file} \label{sec:TheModelFile}

To define a model, the only file we really need to modify/create is the corresponding \textit{model file} specified through the \textcolor{gray}{\texttt{-DMODEL=...}} argument of CMake. In the previous example, the model file used was \texttt{src/models/lphi4.h}. In this section, we will review carefully the contents of such file, so that you can imitate its structure to write a new \texttt{model.h} file, for the simulation of any other scenario with (canonically normalized) interacting (singlet) scalar fields.

\subsubsection{Definition and declaration of the model}
\label{subsubsec:DefAndDeclModel}

The first thing we need to do is to specify the matter content of our theory. In our model example we have two scalar fields, with a potential composed by the sum of two terms: the quartic potential of the inflaton, and the quadratic interaction between the inflaton and the preheat field, see Eq.~(\ref{eq:potentialExampleI}). This is indicated in the following extract of code:

\insertcppcode{src/models/lphi4.h}{16}{35}{code_files/lphi4.h}

If we want to include mode fields with further potential interactions, we simply need to modify the values of \mintinline{C++}{NScalars} and \mintinline{C++}{NPotTerms} of the \textcolor{blue}{\texttt{ModelPars}} structure accordingly. We can also include other types of matter fields by adding extra parameters in this structure, such as gauge fields or complex singlets/doublets, but we wait for Section \ref{sec:MyFirstModelGauge} to explain this. Once {\tt NScalars} and {\tt NPotTerms} are fixed, we give a name to our model:

\insertcppcode{src/models/lphi4.h}{36}{37}{code_files/lphi4.h}

\textbf{The name of the model must match the one of the file (without the .h extension)}. Following this prescription, the name of our example model is \texttt{lphi4}. This information is then passed to a macro \texttt{MakeModel} to generate a customizable skeleton class:


\insertcppcode{src/models/lphi4.h}{39}{42}{code_files/lphi4.h}

Our customized model is then derived from this skeleton, as follows:

\insertcppcode{src/models/lphi4.h}{46}{49}{code_files/lphi4.h}

\subsubsection{Setting-up the model}

The next step is to declare and define some \textbf{model specific parameters}, which can be used for example as an input for the different potential terms. In our example, these are mainly the inflaton self-coupling $\lambda$, and the coupling constant of the interaction $g$. This scenario is characterized by parametric resonance of the preheat field, so it is also convenient to introduce an additional third parameter called the \textit{resonance parameter}, defined in terms of the other two as $q \equiv g^2/\lambda$. Parameters are declared in the model file as follows:

\insertcppcode{src/models/lphi4.h}{50}{54}{code_files/lphi4.h}

We now need to assign values to these parameters. We will do it inside the constructor of our model, namely the function in charge of its initialization:

\insertcppcode{src/models/lphi4.h}{65}{85}{code_files/lphi4.h}

Lines \codeline[V2]{67} and \codeline[V2]{68} are simply the declaration of our constructor. The argument \mintinline{C++}{parser} is the {\it parameter parser} which we will use to add and get model specific arguments. The argument  \mintinline{C++}{runPar} contains generic parameters such as the lattice spacing and the box size, see Section \ref{App:TableParameters} for more information. The \mintinline{C++}{toolBox} is an object that contains information about the internal mechanics of the library, and of which any model needs to be aware. For example, it is used to instantiate the field variables and perform iterations over the lattice, see Section \ref{sec:graybox} for more information. Anyhow, these two lines should not be modified, as they are only there to declare the constructor.

Customization starts on line \codeline[V2]{75}, where we declare a new parameter to be read either from the input file or the command line. To do so, we use the \mintinline{C++}{parser} object and its \mintinline{C++}{get<double>} function. The specification of \mintinline{C++}{double} means that we are expecting a number with double precision. The argument `\texttt{lambda}' is the name of the parameter, which is specified as \texttt{lambda=...} in the input. The parameter \texttt{"q"} on line \codeline[V2]{81} is defined in the same way. On line \codeline[V2]{84} we  compute $g$ as a function of $\lambda$ and $q$.

We now need to initialize some generic variables of the skeleton model \mintinline{C++}{Model}. In particular, we need to specify all the (non-zero) \textit{initial homogeneous components} of the different scalar fields, the variables \{$\alpha$, $f_*$, $\omega_*$\} that will define our program variables, as well as the initial effective masses of the fields.

Regarding the initial homogeneous components of the fields, we also read them from the input file:

\insertcppcode{src/models/lphi4.h}{88}{102}{code_files/lphi4.h}

The variables \mintinline{C++}{fldS0} and \mintinline{C++}{piS0} are arrays of doubles containing the homogeneous values of the scalar fields and their initial velocities. They are variables declared in the skeleton \mintinline{C++}{Model<MODELNAME>}, see Section \ref{sec:graybox} for more information. We are using the same syntax as above to retrieve parameters from the input parameter file. The only novelty is the fact that now we read parameters that take multiple values. The size of the parameter is passed after the argument \mintinline{C++}{double}, \mintinline{C++}{2} in this case. The parameter \mintinline{C++}{"initial_amplitudes"} is mandatory while \mintinline{C++}{"initial_momenta"} is optional, as by default it takes the value zero in its entries. We note that the \textbf{field initial amplitudes and initial velocities must be introduced in the parameter file in units of GeV and GeV$^2$, respectively}. In the given example, we will consider from now on the field ``0" to be the inflaton $\phi$, and the field ``1" to be the daughter or preheat field $\chi$.

Next we set up the re-scaling, as described in Eqs.~(\ref{eq:FieldSpaceTimeNaturalVariables}),(\ref{eq:Alpha-PowLaw}),(\ref{eq:lphi4-ProgVar}):

\insertcppcode{src/models/lphi4.h}{105}{114}{code_files/lphi4.h}

The code is self-explanatory. The parameters \mintinline{C++}{fStar}, \mintinline{C++}{omegaStar} and \mintinline{C++}{alpha} are declared in the skeleton class.
The last step is to set up the masses.

\insertcppcode{src/models/lphi4.h}{117}{127}{code_files/lphi4.h}

Here, we use the default function which sets the masses from the second derivatives of the potential evaluated on the initial homogeneous values of the field, see Section \ref{sec:PotDerivs} below. This function also computes the initial value of the potential, which is useful to initialise the Hubble rate. Were we want provide explicitly the initial field masses, the relevant parameter to be set is \mintinline{C++}{masses2S}, which represents an array containing the square masses of the scalar fields.

\subsubsection{The potential and its derivatives}\label{sec:PotDerivs}

The last and arguably the most important piece of information missing to be specified is the potential under consideration and its associated field derivatives. Let us start by defining the potential. We split it in \mintinline{C++}{NPotTerms}, as specified on line \codeline[V2]{26} of the model file, two in our case.

\insertcppcode{src/models/lphi4.h}{130}{157}{code_files/lphi4.h}

Here we wrote in dimensionless units, the potential defined in Eq.~(\ref{eq:PotNat}), in this example split in two terms. For each term, we need to define a function called \mintinline{C++}{auto potentialTerms(Tag<r>)}, with \texttt{r} an integer between $0$ and \mintinline{C++}{NPotTerms}. In our case, we define on line \codeline[V2]{143} as the first potential term (numbered as the 'zeroth' term) the inflaton potential $\frac{1}{4}{\tilde \phi}^4$, whereas and on line \codeline[V2]{156} we define the second potential term describing the interaction between the inflaton and the daughter field $\frac{1}{2}q\tilde\phi^2\tilde\chi^2$. The object \mintinline{C++}{fldS} is the object which contains the scalar fields. Individual fields are accessed using by calling \mintinline{C++}{fldS(s)} with \texttt{s} the number of the scalar field species, running from $0$ to \texttt{NScalars - 1}. Fields constitute an object of their own in \CLns, so they can be manipulated with many functions and (differential) operators, see Appendix~\ref{app:ImplemFunc} for an exhaustive list. Here we use two different such functions, namely multiplication (which can be used between two fields to represent their site-by-site multiplication, or between a field and a number) and an integer power function \mintinline{C++}{pow<n>}, which computes locally the $n^{th}$ integer power of the field.

The \mintinline{C++}{auto} keyword allows the compiler to automatically deduce the return type. In our case, it is essential as the expression we return are symbolic expression encoded inside the type, through the mechanism known as ``expression templates". We defer the interested reader to Appendix~\ref{app:ExprTemp} for more information. For related reasons, the syntax \mintinline{C++}{1_c} with the unusual \mintinline{C++}{"_c"} is needed as it allows to simply defined compile-time integer.

In exactly the same manner, we introduce the potential derivatives:

 \insertcppcode{src/models/lphi4.h}{171}{187}{code_files/lphi4.h}

Let us highlight the fact that the numbering of these functions needs to be consistent with your numbering of the fields. By this we mean that the function \mintinline{C++}{auto potentialTerms(Tag<0>)} corresponds to the derivative of the potential with respect to the numbered $0th$ field, \mintinline{C++}{fldS(0_c)} in the code, and so on. The derivatives of the potential are used in the equations of motion.

Finally, we also provide the second derivative of the potential with respect to the scalar field (these are needed to compute the effective masses of the fields):

 \insertcppcode{src/models/lphi4.h}{189}{205}{code_files/lphi4.h}

With this we end our presentation of the model file. Any model consisting of canonically normalized interacting singlet scalar fields can be constructed in a similar manner.

\subsection{The physics implemented in \CL} \label{sec:WhatHappensAuto}

We discuss now what actions the code executes when running a simulation with a model we have just set up. Our aim here is to provide a short overview of the different parts of the code automatically called when running a simulation, so that the user can have a full picture of what is happening at the physical level. For a deeper understanding on the implementation details of this, we refer the reader to Section~\ref{sec:graybox}.

\subsubsection{Initialization of fluctuations} \label{sec:InitScalar}

In most applications, on top of the corresponding initial homogeneous modes set up before, we require to initiate as well  a set of fluctuations for each of the simulated scalar fields. An extensive description of how to set initial conditions for scalars field in a lattice can be found in Section~7 of \cite{Figueroa:2020rrl}. Here we basically summarize the most important results, as well as explain their implementation in \CLns.

Whenever considering initial quantum vacuum fluctuations, these can be written in the continuum as
\begin{eqnarray}\label{eq:SpectrumContinuum}
\left\langle \delta \phi^2 \right\rangle = \int d\log k~\Delta_{\delta \phi}(k)\,,\hspace{0.6cm} \Delta_{\delta \phi}(k) \equiv {k^3\over 2\pi^2} \mathcal{P}_{\delta \phi}(k)\,,\hspace{0.6cm} \left\langle {\delta \phi}_{\bf k}{\delta \phi}_{{\bf k}'} \right\rangle \equiv (2\pi)^3\mathcal{P}_{\delta \phi} (k)\delta(\bf{k}-\bf{k}') \ ,
\end{eqnarray}
where $\langle \cdots \rangle$ represents an ensemble average, and the power spectrum is given by
\begin{eqnarray}
\Delta_{\delta \phi}(k) \equiv {k^3\over 2\pi^2} \mathcal{P}_{\delta \phi}(k)\,,~~~~ \mathcal{P}_{\delta \phi} (k) \equiv {1\over 2a^2\omega_{k,\phi}}\,,~~~~ \omega_{k,\phi} \equiv \sqrt{k^2 + a^2m_{\phi}^2} \,,~~~~ m_{\phi}^2 \equiv \frac{\partial^2 V}{\partial \phi^2}\Big|_{\phi = \bar{\phi}_*} \ . \label{eq:QuantumFlucts}
\end{eqnarray}
In this expression, $\omega_{k,\phi}$ is the comoving frequency of the mode, and $m_{\phi}$ is the effective mass of the field, evaluated in terms of the initial homogeneous amplitude $\bar{\phi}_*$ of the field.

In \CLns, this is mimicked by imposing the following sum of left- and right-moving waves to the field amplitude at each lattice point in momentum space,
\bea
  \delta \tilde \phi ({  \bf \tilde{n}}) &=& \frac{1}{\sqrt{2}} (|\delta \tilde \phi_1 ({  \bf \tilde{n}})|  e^{i \theta_1 ({   \bf \tilde{n}}) } + |\delta \tilde \phi_2 ({   \bf \tilde{n}})| e^{i \theta_2 ({   \bf \tilde{n}}) }   ) \label{eq:fpr_influct} \ , \\
 \delta \tilde {\phi}' ({   \bf \tilde{n}}) &=& {1\over a^{1-\alpha}}\left[\frac{i \tilde{\omega}_k}{\sqrt{2}}  \left(|\delta \tilde \phi_1 ({   \bf \tilde{n}})| e^{i \theta_1 ({   \bf \tilde{n}}) } - |\delta \tilde \phi_2  ({   \bf \tilde{n}})| e^{i \theta_2 ({   \bf \tilde{n}}) }   \right)\right]  - \tilde{\mathcal{H}}  \delta \tilde \phi ({   \bf \tilde{n}})\ , \label{eq:fpr_influct2} \eea
where $\tilde{\omega}_k \equiv \sqrt{\tilde{k}^2(\tilde{\bf n}) + a^2\tilde{m}_{\phi}^2}$, and $\tilde{\mathcal{H}} \equiv a^\alpha H / \omega_*$. In these expressions, $\theta_1 ({\bf \tilde{n}})$ and $\theta_2 ({\bf \tilde{n}})$ are two random independent phases (at each Fourier site) drawn from a uniform distribution in the range $[0, 2\pi)$, and $|\delta \tilde{\phi}_1 ({\bf \tilde{n}})|$ and $|\delta \tilde{\phi}_2 ({\bf \tilde{n}})|$ are random amplitudes (set also at each Fourier site), drawn from a {\it Rayleigh} distribution with expected square amplitude given by
\begin{eqnarray} \label{eq:QuantumFlucts2}
    \left\langle | \delta \tilde \phi (\tilde{\bf n})|^2\right\rangle \equiv \left({\omega_*\over f_*}\right)^2\left({N\over \delta \tilde{x}}\right)^3{1\over 2a^2\sqrt{\tilde{k}^2(\tilde{\bf n}) + a^2\tilde{m}_{\phi}^2}} \ , \hspace{0.6cm} \tilde{m}_{\phi}^2 \equiv \frac{\partial^2 \tilde{V}}{\partial \tilde{\phi}^2 } (\tilde{\phi} = \tilde{\bar{\phi}}_* )  \ .
\end{eqnarray}
Drawing both phases and modulus amplitudes as above is mathematically equivalent to drawing $\delta \tilde \phi ({  \bf \tilde{n}})$ and $\delta \tilde \phi'({  \bf \tilde{n}})$ as Gaussian random fields. The way in which this initialization is implemented in the code is discussed in more detail in Section \ref{subsec:Initializers}.

\subsubsection{Evolution of the system} \label{eq:evolution-sc}

Let us now explain how our evolution algorithms solve the field EOM and the evolution of the scale factor. For the purpose of this discussion, we will assume self-consistent expansion of the universe, as the cases of no expansion and fixed background are just a particularization of this. We also consider the case of one single scalar field, as the generalization to multiple fields is immediate.

The equations of motion are solved by using a Hamiltonian evolution scheme, in which we conveniently define the conjugate momenta of $\phi$ and $a$ as $\piSc \equiv  a^{3-\alpha}\tilde\phi'$ and $b \equiv a'$ respectively. Note that a a given time, $\piSc$ varies from point to point in the lattice, while $a$ is just a number. The equations of motion (\ref{eq:EOMscalarContinuumNat}) and (\ref{eq:NewFriedmannEQsII}) can then be written as the following set of four first-order differential equations,
\begin{alignat}{2}
\phi' &\,\, =\,\,  a^{-(3 - \alpha)} \piSc \ , && \\
a' & \,\, = \,\, b \ ,&&  \\
(\piSc)' &\,\,=\,\, \mathcal{K}_{\phi}[a,\tilde\phi] \,\, && \,\,\equiv \,\,  - a^{3 + \alpha} \widetilde V_{,\tilde\phi}  + a^{1 + \alpha} {\widetilde \nabla}^{2} \tilde\phi \ , \\
b' &\,\,=\,\,  \mathcal{K}_a\hspace*{-1mm}\left[a,{\widetilde E}_K^\phi,{\widetilde E}_G^\phi,{\widetilde E}_V\right]
&& \,\,\equiv\,\,  \frac{a^{2\alpha+1}}{3}{f_*^2\over m_p^2}\left[ (\alpha-2) {\widetilde E}_K^\phi + \alpha {\widetilde E}_G^\phi + (\alpha+1) {\widetilde E}_V  \right] \ , \nonumber
\end{alignat}
where the kinetic, gradient, and potential energies of the field in program variables, are given by
\be {\widetilde E}_K \equiv \frac{1}{2 a^{6} }\sum_{i}\left\langle \tilde\pi_i^2 \right\rangle\,,~~~ {\widetilde E}_G \equiv \frac{1}{2 a^2 }\sum_{i,k} \left\langle ({\widetilde\nabla}_k \tilde \phi_{i})^2 \right\rangle\,, ~~~{\widetilde E}_V \equiv \left\langle \widetilde{V}(\lbrace \tilde\phi_j\rbrace) \right\rangle\, \ee
Above $\mathcal{K}_{\phi}$ and $\mathcal{K}_a$ are the {\it kernels} of $\phi$ and $a$ respectively. Note that the definition of $\piSc$ has been chosen so that the kernels are not functions on the time-derivatives of their respective variables.

The current version of \CL has implemented already two different numerical schemes to solve these equations: {\it staggered leapfrog} and {\it velocity verlet}. The main difference between both schemes is the times at which the fields and momenta are defined during the evolution of the system. In staggered leapfrog, fields and momenta are specified at different times, so they must be synchronized each time an output is printed. On the contrary, in velocity verlet, fields and momenta are obtained at the same time, so no such synchronization is needed. A detailed account on the properties and ins and outs of how these algorithms work, can be found in Section 3 of~\cite{Figueroa:2020rrl}. A derivation of the adaptation of these algorithms to the particular problem of the dynamics of (canonically normalized) interacting (singlet) scalar fields, can be found in Section 4 of~\cite{Figueroa:2020rrl}.

The use of standard $\mathcal{O}(\delta \tilde{\eta}^2)$ accurate staggered leapfrog and velocity verlet algorithms in \CLns, can be specified in the input file as {\tt evolver=LF} or {\tt evolver=VV2}, respectively. Although both have an accuracy of order $\mathcal{O}(\delta \tilde{\eta}^2)$, {\tt LF} only needs two steps by iteration, while {\tt VV2} needs three. Therefore, {\tt VV2} is slower than {\tt LF}, typically by a factor $\sim$ 30$\%$-50$\%$ in our test runs. The velocity verlet algortihm has the advantage, however, that it is really a family of algorithms, which can be implemented with successive improved accuracy, from $\mathcal{O}(\delta \tilde{\eta}^4)$, to $\mathcal{O}(\delta \tilde{\eta}^6)$, $\mathcal{O}(\delta \tilde{\eta}^8)$, and $\mathcal{O}(\delta \tilde{\eta}^{10})$. These improved algorithms are already implemented in \CLns, and to use them you simply need to specify in the parameter file, {\tt evolver=VV4}, {\tt VV6}, {\tt VV8}, or {\tt VV10}, respectively. Such improved algorithms conserve energy much better than {\tt LP} or {\tt VV2}, but they are naturally slower, as they require more steps per iteration (the more the higher the accuracy of the integrator). For a discussion on the construction of all these integrators and dedicated versions of them to the dynamics of (canonically normalized) interacting (singlet) scalar field dynamics, we refer again the reader to Sections~3 and~4 of~\cite{Figueroa:2020rrl}.

\section{My first model of gauge fields}
\label{sec:MyFirstModelGauge}

~~~~~In this section we explain how to simulate in \CL a model containing both charged scalar fields and gauge fields (either Abelian or non-Abelian).

\subsection{Scalar-gauge field dynamics: program variables}

\CL is capable of simulating scalar-gauge field theories in an expanding universe. The action of a generic theory (with canonically normalized scalar fields) that can be simulated by \CLns is the following
\begin{eqnarray}
S \hspace*{-0.18cm}&=&\hspace*{-0.18cm} - \int d^4 x \left\{\frac{1}{2}\partial_{\mu} \phi \partial ^{\mu}\phi + (D_{\mu}^A \varphi)^{*}(D_A^{\mu} \varphi) +  (D_{\mu}\Phi )^{\dagger} (D^{\mu} \Phi) + \frac{1}{4} F_{\mu \nu} F^{\mu \nu} + \frac{1}{2}{\rm Tr}\{G_{\mu \nu}G^{\mu \nu}\} + V(\phi,|\varphi|, |\Phi|) \right\} \ , \nonumber \\ \vspace*{-0.5cm}
\label{eq:Lagrangian}
\end{eqnarray}
which contains several types of scalar and gauge fields. In order to simplify notation, we have only added one copy of each field species, but \CL can also handle multiple fields of the same kind. This theory contains three types of scalar fields: a singlet $\phi$, a $U(1)$-charged scalar (complex field) $\varphi$, and a $SU(2)$ doublet $\Phi$. The latter two can be written in terms of real components as in Eq.~(\ref{eq:ChargedScalars}). The complex field $\varphi$ can be charged under a $U(1)$ gauge symmetry, while the doublet can be charged under both $U(1)$ and $SU(2)$. The scalar potential of the theory is $V = V(\phi, |\varphi|, |\Phi|)$, which depends on $\phi$, as well as on the modulus of the complex and doublet scalars, $|\varphi|$ and $|\Phi|$. The corresponding covariant derivatives and field strengths in action (\ref{eq:Lagrangian}) are defined as
\bea
D_{\mu}^{\rm A} & \equiv &  \partial _{\mu} - i Q_A^{(\varphi)} g_{_A}A_\mu \ ,
\label{eq:AbCovDerivCont}\\
D_{\mu} &\equiv & 
\mathcal{I}D^{\rm A}_\mu
- i g_B Q_B B_{\mu}^a \,T_a  \,\, = \,\, \mathcal{I}\left( \partial _{\mu} - i Q_A^{(\Phi)} g_{_A}A_\mu \right)
- i g_B Q_B B_{\mu}^a \,T_a \ , \label{eq:CovDerivCont}\\
F_{\mu \nu} &\equiv & \partial_{\mu}  A_{\nu} - \partial_{\nu} A_{\mu} \ , \label{eq:FmnAbelian}\\
G_{\mu \nu} & \equiv & \partial_{\mu} B_{\nu} - \partial_{\nu} B_{\mu} - i[B_\mu,B_\nu] \ , \label{eq:GmnNonAb}
\eea
where $Q_{A}^{(\varphi)}$ and $Q_{A}^{(\Phi)}$ are the Abelian charges of $\varphi$ and $\Phi$ respectively, $Q_B$ is the non-Abelian charge of $\Phi$, $g_A$ and $g_B$ are the corresponding gauge couplings, and $\mathcal{I}$ is the 2$\times$2 identity matrix. Note that, using the properties of the $SU(N)$ generators, $G_{\mu \nu}$ can be written as
\begin{eqnarray}
G_{\mu \nu} \equiv G_{\mu \nu}^a T_a  \ , \hspace{0.4cm} G_{\mu \nu}^a \equiv \partial_{\mu} B_{\nu}^a - \partial_{\nu} B_{\mu}^a + f^{a b c} B_{\mu}^b B_{\nu}^c \ ,
\end{eqnarray}
where $f_{abc}$ are the structure constants of the SU(N) group, determined by the relation $[T_a, T_b] = i f_{abc} T_c$. \CL is only implemented (at least for the time being) for $SU(2)$, for which we simply have $T_a \equiv \sigma_a /2$, with $\sigma_a$ the Pauli matrices. In \CL we evolve the fields in the temporal gauge, so $A_{0} = B_0^a = 0$. Furthermore, we define the Abelian and non-Abelian electric and magnetic fields as follows,
\be\label{eq:ElectricMagneticDefs}
\mathcal{E}_i \equiv F_{0i} , \,\,\,\,\,\,\,\,  \mathcal{B}_i = \frac{1}{2} \epsilon_{i j k} F^{j k} , \,\,\,\,\,\,\,\,   \mathcal{E}_i^a \equiv G_{0i}^a , \,\,\,\,\,\,\,\,  \mathcal{B}_i^a = \frac{1}{2} \epsilon_{i j k} G^{j k}_a \ , \ee
with $\epsilon_{ijk}$ the Levi-Civita symbol. These expressions represent gauge-invariant physical quantites.


\begin{mdframed}
{\bf Important Note -.} As described in Section \ref{subsec:LatticeScalars} in the context of scalar theories, in the lattice we operate in a set of dimensionless spacetime and field variables called \textbf{program variables}. For scalar theories, these were defined in Eq.~(\ref{eq:FieldSpaceTimeNaturalVariables}) in terms of the three constants $\{f_*,\omega_*,\alpha\}$, that must be judiciously chosen for each specific model. The same definitions hold for the scalar sector(s) of scalar-gauge theories, where we also introduce new dimensionless program variables for the gauge fields. Putting all program variables together, we have
\begin{align}
  d\tilde\eta \equiv a^{- \alpha} \omega_* dt\ , \hspace{0.4cm}
  d\tilde x^i \equiv \omega_* dx^i\ ,
  \hspace{0.4cm}
 \tilde\phi = \frac{\phi}{f_*} \ , \hspace{0.4cm}
  \tilde\varphi = \frac{\varphi}{f_*} \ , \hspace{0.4cm} \widetilde{\Phi} = \frac{\Phi}{f_*} \ , \hspace{0.4cm}  \widetilde{A}_\mu=\frac{A_\mu }{\omega_*} \ , \hspace{0.4cm} \widetilde B_{\mu}^a = \frac{B_{\mu}^a}{\omega_*} \ . \label{eq:GaugeProgramVar}
\end{align}
\end{mdframed}

\begin{mdframed}
{\bf Important Note -.} \CL can run with an arbitrary number of scalar singlets, $U(1)$ complex scalars and $SU(2)$ doublets. \CL has been only tested however when considering a single $U(1)$ gauge field and a single $SU(2)$ gauge field. In principle, the code also works with multiple $U(1)$ gauge fields (coupled or not to scalars), but this feature has not thoroughly tested, so it is deactivated by default: the program crashes when a model is written with more than one $U(1)$ gauge field. It can be re-activated at one own's risk by commenting out line \codeline[V2]{216} in the file  \texttt{src/include/CosmoInterface/abstractmodel.h}. In the case of $SU(2)$ gauge fields, only one of such fields can be considered at once.
\end{mdframed}

We can see that the U(1)- and SU(2)-charged scalars are re-scaled in the same way as the singlet scalar fields. The gauge fields, however, are instead re-scaled by the parameter $\omega_*$. Similarly, we define program variables for the field strengths and covariant derivatives as follows:
\bea
\widetilde{F}_{\mu \nu} \equiv F_{\mu \nu} / \omega_*^2\ , \hspace{0.4cm} \widetilde{G}_{\mu \nu}^a \equiv G_{\mu \nu}^a / \omega_*^2\ , \hspace{0.4cm} \widetilde{D}_{\mu}^A \equiv D_{\mu}^A / \omega_*\ , \hspace{0.4cm} \widetilde{D}_{\mu} \equiv D_{\mu} / \omega_*\,.
\eea
The program potential is defined, as before, as
\be\label{eq:ProgramPotMultiScalar}
\tilde{V} (\tilde{\phi}, |\tilde{\varphi}|, |\widetilde{\Phi}|) \equiv \frac{1}{f_*^2 \omega_*^2} V(f_* \tilde \phi, f_* |\tilde \varphi|, f_* |\widetilde \Phi|  ) \ . \ee

\subsection{Scalar-gauge field dynamics: equations of motion}

In terms of the program variables, the field equations can be written as
\begin{eqnarray}
\tilde \phi'' - a^{-2(1 - \alpha)} {\widetilde \nabla}^{\,2} \tilde \phi + (3 - \alpha)\frac{{a'}}{a} {\tilde  \phi'} &=& - a^{2 \alpha} \widetilde V_{,\widetilde \phi} \ , \label{eq:singlet-eom} \\
\tilde \varphi'' - a^{-2(1 - \alpha)} {\widetilde{\vv D}}_{\hspace{-0.5mm}A}^{\,2}\tilde{\varphi} + (3 - \alpha)\frac{{a'}}{a}  {\tilde  \varphi'} &=& - \frac{a^{2 \alpha}}{2} \widetilde V_{,|\widetilde \varphi|} \cdot \frac{\tilde  \varphi}{ |\tilde  \varphi |}  \ , \label{eq:higgsU1-eom}\\
\widetilde \Phi'' - a^{-2(1 - \alpha)} {\widetilde{\vv D}}^{\,2}\widetilde \Phi + (3 - \alpha)\frac{{a'}}{a}  {\widetilde \Phi'} &=& - \frac{a^{2 \alpha}}{2} \widetilde V_{,|\widetilde \Phi|} \cdot \frac{\widetilde \Phi}{ |\widetilde \Phi |} \ , \label{eq:higgsSU2-eom}
\\
\tilde \partial_0 \widetilde F_{0i} - a^{-2(1 - \alpha )}\tilde  \partial_j \widetilde F_{ji} + (1 - \alpha) \frac{{a'}}{a} \widetilde F_{0i} &=& \left( \frac{f_*}{\omega_*} \right)^2
a^{2 \alpha}\widetilde J^A_i \ , \label{eq:U1eom}
\\
(\widetilde{\mathcal{D}}_0 )_{a b} (\widetilde G_{0i})^b - a^{-2(1 - \alpha )} (  \widetilde{\mathcal{D}}_j )_{a b} (\widetilde{G}_{ji} )^b + (1 - \alpha) \frac{{a'}}{a} (\widetilde{G}_{0i} )^b &=& \left( \frac{f_*}{\omega_*} \right)^2 a^{2 \alpha}(\widetilde{J}_i)_a \ , \label{eq:SU2eom}
\\
\tilde \partial_i \widetilde F_{0i} &=& \left( \frac{f_*}{\omega_*} \right)^2 a^2 \widetilde J^A_0 \ , \label{eq:GaussU1-eom}\\
(\widetilde{\mathcal{D}}_i )_{a b} (\widetilde{G}_{0i})^b &=& \left( \frac{f_*}{\omega_*} \right)^2 a^2(\widetilde{J}_0)_a \,, \label{eq:GaussSU2-eom}
\end{eqnarray}
where $(\widetilde{\mathcal{D}}_{\nu}O)_a = (\widetilde{\mathcal{D}}_{\nu})_{a b}O_b \equiv ( \delta_{a b}  \tilde \partial_{\nu} - f_{abc} \tilde B_{\nu}^c ) O_b$, and the currents are given by
\begin{eqnarray}
\label{eq:AbelianCurrent}
\hspace{1.8cm} \widetilde J_A^\mu & \equiv & 2 g_AQ_A^{(\varphi)} \mathcal{I}m [ \tilde \varphi^{*} ( \widetilde{D}_A^{\mu} \tilde \varphi )] + 2 g_AQ_A^{(\Phi)} \mathcal{I}m [ \widetilde \Phi^\dag (\widetilde D^{\mu} \widetilde \Phi  )]\,,\\
\label{eq:NonAbelianCurrent}
\hspace{1.8cm} \widetilde J_a^\mu & \equiv & 2g_BQ_B\mathcal{I}m [ \widetilde \Phi^{\dag} T_a( \widetilde{D}^{\mu} \widetilde \Phi )]\,.
\end{eqnarray}

Similarly, we define the \textit{program energy density} and \textit{program pressure density} as
\bea \tilde{\rho} \equiv \frac{\rho}{f_*^2 \omega_*^2}  \hspace{-0.2cm} &=& \hspace{-0.2cm} \tilde{K}_{\phi} + \tilde{K}_{\varphi} + \tilde{K}_{\Phi} + \tilde{G}_{\phi} + \tilde{G}_{\varphi} + \tilde{G}_{\Phi} +  \tilde{K}_{U(1)} + \tilde{G}_{U(1)} + \tilde{K}_{SU(2)} + \tilde{G}_{SU(2)} + \tilde{V} \ , \\
\tilde{p} \equiv \frac{p}{f_*^2 \omega_*^2} \hspace{-0.2cm} &=& \hspace{-0.2cm} \tilde{K}_{\phi} + \tilde{K}_{\varphi} + \tilde{K}_{\Phi} -{1\over3}(\tilde{G}_{\phi} + \tilde{G}_{\varphi} + \tilde{G}_{\Phi}) + {1\over3}  (\tilde{K}_{U(1)} + \tilde{G}_{U(1)} + \tilde{K}_{SU(2)} + \tilde{G}_{SU(2)} ) - \tilde{V} \ ,  \nonumber
\eea
where each of the individual kinetic, gradient, and potential energy contributions are
\bea
\begin{array}{lcl} \label{eq:energy-contrib}
\widetilde {K}_{\phi} &=& \frac{1}{2 a^{2\alpha} } \tilde \phi'^2 \vspace{0.1cm}\\
\widetilde {K}_{\varphi} &=& \frac{1}{a^{2\alpha} } (\widetilde D_0^A \widetilde  \varphi)^*(\widetilde D_0^A \widetilde \varphi)
\vspace{0.1cm}\\
\widetilde {K}_{\Phi} &=& \frac{1}{a^{2\alpha} } (\widetilde D_0 \widetilde \Phi )^\dag(\widetilde D_0 \widetilde \Phi)
\vspace{0.1cm}\\
\end{array}
\hspace{0.1cm};\hspace{0.3cm}
\begin{array}{lcl}
\widetilde {G}_{\phi} &=& \frac{1}{2 a^2} \sum_i (\tilde \partial_i \tilde \phi)^2
\vspace{0.1cm}\\
\widetilde {G}_{\varphi} &=& \frac{1}{a^2} \sum_i (\widetilde D_i^A \varphi)^*(\widetilde D_i^A \widetilde \varphi)
\vspace{0.1cm}\\
\widetilde {G}_{\Phi} &=& \frac{1}{a^2} \sum_i (\widetilde  D_i\widetilde  \Phi)^\dag(\widetilde D_i \widetilde \Phi)
\vspace{0.1cm}\\
\end{array}
\hspace{0.1cm};\hspace{0.3cm}
\begin{array}{lcl}
\widetilde {K}_{U(1)} &=& \frac{1}{2 a^{2 + 2 \alpha}} ( {\omega_* \over f_*} )^2  \sum_{i} \widetilde  F_{0i}^2
\vspace{0.1cm}\\
\widetilde {K}_{SU(2)} &=& \frac{1}{2 a^{2 + 2 \alpha}} ( {\omega_* \over f_*} )^2 \sum_{a,i} (\widetilde G_{0i}^a)^2
\vspace{0.1cm}\\
\widetilde {G}_{U(1)} &=& \frac{1}{2 a^4} ( {\omega_* \over f_*} )^2 \sum_{i,j<i} \widetilde F_{ij}^2
\vspace{0.1cm}\\
\widetilde {G}_{SU(2)} &=& \frac{1}{2 a^4} ( {\omega_* \over f_*} )^2 \sum_{a,i,j<i}  (\widetilde G_{ij}^a)^2  \, . \vspace*{0.2cm}\\
\end{array}
\\
\text{(Kinetic-Scalar)} \hspace{2.5cm} \text{(Gradient-Scalar)} \hspace{2.75cm} \text{(Electric \& Magnetic)} \hspace{1cm} \nonumber
\eea

If the expansion of the Universe is self-consistent, i.e.~it is sourced by volume averages of the energy and pressure densities of the simulated fields, the scale factor evolution can be obtained from the Friedmann equations
\begin{eqnarray}\label{eq:FriedmannHubble}
\mathcal{H}^2 \,\, \equiv \,\, \frac{a'^{\,2}}{a^2} &\hspace{-0.1cm}=\hspace{-0.1cm}&  \frac{a^{2 \alpha}}{3} \left( \frac{f_*}{m_p}\right)^2 \left[ \widetilde E_K^{\phi} + \widetilde E_K^{\varphi} + \widetilde E_K^{\Phi} + \widetilde E_G^{\phi} + \widetilde E_G^{\varphi} + \widetilde E_G^{\Phi} + \widetilde E_K^A + \widetilde E_K^B + \widetilde E_G^A + \widetilde E_G^B + \widetilde E_V \right] \,,
\\
\label{eq:FriedmannDDa}
{a''\over a} &\hspace{-0.1cm}=\hspace{-0.1cm}& \frac{a^{2 \alpha}}{3} \left( \frac{f_*}{m_p}\right)^2 \left[ (\alpha-2)(\widetilde E_K^{\phi} + \widetilde E_K^{\varphi} + \widetilde E_K^{\Phi}) + \alpha(\widetilde E_G^{\phi} + \widetilde E_G^{\varphi} + \widetilde E_G^{\Phi}) + (\alpha + 1)\widetilde E_V \right.\\
&& \hspace{5.1cm}\left. +~ (\alpha-1)(\widetilde E_K^A + \widetilde E_K^B + \widetilde E_G^A + \widetilde E_G^B) \right] \ , \nonumber
\end{eqnarray}
where we have defined the following volume-average energy contributions: $E_{K}^{f} = \langle \widetilde{K}_{f} \rangle$ and $E_{G}^{f} = \langle \widetilde{G}_{f} \rangle$ for the scalar fields $f=\phi,\varphi,\Phi$; $\widetilde E_{K}^{A} = \langle \widetilde{K}_{U(1)} \rangle$, $\widetilde E_{G}^{A} = \langle \tilde{G}_{U(1)} \rangle$, $\widetilde E_{K}^{B} = \langle \widetilde{K}_{SU(2)} \rangle$, and $\widetilde E_{G}^{B} = \langle \widetilde{G}_{SU(2)} \rangle$ for the gauge fields, and $\widetilde{E}_V = \langle \widetilde{V} \rangle$ for the potential energy. Instead, if the expansion is sourced by an external energetically-dominant fluid with constant equation of state $w$, $a(\eta)$ is given by the power-law function (\ref{eq:ScaleFactorPowerLaw}).

The evolution algorithms implemented in \CL use a discretized versions of Eqs.~(\ref{eq:singlet-eom})-(\ref{eq:SU2eom}) to solve for the field dynamics, and a lattice version of Eq.~(\ref{eq:FriedmannDDa}) to solve for the scale factor. Eqs.~(\ref{eq:GaussU1-eom}) and (\ref{eq:GaussSU2-eom}) are the Gauss constraints of the U(1) and SU(2) gauge sectors respectively, which must be satisfied (by their lattice version counterparts) during all times during the simulation. Analogously, Eq.~(\ref{eq:FriedmannHubble}), which represents the Hubble constraint, must also be satisfied (again by its lattice analogue) all throughout the simulation. In a sense, this checks the ability of a given integrator to conserve energy). \CL monitors in particular the degree of conservation of both the Gauss and Hubble constraints, providing in this way a procedure for checking the validity of the numerical integration of the EOM. Our discretization techniques guarantee that the Gauss constraints are obeyed up to machine precision, see Ref.~\cite{Figueroa:2020rrl} for details. On the other hand, the Hubble constraint (\ref{eq:FriedmannHubble})  holds numerically only to a certain degree of approximation, possibly reaching down to machine precision (depending on the model) only in the case of the highest order integrators like {\tt VV10}.

\subsection{The model and input files for gauge field theories}

Let us now explain how to implement a model with Abelian and non-Abelian gauge fields in \CLns. Two gauge models are already implemented in \CLns: {\tt lphi4U1}, which includes a complex scalar charged under a $U(1)$ gauge symmetry and one Abelian gauge field (like in scalar-electrodynamics), and {\tt lphi4SU2U1}, which contains a scalar field charged under $U(1)\times SU(2)$, and hence with both one Abelian and one non-Abelian gauge field (similar to the electroweak sector of the Standard Model). These are models ready-to-use as templates for your own models. In this manual, we will explain the implementation of the second model as an example, as it contains all possible field species and interactions that can be currently simulated with \CLns.

The model {\tt lphi4SU2U1} consists of a doublet $\Phi$ charged under a $SU(2)\times U(1)$ gauge group, coupled to one Abelian gauge field $A_{\mu}$ and one non-Abelian one $B_{\mu}^a$ via the previously defined covariant derivative. We also couple $\Phi$ to a scalar singlet $\phi$ and to a $U(1)$-charged scalar field $\varphi$ via quadratic interactions. We will consider a scenario in which the doublet $\Phi$ acts as the dominant mother field, and simulate its non-perturbative decay into the gauge fields and the other scalars, which is induced via parametric field excitations of the other fields due to the coherent oscillations of $\Phi$. For the sake of simplicity, we will consider like if this was a preheating scenario where $\Phi$ plays the role of the inflaton field, coupled to the daughter fields $\phi, \varphi, A_\mu$ and $B_\mu = T_aB_\mu^a$. In particular, we will implement the theory described by action (\ref{eq:Lagrangian}) with the following scalar potential,
\be V(\phi,|\varphi|,|\Phi |) = \lambda |\Phi|^4  + g^2 |\Phi |^2 \phi^2 + 2 h^2 |\Phi |^2 |\varphi|^2 \ , \label{eq:PotGauge} \ee
where $\lambda$, $g$, and $h$ are dimensionless coupling constants. The first term is the inflaton potential, and the second and third terms are quadratic interactions between the inflaton and $\phi$ and $\varphi$ respectively. As said, the theory contains one field of each kind. We will assume that at the onset of the simulation (say at the end of slow-roll inflation), the inflaton amplitude has an initial non-zero homogeneous component with modulus $|\Phi| = |\bar{\Phi}_*|$, while the homogeneous components of the rest of scalar and gauge fields are set to zero (this is natural as they are massive during inflation, with their mass induced by the large amplitude of $\Phi$, and hence they are not excited initially). The post-inflationary oscillations of $\Phi$ trigger a resonant growth of $\phi$ and $\varphi$ due to resonant effects via the second and third terms of the potential, and also a parametric excitation of the gauge fields $A_{\mu}$ and $B_{\mu}^a$ due to their coupling to $\Phi$ via the covariant derivative $(D_{\mu}\Phi )^{\dagger} (D^{\mu} \Phi) \in g_A^2 |\Phi|^2 A_{\mu}^2$, $g_B^2 |\Phi|^2 {B_{\mu}^{a}}^2$.

This scenario is implemented in the model file {\tt src/models/lphi4SU2U1.h}, and the corresponding input parameter file is in {\tt src/models/parameter-files/lphi4SU2U1.in}. Most of the parameters defined in the input file are the same as for singlet scalar fields, see Section~\ref{subsec:Input-Scalars}. However, there are several extra parameters that need to be set now. First, let us specify the initial homogeneous components of all scalar fields as follows:

\inserttxtcode{src/models/parameter-files/lphi4-SU2U1.in}{19}{26}{code_files/lphi4SU2U1.in} 

Above {\tt initial\_amplitudes} and {\tt initial\_momenta} contain the initial homogenous amplitudes of the scalar singlet, $\phi_*$ and $\dot{\phi}_*$. If we had more than one singlet, their initial conditions would be specified in a vector form, as explained in Section~\ref{sec:MyFirstModelScalars}. Parameters {\tt cmplx\_field\_initial\_norm} and {\tt cmplx\_momentum\_initial\_norm} contain the initial \textbf{absolute values} of the complex field amplitude and its time-derivative , i.e.~$|\varphi_*| \equiv \sqrt{(\varphi_{0*}^2 + \varphi_{1*}^2)/2}$ and $|\dot{\varphi}_*| \equiv \sqrt{(\dot{\varphi}_{0*}^2 + \dot{\varphi}_{1*}^2)/2}$ respectively (although in this example we set these to zero). Similarly, parameters {\tt SU2Doublet\_initial\_norm} and {\tt SU2Doublet\_initial\_momenta\_norm} contain $|\Phi_*| \equiv \sqrt{ \sum_{n=0}^3 \varphi_{n*}^2 / 2}$ and $ |\dot{\Phi}_*| \equiv \sqrt{ \sum_{n=0}^3 \dot{\varphi}_{n*}^2 / 2 }$ respectively. \textbf{As before, initial amplitudes must be introduced in $\text{GeV}$, and initial derivatives in $\text{GeV}^2$}.

We also need to specify the parameters that deal with the gauge couplings and charges in the covariant derivatives. This is done as follows:

\inserttxtcode{src/models/parameter-files/lphi4-SU2U1.in}{28}{33}{code_files/lphi4SU2U1.in} 

Here, {\tt gU1s} fixes $g_A$, {\tt gSU2s} fixes $g_B$, {\tt CSU1Charges} fixes $Q_A^{(\varphi)}$, {\tt SU2DoubletU1Charges} fixes $Q_A^{(\Phi)}$, and {\tt SU2DoubletSU2Charges} fixes $Q_B$. Although in our example we only consider one field for each species, multiple couplings and charges can also be specified in vector form.

Finally, we also want to specify the model parameters that appear in the potential (\ref{eq:PotGauge}). This is done as follows,

\inserttxtcode{src/models/parameter-files/lphi4-SU2U1.in}{35}{38}{code_files/lphi4SU2U1.in} 
where we have defined three parameters: {\tt lambda} (which represents $\lambda$), {\tt qG} (which contains $q_G \equiv g^2 / \lambda$, and {\tt qH} (which contains $q_H \equiv h^2 / \lambda$).

Let us now analyze the model file {\tt src/models/lphi4-SU2U1.h}, which can be used as a template to simulate different gauge field theories. We start by specifying the field content of our theory:

\insertcppcode{src/models/lphi4SU2U1.h}{16}{42}{code_files/lphi4SU2U1.h}

In lines \codeline[V2]{24}-\codeline[V2]{28}, we have specified the number of fields of each species: {\tt NScalars} refers to $\phi$, {\tt NCScalars} refer to $\varphi$, {\tt NU1Flds} refer to $A_{\mu}$, {\tt NSU2Doublet} refers to $B_{\mu}^a$, and {\tt NSU2Flds} refers to $\Phi$. In line \codeline[V2]{29} we specify the number of terms in the potential, which is $3$ in our case. Finally, in lines \codeline[V2]{34}-\codeline[V2]{38} we defined three types of \textit{coupling managers}, which deal with the couplings between the scalar and gauge fields in the covariant derivatives. {\tt U1CsCouplings} must be defined if $\varphi$ couples to $A_{\mu}$, {\tt U1SU2DoubletCouplings} must be defined if $\Phi$ couples to $A_{\mu}$, and {\tt SU2SU2DoubletCouplings} must be defined if $\Phi$ couples to $B_{\mu}^a$. Finally, in line \codeline[V2]{42}, we have specified that the name of our model is {\tt lphi4SU2U1}, in agreement with the name of the header file.

After that, inside the template model, we declare several model parameters ({\tt g}, {\tt h}, {\tt lambda}, {\tt qG}, {\tt qH}) as private variables:

\insertcppcode{src/models/lphi4SU2U1.h}{45}{58}{code_files/lphi4SU2U1.h}

We then use the parser to read the initial homogeneous components of the scalar field amplitudes and derivatives as follows:

\insertcppcode{src/models/lphi4SU2U1.h}{62}{91}{code_files/lphi4SU2U1.h}

We can see that for the scalar singlet $\phi$, the initial amplitude and derivative are read by the parser from the input file in lines \codeline[V2]{72}-\codeline[V2]{73}, in the same way as described in Section \ref{sec:MyFirstModelScalars}. As explained, these values must be stored in the variables {\tt fldS0(0\_c)} and {\tt piS0(0\_c)} respectively, where {\tt 0} denotes the label of the field. Regarding $\varphi$, in lines \codeline[V2]{76}-\codeline[V2]{77} we use the same technique to read the initial values of $|\varphi_*|$ and $|\dot{\varphi}_*|$. For convenience, we stored them in the local variables {\tt normCmplx0} and {\tt normPiCmplx0} respectively. Then, in lines \codeline[V2]{81}-\codeline[V2]{82} we specify how these amplitudes are distributed between the two components of $\varphi$, i.e.~$\varphi_0$ and $\varphi_1$ [see Eq.~(\ref{eq:ChargedScalars})]. In the example, this is done with the {\tt Complexify} function, which creates a two-component vector storing the real and imaginary parts of a complex number. For consistency, one must always guarantee that the initial components satisfy $\sqrt{(\varphi_{0*}^2 + \varphi_{1*}^2) /2} = |\varphi_*|$ and $\sqrt{(\dot{\varphi}_{0*}^2 + \dot{\varphi}_{1*}^2) /2} = |\dot{\varphi}_*|$. In the example, we have decided to set the same initial power to all components, so that $\varphi_{0*} = |\varphi_*| $ and $\varphi_{1*} = |\varphi_*|$, as well as $\dot{\varphi}_{0*} = |\dot{\varphi}_*|$ and $\dot{\varphi}_{1*} = |\dot{\varphi}_*| $. In any case, the created vectors must be stored in the model variables {\tt fldCS0(0\_c)} (for the amplitude) and {\tt piCS0(0\_c)} (for the time-derivative), where {\tt 0} is the field label.

We use a similar technique in lines \codeline[V2]{85}-\codeline[V2]{91} to specify the initial conditions of $\Phi_*$ and $\dot{\Phi}_*$. First, in lines 53-54 we store the values of $|\Phi_*|$ and $|\dot{\Phi}_*|$ specified in the input file in the local variables {\tt normDoublet0} and {\tt normPiDoublet0} respectively. We then need to indicate how these are distributed between the four components of the doublets $\Phi_*$ and $\dot{\Phi}_*$, i.e.~$\varphi_{n*}$ and $\dot{\varphi}_{n*}$ for $n=0,1,2,3$ [see Eq.~(\ref{eq:ChargedScalars})]. For consistency, we must always ensure that $\sqrt{ \sum_{n=0}^3 \varphi_{n*}^2 / 2} = |\Phi_*|$ and $\sqrt{ \sum_{n=0}^3 \dot{\varphi}_{n*}^2 / 2 } = |\dot{\Phi}_*|$. In the example, this is done in lines \codeline[V2]{90}-\codeline[V2]{91} with the {\tt MakeSU2Doublet} function, which creates a SU(2) doublet with the same amplitude for all the components (in this case, $\varphi_{n*} = |{\Phi}_*| /\sqrt{2}$ and $\dot{\varphi}_{n*} = |\dot{\Phi}_*| / \sqrt{2}$. Finally, the corresponding initial SU(2) doublets must be stored in the variables {\tt fldSU2Doublet0(0\_c)} and {\tt piSU2Doublet0(0\_c)}, with {\tt 0} again denoting the field label.

We now proceed to read model parameters {\tt qG}, {\tt qH}, and {\tt lambda} from the input file in the usual way, as well as to compute new parameters {\tt g} and {\tt h} as follows,

\insertcppcode{src/models/lphi4SU2U1.h}{93}{103}{code_files/lphi4SU2U1.h}

The next step is to define appropriate program variables for the model, as well as set the initial masses. The potential of the dominating oscillatory field is quartic, similar to the scalar case considered in Section \ref{sec:MyFirstModelScalars}, so mimicking Eq.~(\ref{eq:lphi4-ProgVar}), we choose them as
\begin{align}
  f_*=|\overline{\Phi}_{*} |\,,~~~~ \omega_*=\sqrt{\lambda} | \overline{\Phi}_* |,~~~~ \alpha=1 \ .
\end{align}
This is done in the code as follows,

\insertcppcode{src/models/lphi4SU2U1.h}{105}{115}{code_files/lphi4SU2U1.h}

Finally we call the generic function responsible to set the masses of the matter fields together with the initial potential
\insertcppcode{src/models/lphi4SU2U1.h}{117}{122}{code_files/lphi4SU2U1.h}

We now need to specify the scalar potential of our field theory. As for scalar singlet theories, any gauge field theory in \CL is implemented by means of the \textit{program potential}, defined in Eq.~(\ref{eq:ProgramPotMultiScalar}). In our example, it is given by
\bea
\widetilde V( \tilde\phi, |\tilde\varphi|, |\tilde\Phi| ) \equiv \frac{1}{f_*^2 \omega_*^2}V(f_*\tilde \phi,f_*|\tilde \varphi |, f_*|\tilde \Phi |) = |\tilde\Phi|^4 + \frac{g^2}{\lambda}|\tilde\Phi|^2\tilde\phi^2 + 2 \frac{h^2}{\lambda}|\tilde\Phi|^2|\tilde\varphi|^2  \ .
\eea

The potential is composed of three different terms: the quartic potential of the inflaton, the quartic coupling between the inflaton and $\phi$, and the quartic coupling between the inflaton and $\varphi$. We label them as terms 0, 1, and 2 respectively. The different terms are implemented in the model file with the {\tt potentialTerms} function as described in Section \ref{sec:MyFirstModelScalars}. Scalars are given by the variable {\tt fldS(X\_c)} as before, with {\tt X} the field label. Fields $\varphi$ and $\Phi$ are given instead by the variables {\tt fldCS(X\_c)} and {\tt fldSU2Doublet(X\_c)} respectively. Of course, the potential only depends on the moduli of these fields, which we can obtain with the {\tt norm} function as {\tt norm(fldCS(X\_c))} and {\tt norm(fldSU2Doublet(X\_c))} respectively. The three terms of the potential are then specified as follows:

\insertcppcode{src/models/lphi4SU2U1.h}{126}{149}{code_files/lphi4SU2U1.h}

We now need to specify the first derivatives of $\tilde{V}$ with respect $\tilde\phi$, $|\tilde\varphi|$, and $|\tilde\Phi|$. These must be specified in the functions {\tt potDeriv(Tag<0>)}, {\tt potDerivNormCS(Tag<0>)} and {\tt potDerivNormSU2Doublet(Tag<0>)} respectively, with {\tt Tag<X>} indicating the corresponding field label (there is only one copy for each species, so it is {\tt Tag<0>} in the three cases). This is done as follows:

\insertcppcode{src/models/lphi4SU2U1.h}{151}{171}{code_files/lphi4SU2U1.h}

Finally, we need to specify the second derivatives of $\tilde{V}$ with respect $\tilde{\phi}$, $\tilde{\varphi}$, and $\tilde{\Phi}$. These are implemented in the functions {\tt potDeriv2(Tag<X>)}, {\tt potDeriv2NormCS(Tag<X>) } and {\tt potDeriv2NormSU2Doublet(Tag<X>)} as follows:

\insertcppcode{src/models/lphi4SU2U1.h}{173}{193}{code_files/lphi4SU2U1.h}

\subsection{Output files}

We indicate here the different output files:

\begin{itemize}

    \item {\tt average\_scalar\_[nfld].txt}: $\tilde{ \eta}$, $\langle \tilde{\phi} \rangle$, $\langle \tilde{\phi}' \rangle$, $\langle \tilde{\phi}^2 \rangle$, $\langle \tilde{\phi}^{'2} \rangle$, $\text{rms} (\tilde{\phi})$, $\text{rms} (\tilde{\phi}')$

    \item {\tt average\_norm\_cmplx\_scalar\_[nfld].txt}:  $\tilde{ \eta}$, $\langle |\tilde{\varphi} |\rangle$, $\langle | \tilde{\varphi}' |\rangle$, $\langle |\tilde{\varphi} |^2 \rangle$, $\langle |\tilde{\varphi}'|^{2} \rangle$, $\text{rms} (|\tilde{\varphi}|)$, $\text{rms} (|\tilde{\varphi}'|)$

    \item {\tt average\_[Re/Im]\_cmplx\_scalar\_[nfld].txt}: $\tilde{ \eta}$, $\langle \tilde{\varphi}_n \rangle$, $\langle \tilde{\varphi}'_n \rangle$, $\langle \tilde{\varphi}_n^2 \rangle$, $\langle \tilde{\varphi}^{'2}_n \rangle$, $\text{rms} (\tilde{\varphi}_n)$, $\text{rms} (\tilde{\varphi}'_n)$

    \item {\tt average\_norm\_SU2Doublet\_[nfld]\.txt}:  $\tilde{ \eta}$, $\langle |\tilde{\Phi} |\rangle$, $\langle | \tilde{\Phi}' |\rangle$, $\langle |\tilde{\Phi} |^2 \rangle$, $\langle |\tilde{\Phi}'|^{2} \rangle$, $\text{rms} (|\tilde{\Phi}|)$, $\text{rms} (|\tilde{\Phi}'|)$

    \item {\tt average\_SU2Doublet\_[nfld]\_[n].txt}: $\tilde{ \eta}$, $\langle \tilde{\varphi}_n \rangle$, $\langle \tilde{\varphi}'_n \rangle$, $\langle \tilde{\varphi}_n^2 \rangle$, $\langle \tilde{\varphi}^{'2}_n \rangle$, $\text{rms} (\tilde{\varphi}_n)$, $\text{rms} (\tilde{\varphi}'_n)$

    \item {\tt average\_norm\_[U1]\_[nfld].txt}: $\tilde{ \eta}$, $\langle {|\vec{\widetilde{ \mathcal E}}|} \rangle$, $\langle {|\vec{\widetilde{\mathcal B}}|} \rangle$, $\langle {|\vec{\widetilde{ \mathcal E}}|^2} \rangle$, $\langle {|\vec{\widetilde{\mathcal B}}|^2} \rangle$, $\text{rms} (|\vec{\widetilde{ \mathcal E}}|)$, $\text{rms} (|\vec{\widetilde{\mathcal B}}|)$

    \item {\tt average\_norm\_[SU2]\_[nfld].txt}:

    $\tilde{ \eta}$, $\sum_a \langle {|\vec{\widetilde{ \mathcal E^a}}|} \rangle$, $\sum_a \langle {|\vec{\widetilde{\mathcal B^a}}|} \rangle$, $ \sum_a \langle {|\vec{\widetilde{ \mathcal E^a}}|^2} \rangle$, $\sum_a \langle {|\vec{\widetilde{\mathcal B^a}}|^2} \rangle$, $\sum_a \text{rms} (|\vec{\widetilde{ \mathcal E^a}}|)$, $\sum_a \text{rms} (|\vec{\widetilde{\mathcal B^a}}|)$

    \item {\tt average\_energies.txt}:

    $\tilde{\eta}$,
    $\tilde{E}_K^{(\phi, 0)}$,
    $\tilde{E}_G^{(\phi, 0)}$, ... ,
    $\tilde{E}_K^{(\phi, N_s-1)}$,
    $\tilde{E}_G^{(\phi, N_s-1)}$,
    $\tilde{E}_K^{(\varphi, 0)}$,
    $\tilde{E}_G^{(\varphi, 0)}$, ... ,
    $\tilde{E}_K^{(\varphi, N_c-1)}$,
    $\tilde{E}_G^{(\varphi, N_c-1)}$,  \newline
    $\tilde{E}_K^{(\Phi, 0)}$,
    $\tilde{E}_G^{(\Phi, 0)}$, ... ,
    $\tilde{E}_K^{(\Phi, N_d-1)}$,
    $\tilde{E}_G^{(\Phi, N_d-1)}$,
    $\tilde{E}_K^{(A, 0)}$,
    $\tilde{E}_G^{(A, 0)}$, ... ,
    $\tilde{E}_K^{(A, N_{u1}-1)}$,
    $\tilde{E}_G^{(A, N_{u1}-1)}$,  \newline
    $\tilde{E}_K^{(B, 0)}$,
    $\tilde{E}_G^{(B, 0)}$, ...,
    $\tilde{E}_K^{(B, N_{s2}-1)}$,
    $\tilde{E}_G^{(B, N_{s2}-1)}$,
    $\tilde{E}_V^{(0)}$, ...,
    $\tilde{E}_V^{(N_p-1)}$,
    $\langle \tilde{\rho} \rangle$

    \item {\tt average\_energy\_conservation.txt}:

    \begin{itemize}
    \item If no expansion: $\tilde{\eta}$, $1 - \frac{\langle \tilde{\rho} (\tilde{\eta} ) \rangle}{\langle \tilde{\rho} (\tilde{\eta}_*  ) \rangle}$

    \item If self-consistent expansion: $\tilde{\eta}$, $\frac{\langle\text{LHS} - \text{RHS}\rangle}{\langle \text{LHS} + \text{RHS}\rangle}$, $\langle  \text{LHS} \rangle$, $\langle \text{RHS} \rangle$, \newline
    where LHS and RHS are the left and hand sides of Eq.~(\ref{eq:FriedmannHubble}).

    \end{itemize}

    \item {\tt average\_gauss\_[U1/SU2]\_[nfld].txt}: $\tilde{\eta}$, $\frac{\langle \sqrt{(\text{LHS} - \text{RHS})^2} \rangle}{\langle \sqrt{(\text{LHS} + \text{RHS})^2} \rangle}$,  $\langle \sqrt{(\text{LHS} - \text{RHS})^2} \rangle$ , $\langle \sqrt{(\text{LHS} + \text{RHS})^2} \rangle$.

    where LHS and RHS are the left and hand sides of Eqs.~(\ref{eq:GaussU1-eom}) (for the U(1) sector) and Eqs.~(\ref{eq:GaussSU2-eom}) (for the SU(2) sector).

    \item {\tt average\_scale\_factor.txt}: $\tilde \eta$, $a$, $a'$, $a' \over a$

    \item {\tt spectra\_scalar\_[nfld].txt}:  $\tilde{k}$,  $\widetilde{\Delta}_{\tilde \phi} (\tilde k)$, $\widetilde{\Delta}_{\tilde \phi'} (\tilde k)$, ${\tilde n}_{\tilde k}$, $\Delta n_{bin}$
    \item {\tt spectra\_norm\_cmplx\_scalar\_[nfld].txt}:
    $\tilde{k}$,  $\widetilde{\Delta}_{\widetilde\varphi} (\tilde k)$, $\widetilde{\Delta}_{\widetilde\varphi'} (\tilde k)$, ${\tilde n}_{\tilde k}$, $\Delta n_{bin}$
    \item {\tt spectra\_norm\_SU2Doublet\_scalar\_[nfld].txt}: $\tilde{k}$, $\widetilde{\Delta}_{\widetilde\Phi} (\tilde k)$, $\widetilde{\Delta}_{\widetilde\Phi'} (\tilde k)$, ${\tilde n}_{\tilde k}$, $\Delta n_{bin}$
    \item {\tt spectra\_norm\_[U1/SU2]\_[nfld].txt}:
    $\tilde{k}$, $ \widetilde{\Delta}_{\widetilde{\mathcal{E}}}(\tilde k)$
    $ \widetilde{\Delta}_{\widetilde{\mathcal{B}}}(\tilde k)$, $\Delta n_{bin}$
\end{itemize}

\subsection{The physics implemented in \CL}

\subsubsection{Initial conditions}
\label{subsubsec:initialConditionsNonAb}

In Section \ref{sec:InitScalar} we summarized how the initial conditions for scalar singlets are imposed.
Here we explain now how we set the initial conditions to complex scalars and SU(2) doublets, as well as to the Abelian and non-Abelian gauge fields. We denote the time at which the initial conditions are imposed (i.e.~the initial time of the simulation) as $t_*$, and all quantities with a $*$ subindex must be understood to be evaluated at that time: for example, $\varphi_* \equiv \varphi ({\bf x}, t_* )$ for complex scalars, and $\dot{\varphi}_* \equiv \dot{\varphi} ({\bf x}, t_* )$ for the doublets.

The initialization of the complex scalars and $SU(2)$ doublets is very similar to the scalar singlets: they consist in a homogeneous amplitude chosen by the user, over which a set of fluctuations is imposed. However, we must take into account that these fields have multiple components. As described above, in \CL the user can specify the initial absolute values $|\varphi_*|$ and $|\Phi_*|$ in the input file, and then decide how to distribute this power between the different components in the model file. However, as the scalar potential only depends on $|\varphi|$ and $|\Phi|$, we can always rotate the system so that all components have the same initial homogeneous amplitudes. Therefore, for the complex scalars we can impose
\bea \varphi_* &=& \frac{1}{\sqrt{2}} (\varphi_{0*} + i \varphi_{1*} ) \hspace{0.4cm} \Longrightarrow \hspace{0.4cm}
 \varphi_{n*} \equiv |\varphi_*|  +  \delta \varphi_{n*} ({ \bf x})  \ , \hspace{0.4cm} [n=0,1] \ ,  \\
 \dot{\varphi}_{*} &=& \frac{1}{\sqrt{2}} (\dot{\varphi}_{0*} + i \dot{\varphi}_{1*} ) \hspace{0.4cm} \Longrightarrow \hspace{0.4cm}
  \dot{\varphi}_{n*}  \equiv |\dot{\varphi}_*|  +  \delta \dot{\varphi}_{n*} ({ \bf x})  \ , \hspace{0.4cm} [n=0,1] \ ,
\eea
where $\delta \phi_{n*} (\vec{x})$ and $\delta \dot{\phi}_{n*} (\vec{x})$ are the initial spectrum of fluctuations of the field components and their corresponding time-derivatives (we present the functional form of these functions below). Similarly, for the complex doublets we impose
\bea \Phi_*  &=&
{1\over\sqrt{2}}
\left(
\begin{array}{c}
\varphi_{0*} +i\varphi_{1*} \vspace*{0.1cm}\\ \varphi_{2*} +i\varphi_{3*}
\end{array}
\right)  \hspace{0.4cm}  \Longrightarrow  \hspace{0.4cm}
 \varphi_n ({\bf x}, t_* ) \equiv \frac{|\Phi_*|}{\sqrt{2}}  +  \delta \varphi_{n*} ({ \bf x})   \ , \hspace{0.4cm} [n=0,1,2,3] \ , \\
 \dot{\Phi}_*  &=&
{1\over\sqrt{2}}
\left(
\begin{array}{c}
\dot \varphi_{0*} +i \dot \varphi_{1*} \vspace*{0.1cm}\\ \dot \varphi_{2*} +i \dot \varphi_{3*}
\end{array}
\right)  \hspace{0.4cm} \Longrightarrow  \hspace{0.4cm}
 \dot \varphi_n ({\bf x}, t_* ) \equiv \frac{|\dot{\Phi}_*|}{\sqrt{2}}  +  \delta \dot{\varphi}_{n*} ({ \bf x})    \ , \hspace{0.4cm} [n=0,1,2,3] \ .
 \eea

 On the other hand, for the Abelian and non-Abelian gauge field modes we impose
 \bea
A_i ({\bf x}, t_* ) & \equiv & 0 \ , \label{eq:Inflc1}\\
B_i^a ({\bf x}, t_* ) & \equiv & 0 \ ,  \label{eq:Inflc2} \\
\dot{A}_i ({\bf x}, t_* ) & \equiv & \delta \dot{A}_{i*} ({\bf x}) \ ,  \label{eq:Inflc3} \\
\dot{B}_i^a ({\bf x}, t_* ) & \equiv & \delta \dot{B}_{i*}^a ({\bf x}) \ ,  \label{eq:Inflc4} \eea
i.e.~the initial amplitude of the gauge fields is set \textit{exactly} to zero at all lattice points, while we only impose an initial spectrum of fluctuations to their time-derivatives (over vanishing homogeneous values). Due to this, the initial magnetic energy will be exactly zero, while a small amount of electric energy will be initially present due to the fluctuations of the time-derivatives of the gauge fields.

The initial spectrum of fluctuations for both the charged scalars and gauge fields must be imposed so that the Gauss constraint is verified initially. As long as this is true, the Gauss constraint will remain preserved during the entire dynamical evolution of the system. More specifically, let us Fourier transform the Gauss constraints (\ref{eq:GaussU1-eom}) and (\ref{eq:GaussSU2-eom}) at the initial time $t=t_*$. We get
\be {k}^i \widetilde{A}'_i ({\bf k}) = \frac{f_*^2}{\omega_*^2} \widetilde{J}_0^A ({\bf k}) \ , \hspace{0.4cm} {k}^i \widetilde{B}_i^{a'} ({\bf k}) = \frac{f_*^2}{\omega_*^2} \widetilde{J}_0^a ({\bf k}) \ , \label{eq:kAi1} \ee
where $J_0^A ({\bf k})$ and $J_0^a ({\bf k})$ are the Fourier transforms of each current. A solution of these equations is, for $\bf k \neq \bf 0$,
\be \widetilde{A}'_i ({\bf k}) = i \frac{{k}_i}{{k}^2} \frac{f_*^2}{\omega_*^2} \widetilde{J}_0^A ({\bf k}) \ , \hspace{0.4cm} \widetilde{B}^{a'}_i ({\bf k}) = i \frac{{k}_i}{{k}^2} \frac{f_*^2}{\omega_*^2} \widetilde{J}_0^a ({\bf k}) \label{eq:GaugeCurrentFluc} \ .\ee
The way in which we proceed to set fluctuations is the following. First,
we impose in the lattice the following fluctuations to the components of the charged scalars (in program units), mimicking the spectrum of fluctuations of the scalar singlets (\ref{eq:fpr_influct})-(\ref{eq:fpr_influct2}):
\begin{eqnarray}
\delta \tilde{\varphi}_{n*}({  \bf \tilde n}) &=& \frac{1}{\sqrt{2}} \left(|\delta \tilde{\varphi}_{n1} ({  \bf \tilde n})|  e^{i \theta_{n1} ({  \bf \tilde n}) } + |\delta \tilde{\varphi}_{n2} ({  \bf \tilde n})| e^{i \theta_{n2} ({  \bf \tilde n}) }   \right) \label{eq:fpr_influct3} \ , \\
\delta \tilde{\varphi}'_{n*} ({  \bf \tilde n}) &=& a^{1-\alpha}\left[\frac{1}{\sqrt{2}} i \tilde{\omega}_{k,n} \left(|\delta \tilde{\varphi}_{n1} ({  \bf \tilde n})| e^{i \theta_{n1} ({  \bf \tilde n}) } - |\delta \tilde{\varphi}_{n2}  ({  \bf \tilde n})| e^{i \theta_{n2} ({  \bf \tilde n}) }  \right)\right]  - \tilde{\mathcal{H}} \delta \tilde{\varphi}_{n} ({  \bf \tilde n})\ ,  \label{eq:fpr_influct4}
\end{eqnarray}
where $\tilde{\omega}_{k,n}  \equiv \omega_{k,n} /\omega_* =  \sqrt{\tilde{k}^2 + a^2 (\partial^2 \tilde{V} / \partial \tilde{\varphi}_n^2)}$ is the initial effective frequency of the modes of each field component in program units. Then these fluctuations generate fluctuations on the currents $\widetilde{J}_0^A ({\bf x})$, and $\widetilde{J}_0^a ({\bf x})$, which can be used to compute the corresponding fluctuations to the gauge fields in momentum space via Eqs.~(\ref{eq:GaugeCurrentFluc}). Finally, transforming back to position space we obtain $\delta \widetilde{A'}_{i*} ({\bf x})$, $\delta \widetilde{B}_{i*}^{a'} ({\bf x}) $.

However, in order for this procedure to work, we need to slightly modify the initialization of the charged field components (\ref{eq:fpr_influct3})-(\ref{eq:fpr_influct4}) with respect to the prescription for scalar singlets. In the case of scalar singlets, we would have $\theta_1 ({\bf \tilde{n}})$ and $\theta_2 ({\bf \tilde{n}})$ as two random independent phases which vary uniformly in the range $[0, 2\pi)$ from point to point, while $|\delta \tilde{\phi}_1 ({\bf \tilde{n}})|$ and $|\delta \tilde{\phi}_2 ({\bf \tilde{n}})|$ would be two amplitudes that vary from point to point according to a {\it Rayleigh} distribution with expected square amplitude given by Eq.~(\ref{eq:QuantumFlucts2}). However, in the case of charged fields we need to set the homogeneous mode of the currents to zero, i.e.~$J_0^A ({\bf k} ={\bf 0} ) = J_0^a ({\bf k} ={\bf 0} ) = 0$. As shown in Ref.~\cite{Figueroa:2020rrl}, this can be ensured if
\bea  |\delta \varphi_{n1} ({\bf k})| &=& |\delta \varphi_{n2}  ({\bf k})| \ , \hspace{3.5cm} n=0,1(,2,3) \ , \label{InConstr:1} \\
\theta_{n2} ({\bf k}) &=& \theta_{02} ({\bf k}) + \theta_{n1} ({\bf k}) - \theta_{01} ({\bf k}) \ , \hspace{0.85cm} n=1(,2,3) \ . \label{InConstr:2} \eea
In \CLns, we then generate randomly only $\theta_{01}$, $\theta_{02}$, as well as $\theta_{n1}$, $\varphi_{n1}$ for $n=1(,2,3)$, and let the other functions be imposed via constraints (\ref{InConstr:1})-(\ref{InConstr:2}).

\subsubsection{Evolution equations}

In Section \ref{eq:evolution-sc} we wrote a Hamiltonian scheme for the equations of motion of a system of scalar singlets in an expanding universe. As explained, we conveniently defined a set of conjugate momenta $\{\pi_{\phi},b\}$ for the scalar field(s) $\phi$ and the scale factor $a$, in a manner that allowed us to write the field and Friedmann equations of motion as a set of four first-order differential equations. Thanks to our definitions, the kernels of the conjugate momenta do now depend on the time-derivatives of the corresponding fields, which then allow for solving the equations of motion (their discretized version) using algorithms such as staggered leapfrog or velocity verlet.

The same idea can be applied in our present case of a scalar-gauge theory that contains both scalar and gauge fields. In particular, we can define the following momenta for each of the five field species $\{\phi,\varphi,\Phi,A_i,B_i^a\}$ and for the scale factor $a(\eta)$, as
\bea
\piSc &\equiv & a^{3-\alpha}\tilde\phi' \, ,  \\
\piSingl&\equiv &  a^{3-\alpha}\tilde\varphi' \, ,   \label{eq:momU1singlet}\\
\piDoubl&\equiv & a^{3-\alpha}\widetilde\Phi' \, ,    \label{eq:momSU2doublet}\\
\piApar_i &\equiv & a^{1-\alpha}\widetilde F_{0i} \, , \label{eq:momU1vec}  \\
\piBpar^{a}_i &\equiv & a^{1-\alpha}\widetilde G^a_{0i} \, , \label{eq:momSU2vec} \\
b &\equiv & a' \, .\eea
With these definitions the equations of motion can be then written as
\begin{alignat}{2}
(\piSc)' \,\, &=\,\, \mathcal{K}_{\phi}[a,\tilde\phi,|\tilde{\varphi}|,|\widetilde{\Phi}|] & & \,\,\equiv \,\,  - a^{3 + \alpha} \widetilde V_{,\tilde\phi}  + a^{1 + \alpha} {\widetilde \nabla}^{2} \tilde\phi  \,\ ,\\
(\piSingl)'  \,\,&=\,\, \mathcal{K}_{\varphi}[a,\tilde\phi,\tilde\varphi,|\widetilde{\Phi}|,\widetilde A_j]  & & \,\,\equiv \,\,  a^{3 + \alpha} \widetilde V_{,|\tilde\varphi|} \frac{1}{2} \frac{\tilde\varphi}{|\tilde\varphi |} + a^{1 + \alpha} {\vec{\widetilde D}}_{\hspace{-0.5mm}A}^{\,2}\tilde\varphi  \,\ , \label{eq:kernelcomplexscalar} \\
(\piDoubl)'  \,\,&=\,\, \mathcal{K}_{\Phi}[a,\tilde\phi,|\tilde\varphi|,\widetilde\Phi,\widetilde B_j^a] & & \,\, \equiv \,\,  - a^{3 + \alpha} \widetilde V_{,|\widetilde\Phi|} \frac{1}{2} \frac{\widetilde\Phi}{|\widetilde\Phi |} + a^{1 + \alpha} {\vec{\widetilde{D}}}_{\hspace{-0.5mm}A}^{\,2}\widetilde\Phi  \label{eq:kernelsSU21}  \,\ ,\\
\piApar'_i  \,\,&=\,\, \mathcal{K}_{A_i}[a,\tilde\varphi,\widetilde \Phi,\widetilde A_j] & & \,\,\equiv \,\,  a^{1+ \alpha}\widetilde J^A_i + a^{\alpha - 1}\tilde{\partial}_j \widetilde F_{ji}  \, \ ,   \\
\left(\piBpar_i^a\right)' \,\,&=\,\, \kersutwoComp[a,\widetilde\Phi,\widetilde A_j,\widetilde B_j^a] & & \,\, \equiv \,\,  a^{1+ \alpha}\widetilde J^a_i + a^{\alpha - 1}( \mathcal{\widetilde D}_j )_{a b} (\widetilde G_{ji} )^b    \,\ .
\end{alignat}
On the other hand, the equation for the evolution of the scale factor can be written as
\bea
b' & \hspace{-0.2cm}=\hspace{-0.2cm}& \mathcal{K}_a\hspace*{-1mm}\left[a,{\widetilde E}_K^\phi,{\widetilde E}_K^\varphi,{\widetilde E}_K^\Phi,{\widetilde E}_G^\phi,{\widetilde E}_G^\varphi,{\widetilde E}_G^\Phi,{\widetilde E}_K^A,{\widetilde E}_G^A,{\widetilde E}_K^B,{\widetilde E}_G^B,{\widetilde E}_V\right]  \\
&\hspace{-0.2cm} \equiv \hspace{-0.2cm} & \frac{a^{2\alpha+1}}{3}{f_*^2\over m_p^2}\left[ (\alpha-2)({\widetilde E}_K^\phi  + {\widetilde E}_K^\varphi + {\widetilde E}_K^\Phi )  +
 \alpha ({\widetilde E}_G^\phi  + {\widetilde E}_G^\varphi + {\widetilde E}_G^\Phi )+ (\alpha-1)
({\widetilde E}_K^A+{\widetilde E}_G^A + {\widetilde E}_K^B+{\widetilde E}_G^B) + (\alpha+1) {\widetilde E}_V  \right] \ , \nonumber
\eea
where $\mathcal{K}_f$ with $f=\phi,\varphi,\Phi,A_i,B_i^a$ are the kernels for the different field species. As for singlet scalar theories, \CL provides already implemented two different evolution algorithms to solve these equations: staggered leapfrog (of accuracy 2) and velocity verlet (with accuracy of order 2, 4, 6, 8, and even 10). The details of how these algorithms work, specialized for Abelian and non-Abelian gauge theories, can be found in Sections 5 and 6 of Ref.~\cite{Figueroa:2020rrl}, respectively. In the lattice, the different kernels can be discretized using the toolkits of Section~\ref{subsec:LGT}, or Section 3 of Ref.~\cite{Figueroa:2020rrl}. The details of the lattice version EOM can they all found in Sections 5 (for Abelian gauge theories) and 6 (non-Abelian gauge theories) of Ref.~\cite{Figueroa:2020rrl}.

\section{What \CL does in detail}
\label{sec:graybox}

In this section, our aim is to provide the reader with a better understanding of our physics interface, such as how the fields are actually initialized, how they are evolved, or how the measurements are actually done. In order to do this, we guide the reader through the \mintinline{C++}{main} function of \CLns, highlighting the important class/routines responsible for some specific tasks. We will then explain some of these classes in detail, namely the ones responsible for initializing, evolving and measuring the fields. These are the ones that a user will most likely want to modify to suit their specific purposes.

The \mintinline{C++}{main} function is located in the file \texttt{src/cosmolattice.cpp}, which we present now:

\insertcppcode{src/cosmolattice.cpp}{5}{17}{code_files/cosmolattice.cpp}
For completeness, we explain first the top lines in the file, which take care of the model selection. The variable \mintinline{C++}{MODELINCLUDE} is passed at compilation through the \texttt{CMake} and contains the path to the model file. It is transformed into a string by the macro \mintinline{C++}{STRINGIFY}, declared in the file  \mintinline{C++}{"TempLat/util/stringify.h"}. Next, on line \codeline[V2]{10}, we declare for convenience that we are using the namespace \texttt{TempLat}, to which all the \CL functions belong to. Lastly, we redeclare the custom model passed through CMake \texttt{MODELTYPE} to \texttt{ModelType} in line \codeline[V2]{15}. After this, we are ready to go into the main function which is executed when the software runs:

\insertcppcode{src/cosmolattice.cpp}{19}{26}{code_files/cosmolattice.cpp}
Line \codeline[V2]{21} is the standard \texttt{C++} main declaration that accepts arguments from the command line. We then instantiate a \texttt{SessionGuard} object, which is in charge of allocating and deallocating internal memory space needed by the library. The file follows as:

 \insertcppcode{src/cosmolattice.cpp}{28}{32}{code_files/cosmolattice.cpp}
Here we have created a \mintinline{C++}{ParameterParser} called \mintinline{C++}{parser}. As we saw in the previous sections, this object is used throughout the library to declare and read parameters which are parsed from the input file and the command line. The file then follows like:

\insertcppcode{src/cosmolattice.cpp}{34}{40}{code_files/cosmolattice.cpp}
Here we have created an object \mintinline{C++}{SimulationManager} called \mintinline{C++}{manager}. This object is used to deal, for instance, with printing the simulation-related information file, or with the backing up of the simulation. It also deals with the fact of whether or not the simulation is restarting from a previous one. In the first case, parameters are retrieved from the restarting simulations, as illustrated on line \codeline[V2]{39}.

\insertcppcode{src/cosmolattice.cpp}{42}{49}{code_files/cosmolattice.cpp}
We collect all the relevant simulation parameters in a \texttt{RunParameters} object called \texttt{runParams}. This object receives the information on the parameters, such as the number of points in the lattice, the frequency at which measurements are performed, etc. The file then follows as:

\insertcppcode{src/cosmolattice.cpp}{51}{68}{code_files/cosmolattice.cpp}
Here we went on defining quantities and objects useful to run the simulation. The first of these is the so-called number of ghost-cells (see Section~\ref{subsec:para} for the technical details on this). In the current version of \CLns, we use a spatial discretization that requires only one layer of the so called `ghost-cells' for parellization purposes. However, if you were to implement and use a spatial lattice derivative that requires to call lattice sites beyond the nearest neighbor positions, you would need to change this number, say to e.g.~{\tt nGhost = 2} if your derivatives require the next to nearest neighbour.

Following we create a \texttt{MemoryToolBox} object, or more specifically, a pointer to such an object. It holds many internal functions related to memory. For instance, this is where the objects that allow to iterate over the lattice are stored. This object is needed to create fields and some other objects, and as a result, it will be passed around to many functions.

We also retrieve on line \codeline[V2]{66} whether or not the current process is the root process when running in parallel. The only reason why we do this here is to be able to output in a clear manner some information about the run (typically the running time) in the command line while running in parallel. The file then follows with:

\insertcppcode{src/cosmolattice.cpp}{70}{77}{code_files/cosmolattice.cpp}
There we just created our model as defined in the model header file we specified, for instance in the \texttt{src/models/lphi4.h} of Section \ref{subsubsec:DefAndDeclModel}.
We also print its name, so that we can make sure that the model running is actually the model we wanted. We are then ready to initialize the files in the model, as described in Section \ref{sec:InitScalar} and \ref{subsubsec:initialConditionsNonAb}:

\insertcppcode{src/cosmolattice.cpp}{82}{95}{code_files/cosmolattice.cpp}
Here the initialization was taken care by an object of the type \mintinline{C++}{ModelInitializer<double>}, called {\tt initializer}, which we created on line \codeline[V2]{85} after having checked that we were starting a new simulation and not continuing another one. The \mintinline{C++}{ModelInitializer<double>} class will be explained in detail in Section~\ref{subsec:Initializers}. We also set the initial time. The file now follows with:

\insertcppcode{src/cosmolattice.cpp}{96}{105}{code_files/cosmolattice.cpp}
This is in case we are restarting from some previous simulation, so the model is reloaded from the saved state. This is taken care of by the simulation manager.

From the point of view of the physics, we have two more classes to instantiate. This is what we do next:

\insertcppcode{src/cosmolattice.cpp}{107}{114}{code_files/cosmolattice.cpp}
Above, we first create an object called \texttt{evolver} of type \mintinline{C++}{Evolver<double>}. This class, as further explained in Section \ref{subsec:Evolvers}, is in charge of calling the appropriate evolution algorithms. We also create in line \codeline[V2]{112} a \mintinline{C++}{Measurer<double>} \texttt{measurer} in charge of doing and outputing all the standard measurements. This class is explained at great length in Section \ref{subsec:Measurers}. Before starting the actual simulation, the file continues with:

\insertcppcode{src/cosmolattice.cpp}{117}{122}{code_files/cosmolattice.cpp}
Here we print in line \codeline[V2]{117} the parameters the program is running with. We also mandate the \texttt{manager} to create an {\it information file} which contains these parameters (so their values can be checked after the simulation is concluded). This file also indicates the time at which the simulation started, as well as the type of parallelization that was used. We are now ready to proceed to the time evolution of the system:

\insertcppcode{src/cosmolattice.cpp}{125}{154}{code_files/cosmolattice.cpp}
Here we started by checking whether we are performing any measurements at the given time step. If so, we ask the evolvers to synchronize themselves. That means that if a given algorithm works with fields which do not live at integer time step (e.g.~in leap frog conjugate momenta live at semi-integer times), before measuring we evolve the corresponding quantities so that they live at integer time steps (together with the rest of field and scale factor variables). Other algorithms like Velocity Verlet do not need this synchronization as, so the previous step is just ignored by them. Next, we let the \texttt{evolver} object to \texttt{evolve} the system, i.e.~this calls the routines to perform one iteration step of the evolution of both the scale factor, the field content, and all their associated derivatives. Finally, we check whether we are require to backup the simulation or not at the given time step. If we do, the \texttt{manager} takes care of it. Once the time evolution is done, we are ready to finalize the simulation:

\insertcppcode{src/cosmolattice.cpp}{158}{1749}{code_files/cosmolattice.cpp}
We checked on line \codeline[V2]{158} whether or not we are required to save the simulation at the end of the execution, in case we desire to restart the simulation from the same time in a future run. If so, the \texttt{manager} does it on line \codeline[V2]{169}. Finally we close the information file, printing the time at which the simulation ended and the total time it ran for.

\subsection{Evolvers}
\label{subsec:Evolvers}

In this section we present the main routines responsible for the evolution of the fields, showing explicitly how different evolution algorithms are implemented. The relevant classes are located in \path{src/include/evolvers/}. As presented above, the class which is instantiated in the main is the \mintinline{C++}{Evolver<double>}. Its sole purpose is to be able to choose at runtime between different algorithms, which are implemented in their respective classes. We show here the full class:

\insertcppcode{src/include/CosmoInterface/evolver.h}{26}{81}{code_files/evolver.h} 
The class holds a pointer to any of the implemented algorithm classes, namely  \texttt{LeapFrog} and \texttt{VelocityVerlet} for the time being, as declared on lines \codeline[V2]{75} and \codeline[V2]{76}. The relevant pointer is initialized in the constructor, depending on which algorithm was requested. This is what happens between lines \codeline[V2]{32} and \codeline[V2]{41}. The argument \texttt{pType} is an \texttt{EvolverType} enumerator. These enumerators are defined in the file \texttt{src/CosmoInterface/evolvers/evolvertype.h}, and are used to differentiate between evolvers. The \mintinline{C++}{VelocityVerlet} pointer is initialized in the default case because all the different higher-order Velocity-Verlet schemes are implemented in the same class.
This structure is repeated in the \texttt{evolve} function on line \codeline[V2]{43}, which uses the appropriate pointer to call the evolution function\footnote{Note than a more oriented object manner to implement this mechanism would have been to use {\it polymorphism}. We decided against this option to make it more accessible to users less familiar with C++.}. Note that the \mintinline{C++}{template<class Model>} template argument of the function is the mechanism we use to be able to have the \mintinline{C++}{Evolver} (or as we will see later on any class) to operate on arbitrary models. The last function we have is the \texttt{sync} function, which is used to synchronize all the fields (when needed) at integer times. It is used before measuring, and it is useful for instance in the case of staggered leapfrog.

To summarize, the \mintinline{C++}{Evolver<double>} class allows us to choose among evolvers. If the user decides to implement their own evolver, they can simply add it here to be able to choose it at run-time.

The actual characterization of each evolution algorithm is in the {\tt evolvers} files \texttt{src/include/CosmoInterface/evolvers/leapfrog.h} and \texttt{src/include/CosmoInterface/evolvers/velocityverlet.h}. In the following, we present in detail the leapfrog evolver as an example\footnote{We invite the reader interested in the details of the velocity-verlet algorithm to look them directly in the code, as the structure is pretty similar to the one of the leapfrog method presented here.}

\insertcppcode{src/include/CosmoInterface/leapfrog.h}{20}{36}{code_files/leapfrog.h} 

We see here that the only parameter that our class takes is a boolean, which specifies whether we are considering an expanding universe or not. The structure of the algorithm is really laid out in the \texttt{evolve} function:

\insertcppcode{src/include/CosmoInterface/leapfrog.h}{38}{86}{code_files/leapfrog.h} 

As explained in detail in Ref.~\cite{Figueroa:2020rrl} and outlined on the comments on lines \codeline[V2]{45-46}, the Leapfrog algorithm consists of two parts. First, we need to evolve the conjugate momenta by computing ``kicks", and then we evolve the field variables by means of the ``drifts", using the previously updated momenta. This also reflects itself in the code. The kick function are first called between line \codeline[V2]{58} and \codeline[V2]{63}, and then the drift functions are called between line \codeline[V2]{73} and \codeline[V2]{81}. In order to evolve the scale factor, we store the averages of the momenta squared after the kicks, and the averages of the fields squared after the drifts, as can be seen on lines \codeline[V2]{66} and \codeline[V2]{83}. We note a small subtlety, which manifests itself in line \codeline[V2]{56}: In order to perform the measurements, we synchronize the momenta to live at integer time steps. After a synchronization, conjugate momenta only need the to be evolved by half a time step, immediately after the call to a measurement. This is what happens when the weight variable is set to $0.5$. The file follows with:

\insertcppcode{src/include/CosmoInterface/leapfrog.h}{88}{113}{code_files/leapfrog.h} 
Here the sync function is responsible of evolving the momenta, so that they live at integer times for a measurement (e.g.~of energy outputs). It also computes the resulting momenta averages. Following, all kicks are constructed in the same way:

\insertcppcode{src/include/CosmoInterface/leapfrog.h}{115}{168}{code_files/leapfrog.h} 
The file calls the appropriate ``kick" functions defined in the folder \texttt{src/include/CosmoInterface/kernels/}. The structure of the kernels and their implementation will be presented in detail in the next section. What is perhaps worth noting here is the use of \mintinline{C++}{ForLoop} to iterate over the field. This is necessary because the iteration needs to happen during compilation, so that the kernel for the appropriate field can be returned by the kernels function.

Following, the drifts are implemented in a very similar way:

\insertcppcode{src/include/CosmoInterface/leapfrog.h}{170}{239}{code_files/leapfrog.h} 
Here we take the opportunity to present some of the versatility of \CL for introducing a new syntax to perform operations on fields. For instance, we see on line \codeline[V2]{191} that we do not iterate over the collection of fields. Indeed, with \CL we can directly operate at the level of collections, and the appropriate \mintinline{C++}{ForLoop} is expanded internally. Of course, we could write all the drifts using \mintinline{C++}{ForLoop} as for the kicks, and the result would be the same. Note also that for the scale factor, we also allow for the possibility of a fixed background expansion. In this case, the scale factor is simply given by a function of time.

The last relevant methods to be defined are the ones that store the appropriate averages, which are needed for the scale factor evolution (in the case of self-consistent evolution):

\insertcppcode{src/include/CosmoInterface/leapfrog.h}{242}{296}{code_files/leapfrog.h} 
Here we call the \mintinline{C++}{Averages} routines where the appropriate averages over fields squared are defined, see next section. It also uses the averaging function \mintinline{C++}{average} on the model potential directly, returned by \mintinline{C++}{Potential::potential(model)} (which computes the sum of the potential terms defined in the model).

The structure of the Velocity-Verlet evolver is exactly the same, so we invite the reader to have a look at the code for themselves.

\subsection{Kernels and other physics formulae}
\label{subsec:Kernels}

\CL has been designed to be as compartmentalized as possible, so that any potential change in the code has only ``local" consequences and does not propagate throughout the whole code. This is particularly true for the main field operations, and we have also attempted this in the implementation of the physics. All physical formulae that can be potentially needed for more than one routine have been extracted out and implemented in their own separate class. We have already encountered one example of this in the {\it kernels} for the evolution algorithms, as different {\it evolvers} may use the same kernels or parts of them. This aspect is also true for the averages of some fields, which are both used in the evolution routines and in the measurements.

Let us first have a look at some of the kernels, to gain a better understanding of the code structure. As an example we will have a look at the complex scalar kernel first:

\insertcppcode{src/include/CosmoInterface/evolvers/kernels/complexscalarkernels.h}{10}{31}{code_files/complexscalarkernels.h} 
As we can see, the \mintinline{C++}{class ComplexScalarKernels} is extremely simple, and its sole purpose is to contain the complex scalar field's evolution kernel defined in Eq.~(\ref{eq:kernelcomplexscalar}). As we can see on lines \codeline[V2]{29} and \codeline[V2]{30}, and correspondingly from the includes on lines \codeline[V2]{10} and \codeline[V2]{11}, the kernel uses two {\tt definitions} classes, namely \mintinline{C++}{GaugeDerivatives} and \mintinline{C++}{Potential}. The first one stores the generic gauge covariant derivative expressions and related expressions, while the later is used to compute the potential derivative with respect to the field components, as well as to compute the potential from the different potential terms defined in the model.

\insertcppcode{src/include/CosmoInterface/definitions/gaugederivatives.h}{26}{57}{code_files/gaugederivatives.h} 
Let us start with the \mintinline{C++}{GaugeDerivatives} class from {\tt definitions}. We see that this class has no internal variable, but it is just used to regroup some functions under one hood. In particular, it holds the functions capable of computing the covariant derivatives and covariant Laplacians of the matter/gauge content of the users' models (for conciseness, we show here only the covariant Laplacians). They rely on the functions \texttt{U1sForCSCovDerivs}, \texttt{U1sForSU2DoubletCovDerivs} and \texttt{SU2sForSU2DoubletCovDerivs}, which compute the appropriate combinations of link variables, that depend on which matter fields couple to which gauge field.

This is a good time to emphasize one of the advantages of having implemented  a `functional' interface through expression templates in \CL. The expressions computed by these Laplacian functions (as well as most of the other functions in \texttt{definitions}) are really abstract expressions (formulae), in the sense that they are left to be evaluated. Calling e.g.~\texttt{covLaplacianCS} does not lead to any actual computation, it simply creates an abstract expression which can be evaluated later on. In particular, it gives the opportunity to perform some ``analytical" simplifications to the expressions at compilation. For example, a redundant multiplication of the type ``$1 \cdot \phi$" with $\phi$ some field, can be detected at compilation time and replaced by $\phi$ only. This mechanism is used for instance in the \texttt{U1sForCSCovDerivs} function.

Following in the same file,

\insertcppcode{src/include
/CosmoInterface/definitions/gaugederivatives.h}{80}{92}{code_files/gaugederivatives.h} 
We encounter the function \texttt{fold\_multiply} in line \codeline[V2]{87}, which takes vectorial objects as an argument, and returns the multiplication of all the elements inside the object. In this case we use it on an array, created on line \codeline[V2]{88} by the macro \texttt{MakeArray}, which contains either the link \texttt{U1Links(model.fldU1(a),i)} or ``$1$", depending on whether the matter field $N$ couples to the $a^{th}$ $U(1)$ gauge field. The constant $1$ is represented by the object \texttt{OneType} on line \codeline[V2]{91}, which is then automatically discarded by the compiler in multiplications. For more information about how this works, we refer the reader to Appendix \ref{app:ExprTemp}.

The rest of the \texttt{GaugeDerivatives} class contains similar functions to compute the other covariant derivatives, as well as functions which implement gradients. As they are all implemented in a very similar fashion, we let the reader explore the code by themselves.

All the other functions in {\tt definitions} are implemented in a similar fashion. Rather than going through all the code, we explain below what they are used for.

\begin{itemize}
  \item \texttt{src/include/CosmoInterface/definitions/averages.h}: Computes the appropriate averages of the momenta squared and the gradient squared, used in the evolution of the scale factor. It relies on the \texttt{FieldFunctionals} class, which defines the correct sum over components.

  \item \texttt{src/include/CosmoInterface/definitions/energies.h}: Contains the correct rescaling of the conjugate momenta and the correct normalization to compute the energy contributions. These methods can be called either with averages, to obtain a single number, or with the \texttt{FieldFunctionals}, to obtain the energy distributions over the whole lattice.

    \item \texttt{src/include/CosmoInterface/definitions/fieldfunctionals.h}: Defines the appropriate sum over component to compute the relevant energy contributions of the different fields species to the Hubble laws.
  \item \texttt{src/include/CosmoInterface/definitions/gaugederivatives.h}: As reviewed in the text, defines the expressions to compute the gauge covariant derivatives.
  \item \texttt{src/include/CosmoInterface/definitions/gausslaws.h}: Defines the expressions to compute the Gauss laws in the Abelian and non-Abelian sectors.
  \item \texttt{src/include/CosmoInterface/definitions/hubblelaws.h}: Defines the expressions of the Hubble constraint.
  \item \texttt{src/include/CosmoInterface/definitions/mattercurrents.h}: Defines the expressions of the matter currents for the equations of motion. It is implemented in the same spirit as the \texttt{GaugeDerivatives}, and it is able to compute the current for generic matter content.
  \item \texttt{src/include/CosmoInterface/definitions/potential.h}: Computes the potential from the \texttt{potentialTerms} defined in the user's model. Also computes, for the complex scalar fields and the $SU(2)$ doublets, the potential derivative with respect to the field's component in terms of the potential derivative with respect the norm.
\end{itemize}

These functional forms from \texttt{definitions} are used all throughout \texttt{CosmoInterface}.

\subsection{Initializers}
\label{subsec:Initializers}

Another important aspect of the lattice simulations is the initialization. In this section, we will have a closer look at how this initialization happens in the code. First, there is a class \texttt{ModelInitializer} that synchronizes the initialization of each field type, while the specific initialization happens in dedicated classes.

\insertcppcode{src/include/CosmoInterface/initializers/modelinitializer.h}{24}{67}{code_files/modelinitializer.h} 
As previously claimed, the only purpose of the \texttt{ModelInitializer} class is to call the specific initializers in the correct order and to initialize the required averages and Hubble laws; this is what happen in the \texttt{initialize} function between lines \codeline[V2]{36} and \codeline[V2]{60}. It also holds a dedicated object \mintinline{C++}{FluctuationsGenerator<T>}, which is used by the specific class to generate random Gaussian fluctuations.

\insertcppcode{src/include/CosmoInterface/initializers/scalefactorinitializer.h}{26}{60}{code_files/scalefactorinitializer.h} 
We start with the the \texttt{ScaleFactorInitializer}. The only method it contains is \texttt{initializeScaleFactor}, shown above. We initialize the scale factor from the homogeneous initial values of the fields, assuming that the gauge fields have zero initial homogeneous components.

\insertcppcode{src/include/CosmoInterface/initializers/scalarsingletinitializer.h}{24}{39}{code_files/scalarsingletinitializer.h} 
The \texttt{ScalarSingletInitializer} is also very simple and holds only the \texttt{initializeScalars} methods. There, on line \codeline[V2]{30}, we set the initial fluctuations of the scalar fields using the \texttt{FluctuationsGenerator}. We then add on lines \codeline[V2]{37} and \codeline[V2]{38} the homogeneous initial conditions.

The \texttt{FluctuationsGenerator} generator class implements Gaussian fluctuations of the type described in Section \ref{sec:InitScalar}:

\insertcppcode{src/include/CosmoInterface/initializers/fluctuationsgenerator.h}{36}{101}{code_files/fluctuationsgenerator.h} 
The most important function here is \texttt{getNormedFluctuations}, defined between lines \codeline[V2]{62} and \codeline[V2]{67}. It first computes the correct normalization factor in momentum space by calling the \texttt{getFluctuationsNorm} function, and returning a correctly normalized \texttt{RandomGaussianField}. This \texttt{RandomGaussianField} operates in Fourier space returning Gaussian distributed random modes at every point of the Fourier lattice.
To facilitate the initialization, it also provides functions to directly compute the amplitudes as described in Eqs.~\eqref{eq:fpr_influct}-\eqref{eq:fpr_influct2}.
The function \texttt{gaussianFluctuations} initializes fluctuations of only one given field, while \texttt{conjugateGaussianFluctuations} sets the fluctuations of a field and its conjugate momentum. As we will see shortly in the rest of the initializers, this class exists because Gaussian fluctuations enter in some way or another in the initialization of all the different type of matter fields.

The next initializer we want to present is the one for the $SU(2)$ sector:

\insertcppcode{src/include/CosmoInterface/initializers/su2initializer.h}{35}{89}{code_files/su2initializer.h} 
The main function above is \texttt{initializeSU2}. It first calls the \texttt{initializeSU2Doublet} functions, which are responsible to initialize the matter sector in a way that is compatible with the Gauss law, see later on. Once this is done, it sets the initial fluctuations of the gauge fields by inverting the Gauss law, as recalled in Section \ref{subsubsec:initialConditionsNonAb} and described in detail in Ref.~\cite{Figueroa:2020rrl}. First, on line \codeline[V2]{67} we compute the matter current in real space. Then, on line \codeline[V2]{75} we invert the Gauss law in Fourier space. The current is computed in Fourier space by calling the \texttt{inFourierSpace()} method of the \texttt{Field} class. Lastly, we set the gauge links to unity on line \codeline[V2]{82}. Let us know look at how the matter fluctuations are imposed:

\insertcppcode{src/include/CosmoInterface/initializers/su2initializer.h}{91}{111}{code_files/su2initializer.h} 
We see that the  \texttt{initializeSU2Doublet} is in charge of first imposing the random fluctuations by calling \texttt{addFluctuationsSU2DoubletFromPhases}. Then, the homogeneous components are added to the fields.

\insertcppcode{src/include/CosmoInterface/initializers/su2initializer.h}{113}{178}{code_files/su2initializer.h} 

What is implemented in the function \texttt{addFluctuationsSU2DoubletFromPhases} simply corresponds to the procedure given in Eqs~(\ref{eq:fpr_influct3})-(\ref{InConstr:2}). The independent amplitudes are created first, and then these are used to construct the dependent ones afterwards. All of this is then used to initialize the fluctuations of the fields and momenta.

Once the $SU(2)$ sector is initialized, the last part which needs to be initialized is the $U(1)$ sector. This is taken care of by the \texttt{U1Initializer}. It works in a similar was a the \texttt{SU2Initializer}, so we let the interested reader to go directly to explore the code located at: \texttt{src/include/CosmoInterface/initializers/u1initializer.h}.

\subsection{Measurers}
\label{subsec:Measurers}

The measurements are built in a similar way as the initializers. All the measurements are synchronized by the \texttt{measurer} class:

\insertcppcode{src/include/CosmoInterface/measurements/measurer.h}{34}{35}{code_files/measurer.h} 
\insertcppcode{...}{67}{135}{code_files/measurer.h} 
It possesses a single method \texttt{measure}, displayed above, whose aim is to coordinate the different measurements and call field specific classes which perform field specific measurements. We start with what we call ``frequent  measurements". These call the different field and energy measurers and ask them to compute their respective observable at a ``frequent" rate defined by the user in the input parameter file. As we will see, it mostly consists in field averages and variances together with the Gauss laws and energy conservation/Hubble constraints. Once this is done, we move on to the ``infrequent measurements", in line \codeline[V2]{101}, and proceed in a similar fashion. Infrequent measurements mostly consists in fields' spectra measured at a more infrequent rate determined by the user (again in the input parameter file). Also, when using \texttt{txt} output, we store the ``infrequent" times in a file, on line \codeline[V2]{120}, to facilitate the data analysis. After this, we proceed with the ``rare measurements", corresponding to the most resource consuming. This is where for instance the three-dimensional snapshots of energy densities are measured.
Finally, the measurer is also used to print out some information to the console, see line \codeline[V2]{129}.

To understand better what measurements are performed, we will inspect specific measurers:

\insertcppcode{src/include/CosmoInterface/measurements/measurer.h}{155}{163}{code_files/measurer.h} 

We show above the different measurers that can be used by \CLns. All the field measurers are similar, so we will only present in details of \texttt{ScalarSingletMeasurer} and \texttt{SU2Measurer}. We will then move on to the \texttt{EnergiesMeasurer} and the \texttt{EnergySnapshotMeasurer}:

\insertcppcode{src/include/CosmoInterface/measurements/scalarsingletmeasurer.h}{23}{49}{code_files/scalarsingletmeasurer.h} 
We start by looking here at how our \texttt{ScalarSingletMeasurer} is initialized. It contains two arrays, namely \texttt{standardOut} and \texttt{spectraOut} which contains objects which can save respectively mean values and spectra, one for each scalar field. Here, the files are created by passing the fields and the file names are automatically generated through the fields' names.

\insertcppcode{src/include/CosmoInterface/measurements/scalarsingletmeasurer.h}{51}{61}{code_files/scalarsingletmeasurer.h} 

The averages are actually computed in the \texttt{measureStandard} function. As the names suggest, the measurements we want to perform are very standard and will be the same for all the fields. As such they are performed by an external class, called \texttt{MeansMeasurer}, whose \texttt{measure} function takes a \texttt{MeasurementsSaver}, a field, the corresponding velocity and the time at which the measurements have to be conducted.

\insertcppcode{src/include/CosmoInterface/measurements/meansmeasurer.h}{26}{41}{code_files/meansmeasurer.h} 

As we see now, this functions simply computes the average of the field, its average square and its variance, as well for the velocity, and adds it to the \texttt{MeasurementsSaver}.

After the averages are measured, we move on to the power spectra.

\insertcppcode{src/include/CosmoInterface/measurements/scalarsingletmeasurer.h}{64}{73}{code_files/scalarsingletmeasurer.h}

This is taken care of by the \texttt{measureSpectra} function of the \texttt{ScalarSingletMeasurer}. There, for each field, we compute its power spectrum as defined in Eq.~(\ref{eq:discretePS}), we compute the power spectrum of its associated velocity and the occupation number, defined in Eq.~(\ref{eq:OccuppationNum}). They are all saved in the same file. As noted in the comment, in the current implementation, it is better to perform the scale factor re-scaling of the momentum outside the \texttt{powerSpectrum} function. Indeed, when called on a \texttt{Field}, the function does not allocate extra memory to perform the Fourier transform. It does do that for any argument which is not purely a \texttt{Field}, as it would be that case had we called for instance \mintinline{C++}{powerSpectrum(piS(i) * pow(model.aI,model.alpha - 3))}.

Let us move briefly to the \texttt{SU2Measurer}, to highlight some features not used in the \texttt{ScalarSingletMeasurer}. First, at initialization, we do not create files for individual quantities. We will store the averages of the norms of the  fields. As such, the \texttt{MeasurementsSaver} are created with customs name. In the case of the $SU(2)$ sector, we are also interested in checking the conservation of Gauss' law and as a result create another file to store them. The spectra are computed also only for the norms.

\insertcppcode{src/include/CosmoInterface/measurements/su2measurer.h}{30}{54}{code_files/su2measurer.h} 
The measurements of the means are performed in the \texttt{measureStandard} function:

\insertcppcode{src/include/CosmoInterface/measurements/su2measurer.h}{56}{76}{code_files/su2measurer.h} 
We see here that we can use again the \texttt{MeansMeasurer} to measure the mean values and variances of the electric and magnetic fields. We also measure the Gauss law and how well it is satisfied. The first component of the \texttt{gaussArr} contains the violation degree of the Gauss law, while the second and third contain respectively the left-hand and right-hand sides of the Gauss law.

Finally, electric and magnetic spectra are computed in the \texttt{measureSpectra} function:

\insertcppcode{src/include/CosmoInterface/measurements/su2measurer.h}{78}{91}{code_files/su2measurer.h} 
Contrary to the scalar case, since we are anyhow measuring composite field expressions, it does not matter where we do the momentum rescaling. For readability we do it in an auxiliary field variable \texttt{ESU2}.

The \texttt{EnergyMeasurer} is also built in very similar way. It contains two \texttt{MeasurementsSaver}, one to store the different energy components and one to store the energy conservation check in the case without expansion or the Hubble law check in the case with expansion. The measurements are then performed in the \texttt{measure} function.

\insertcppcode{src/include/CosmoInterface/measurements/energiesmeasurer.h}{44}{62}{code_files/energiesmeasurer.h} 
\insertcppcode{...}{99}{109}{code_files/energiesmeasurer.h} 

One by one field species, we compute their energy contributions (this is why we do not used the stored averages, as these are summed over field species). We save them one after another in a file. We also compute the total energy of the system. Above we show explicitly only the scalar contributions, the reader can see the other contribution directly in the file of the code. After this, we also compute the contribution from the potential, term by them. We store finally the total energy of the system in the last column of the file.

The last distinct type of measurer is the \texttt{EnergySnapshotMeasurer}. In the current implementation, it can only be used with the \texttt{HDF5} library. When asked, it prints the three-dimensional distribution of the requested energy components. Below we show the code only for the scalar sectors, other being similar.

\insertcppcode{src/include/CosmoInterface/measurements/energysnapshotmeasurer.h}{27}{34}{code_files/energysnapshotmeasurer.h} 
\texttt{...}
\medskip

First, upon initialization, it determines which energy components it needs to save to file.
Then, again shown only for the scalar sector, if needed, the binary files which will contain the snapshots are created.
Finally, the energy snapshots are taken in the \texttt{measure} function:

\insertcppcode{src/include/CosmoInterface/measurements/energysnapshotmeasurer.h}{117}{123}{code_files/energysnapshotmeasurer.h} 
Species by species, we measure their energy contribution and save the three-dimensional distribution to file.

\subsubsection{Measurements Input/Output}

To conclude the section devoted to measurements, we want to briefly present the structure of the classes responsible to output the measurements to file, so that a user who wants to modify them or add its own output format, can do so easily. They are located in the \texttt{src/include/CosmoInterface/measurements/measurementsIO/} folder. They are of two types: the \texttt{MeasurementsSaver}'s, to save mean values, and the \texttt{SpectrumSaver}'s,to save spectra to file. They are designed in a similar way than the \texttt{Evolver}'s, in the sense that they both have an ``interface" class, called \texttt{MeasurementsSaver} and \texttt{MeasurementsSaver}, which redirect to the appropriate ``implementation" class, which is in charge of actually saving the output. In this way, it is straightforward to implement different output format without having to modify any other part of the code. These interfaces work together with a helper class called \texttt{FilesManager}, which has global information over all the measurement files.

Let us first discuss the \texttt{MeasurementsSaver} interface. In the current state of the code, there is only one ``standard" implementation to save the measurements, so strictly speaking there would be no need for an interface; we kept this design for future development purposes. To create a \texttt{MeasurementsSaver}, you have two different options, one which takes a \mintinline{C++}{std::string} as input and names the file this way, and another that takes a \texttt{Field} variable, naming the files according to its name (this constructor can also accept algebraic expression and name the file accordingly).

\insertcppcode{src/include/CosmoInterface/measurements/measurementsIO/measurementssaver.h}{26}{26}{code_files/measurementssaver.h} 
\insertcppcode{src/include/CosmoInterface/measurements/measurementsIO/measurementssaver.h}{35}{35}{code_files/measurementssaver.h} 

Both methods take the same arguments. First, the \texttt{FilesManager}, which is created and stored in the \texttt{Measurer} class. Its purpose is to collect generic information about the outputing procedure. This is where the choice between different interfaces should be stored. Even if it is not implemented yet, it could also be of use to create a folder structure to save the different outputs. The \texttt{amIRoot} parameter tells the \texttt{MeasurementsSaver} whether it belongs to the root \texttt{MPI} process or not, as only the root process is allowed to save to file (in the standard implementation at least). The \texttt{appendMode} specify whether the old measurements files potentially present (by mistake or when restarting a simulation) should be appended to or overwritten. The \texttt{headers} file is an array, by default empty, containing the headers for the file. Lastly, the \texttt{dontCreate} allows to call one of the \texttt{MeasurementsSaver} saver constructors without actually creating a file; this is useful to switch between different physical scenarios which do not need the same output (e.g.~expanding versus non expanding universe).

The use of the \texttt{MeasurementsSaver} is intended to be very straightforward. It relies on the method \texttt{addAverage}, which registers a value to be stored, and a \texttt{save} function which actually saves all the registered values to file.

\insertcppcode{src/include/CosmoInterface/measurements/measurementsIO/measurementssaver.h}{44}{44}{code_files/measurementssaver.h} 

\insertcppcode{src/include/CosmoInterface/measurements/measurementsIO/measurementssaver.h}{49}{49}{code_files/measurementssaver.h} 

The \texttt{SpectrumSaver} interface works in a similar fashion, except that in this case, two interfaces are already implemented: the standard one and the \texttt{HDF5} one (see Section~\ref{subsubsec:hdf5spectra}). Filenames are also constructed either from a \mintinline{C++}{std::string} or a \texttt{Field}'s name.

\insertcppcode{src/include/CosmoInterface/measurements/measurementsIO/spectrumsaver.h}{28}{28}{code_files/spectrumsaver.h} 

\insertcppcode{src/include/CosmoInterface/measurements/measurementsIO/spectrumsaver.h}{45}{45}{code_files/spectrumsaver.h} 
Its use is also straightforward and based on a single \texttt{save} function.

\insertcppcode{src/include/CosmoInterface/measurements/measurementsIO/spectrumsaver.h}{63}{64}{code_files/spectrumsaver.h} 
It takes as arguments a time $t$, and an arbitrary number of spectra. It then forwards them to the correct implementation (standard or \texttt{HDF5}) and save them to a file. The $\dots$ syntax is the modern way of \texttt{C++} to create functions for an arbitrary number of arguments. To learn about the actual implementation, we invite the interested reader explore the files in the \texttt{std} and \texttt{hdf5} sub-folders.

\section{Useful features: parallel support, backing up and others}\label{sec:UsefulFeatures}

There is clearly a vast number of physical scenarios that can be implemented in \CLns, and in order to optimize the ``physics output" from many different scenarios, we have made available a number of powerful technical features in the code. The most relevant one is the possibility of directly running any model written in \CL with multiple processors in parallel, without any modification whatsoever of the code. As we will shortly show, all it takes for the user to run their model on potentially hundreds or even thousands of processors, is a simple flag passed to the \texttt{CMake}. Before explaining this, we give in Section \ref{subsec:para} a brief explanation about what parallelization means, as well as describe what happens technically at the computation level. Any user not interested in these technical details may want to skip directly to Sections~\ref{subsubsec:para1D} and \ref{subsubsec:para2D}, where we simply explain how they can activate the parallelization option in \CLns.

Another useful feature provided by \CL is the possibility of saving up and restarting simulations, as well as the possibility of having an automatic backup every given number of iterations. This feature can also be enabled through a \texttt{CMake} flag; we elaborate about it in Section~\ref{subsec:hdf5spec}. Using the same external library (\texttt{HDF5}), we also provide the user with the possibility of saving spectra in \texttt{HDF5} format, which has the advantage of being more structured than the default text files. We explain this also in Section~\ref{subsec:hdf5spec}.

\subsection{Parallelization}
\label{subsec:para}

As we increase the size of our lattice simulations, we quickly encounter computational limitations. These are of two types. First, we are limited by the real duration (as counted by hours/days/etc by ourselves) that it takes to run a given simulation. Every time the number of points/dimension $N$ of a lattice is doubled, the execution time increases roughly by a factor $\sim 2^d$, with $d$ the number of dimensions. That is, the execution time scales with the volume of the lattice. Secondly, the memory (RAM) needed to perform the simulation also scales with the volume, as every time $N$ is doubled, the amount of required RAM memory increases by a factor $8$. Lack of memory is often a more severe limitation than the execution time, as longer execution times may be compensated by more patience (at least to some extent), while the limit on memory can not.

Both of these hindrances can be sharply mitigated by a simple idea: the use of more than one computer to perform the simulations. This is what we mean by parallelization. In spirit, it works as follows: given $n_c$ ``computers" (or ``cores", as we will refer to them), you can split your lattice into $n_p$ smaller sub-lattices. Then, instead of evolving the whole lattice on a single process, you can evolve the $n_p$ smaller sub-lattices on one or several of your $n_c$ cores, and then combine their results whenever needed. In theory, this would speed up your simulation by a factor $n_p$, and give you access to $n_p$ times more memory.

Of course, most of the problems suitable for lattice simulations involve some spatial derivatives or some kind of finite range interaction, and as a result,
the system evolved over the original lattice is not equivalent to the $n_p$ systems over smaller lattices. In the case of systems of equations which involve an interaction between neighboring sites\footnote{Of course, the idea presented here also works and is practical for interactions between sites that are a few sites apart.}, in order to be able to solve the system over the whole original lattice but splitting the evolution over the smaller lattices, it is enough for every sub-lattice to be aware of the values of the fields in the sites of its neighboring lattices, those directly adjacent to their own sides. This is clarified in Fig.~\ref{fig:1d}, where we consider the one-dimensional case and explain the case with two cores. The physical lattice $\Lambda$ consist of the field values $\phi_0$ to $\phi_7$. To perform the computation, we can subdivide it into two smaller lattices $\Lambda_1$ and $\Lambda_2$. The first one, which contains the field values $\phi_0$ to $\phi_3$, is assigned to the first process. The second, containing the values $\phi_4$ to $\phi_7$, is assigned to the second process. Now imagine that, in order to solve our system of equations, we need to compute a gradient, which we write as a forward derivative [recall Eq.~(\ref{eq:forwardbackwardd})]. When our first process tries to evaluate it around site $3$, it needs to compute $\frac{\phi_4-\phi_3}{\delta x}$, i.e.~it requires the information on the value of $\phi_4$, which belongs however to the adjacent sub-lattice.

To solve this problem, we introduce {\it ghost cells}. When two sub-lattices have a common boundary, the boundary values are stored in both sub-lattices, and whenever they are modified, the new values are communicated to their neighboring sub-lattice; boundaries are ``exchanged". Very explicitly, in our one-dimensional example of Fig.~\ref{fig:1d}, we can add an extra site (the ghost cell) to $\Lambda_1$ containing $\phi_4$, and an extra site to $\Lambda_2$ containing $\phi_3$. Whenever $\phi_3$ is modified in $\Lambda_1$, it needs to be communicated to $\Lambda_2$, and whenever $\phi_4$ is modified in $\Lambda_2$, it needs to be communicated to $\Lambda_1$. Of course, since we use periodic boundary conditions, the same needs are in place with respect the boundaries $\phi_0$ and $\phi_7$.

In higher spatial dimensions, the geometry of the boundaries to be exchanged might become more complicated, but the intrinsic idea remains the same. This is the parallelization idea implemented in \CLns, based on the use of the {\it Message Passing Interface} (\texttt{MPI}), a standard library to program the boundary exchanges. We discuss in the next sections two different parallelization strategies, and how the user can choose between one or another when using \CLns.

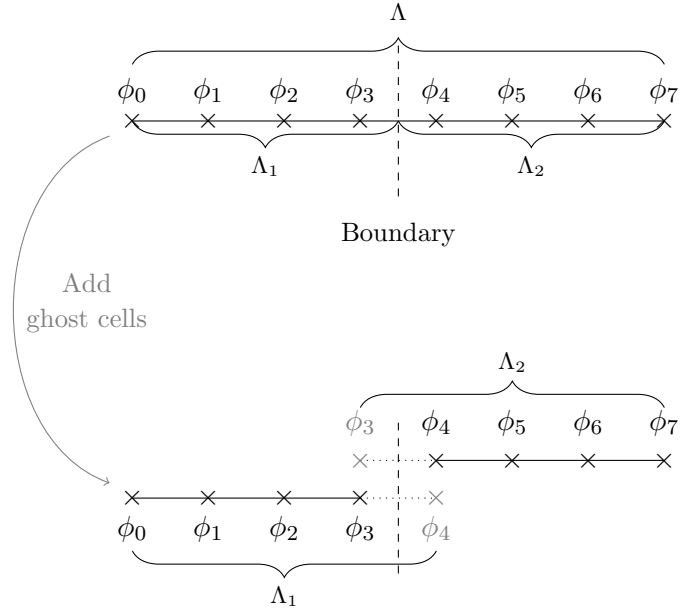
\begin{figure}
  \centering
  \begin{tikzpicture}

  \draw (0,0) -- (7,0);
  \foreach \x in{0,...,7}
  {
      \node at (\x ,0){$\times$};
      \node at (\x ,0.4){$\phi_\x$};
  }
  \draw[dashed](3.5,-1)--(3.5,1);

  \draw [decorate,decoration={brace,mirror,amplitude=10pt},xshift=0cm,yshift=-0.pt]
(0,0) -- (3.5,0) node [black,midway,yshift=-0.6cm]
{\footnotesize $\Lambda_1$};
\draw [decorate,decoration={brace,mirror,amplitude=10pt},xshift=0cm,yshift=-0.pt]
(3.5,0) -- (7,0) node [black,midway,yshift=-0.6cm]
{\footnotesize $\Lambda_2$};
\draw [decorate,decoration={brace,amplitude=10pt},xshift=0cm,yshift=-0.pt]
(0,0.75) -- (7,0.75) node [black,midway,yshift=0.7cm]
{\footnotesize $\Lambda$};

\node at (3.5,-1.5){\small Boundary};

\draw[->, color=gray] (-0.3,-0.2) to[out=200,in=160] (-0.3,-4.8);
\node[color=gray] at (-0.6,-2.15){\small Add};
\node[color=gray] at (-0.6,-2.65){\small ghost cells};

\draw (0,-5) -- (3,-5);
\draw[dotted] (3,-5) -- (4,-5);
\draw (4,-4.5) -- (7,-4.5);
\draw[dotted] (3,-4.5) -- (4,-4.5);
\foreach \x in{0,...,3}
{
    \node at (\x ,-5){$\times$};
    \node at (\x ,-5.4){$\phi_\x$};
}
\node[color=gray] at (4 ,-5){$\times$};
\node[color=gray] at (4 ,-5.4){$\phi_4$};
\draw [decorate,decoration={brace,mirror,amplitude=10pt},xshift=0cm,yshift=-0.pt]
(0,-5.7) -- (4,-5.7) node [black,midway,yshift=-0.6cm]
{\footnotesize $\Lambda_1$};

\draw [decorate,decoration={brace,amplitude=10pt},xshift=0cm,yshift=-0.pt]
(3,-3.8) -- (7,-3.8) node [black,midway,yshift=0.6cm]
{\footnotesize $\Lambda_2$};
\foreach \x in{4,...,7}
{
    \node at (\x ,-4.5){$\times$};
    \node at (\x ,-4.){$\phi_{\x}$};
}
\node[color=gray] at (3 ,-4.5){$\times$};
\node[color=gray] at (3 ,-4.){$\phi_{3}$};
\draw[dashed](3.5,-6)--(3.5,-4);

  \end{tikzpicture}
  \caption{Parallelization of a one-dimensionnal lattice (line). The physical lattice $\Lambda$ is split into two sublattices $\Lambda_1$ and $\Lambda_2$. In order to be able to consistently solve systems of equations involving derivatives, the sublattices need to know about the value of the fields just over the boundary. This is achieved by introducing ``ghost cells" to store this information. Every time a boundary value is modified, the two lattices need to ``exchange their boundary"; $\Lambda_1$ communicates the new value of $\phi_3$ to $\Lambda_2$ while $\Lambda_2$ communicates the new value of $\phi_4$ to $\Lambda_1$. Were we to consider periodic boundary conditions, we would also need ghost-cells and boundary exchange around sites $0$ and $7$.}
  \label{fig:1d}
\end{figure}

\begin{figure}
  \centering
  \begin{tikzpicture}
    \foreach \x in{0,...,3}
    {
        \draw (\x ,0,3) -- (\x ,3,3);
        \draw (\x ,3,3) -- (\x ,3,0);

    }

    \draw (3,0 ,3) -- (3,0 ,0);
    \draw (0,0 ,3) -- (3,0 ,3);
    \draw (0,3 ,3) -- (3,3 ,3);

      \draw (3,0,0 ) -- (3,3,0 );
      \draw (0,3,0 ) -- (3,3,0 );

      \foreach \x in{1,2,3}
      {
        \node at (\x-0.5,1.5,3){$\Lambda_\x$};
      }

      \node at (0.9,4,0){Parallelization in one direction};

    \newcommand{\xshift}{6.5}
  \foreach \x in{0,...,3}
  {   \draw (\xshift,\x ,3) -- (\xshift+3,\x ,3);
      \draw (\xshift+\x ,0,3) -- (\xshift+\x ,3,3);
      \draw (\xshift+3,\x ,3) -- (\xshift+3,\x ,0);
      \draw (\xshift+\x ,3,3) -- (\xshift+\x ,3,0);
  }
    \draw (\xshift+3,0,0 ) -- (\xshift+3,3,0 );
    \draw (\xshift,3,0 ) -- (\xshift+3,3,0 );

    \draw[-] (-2.8,0,2) -- (-2.8,1,2);
    \draw[-] (-2.8,0,2) -- (-1.8,0,2);
    \draw[-] (-2.8,0,2) -- (-2.8,0,3);
    \node at (-2.9,-0.1,3.1) {$z$};
    \node at (-1.6,0,2) {$x$};
    \node at (-2.8,1.2,2) {$y$};

    \foreach \x in{1,2,3}
    {
      \node at (\x+\xshift-0.5,2.5,3){$\Lambda_{1\x}$};
    }
    \foreach \x in{1,2,3}
    {
      \node at (\x+\xshift-0.5,1.5,3){$\Lambda_{2\x}$};
    }
    \foreach \x in{1,2,3}
    {
      \node at (\x+\xshift-0.5,0.5,3){$\Lambda_{3\x}$};
    }

    \node at (\xshift+0.9,4,0){Parallelization in two directions};

  \end{tikzpicture}
  \caption{\textbf{Left:} Parallelization in one direction. The physical lattice is split into slices along the $x$ direction, and each process deals with the fields' evolution in a separate slice. The boundary region between slices needs to be exchanged between cores to be able to perform operations involving neighboring points, such as computing spatial derivatives. In our current implementation, the number of points in the $x$ direction should be a multiple of the number of cores used. \textbf{Right:} Parallelization in two directions. The physical lattice is divided along the $x$ and $y$ directions into parallelepipeds as drawn on the figure. Boundaries also need to be exchanged, for instance to compute spatial derivatives. In the current implementation, the number of points/dimension $N$ needs to be a multiple of the number of cores. }
  \label{fig:parageo}
\end{figure}
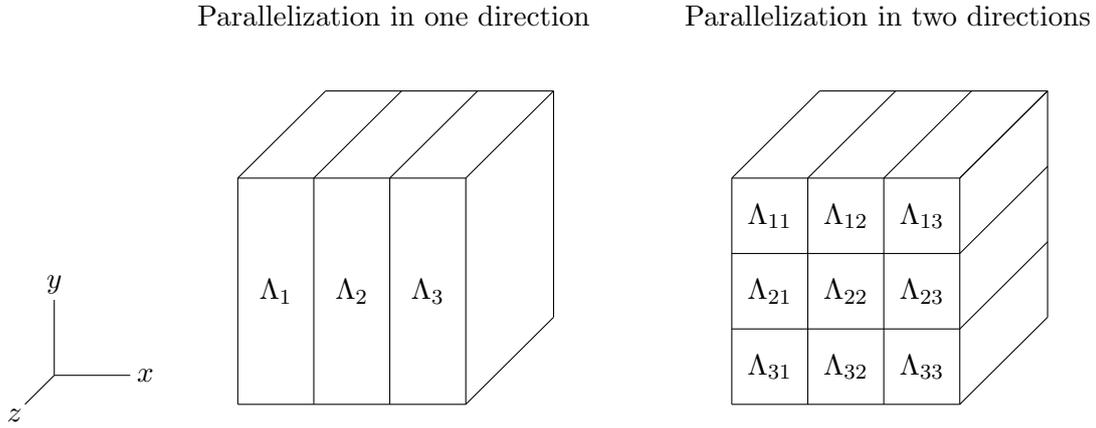

\subsubsection{Parallelization in one direction: \texttt{MPI}}
\label{subsubsec:para1D}

As briefly discussed in the previous section, the parallelization of ``local" operations, like solving finite difference systems,  is relatively straightforward. However, this is not the case for ``non-local" operations such as Fourier transforms. Typical simulations performed through the \texttt{CosmoInterface} require Fourier transforms, in order to e.g.~setting the initial fluctuations of the fields, or computing their spectra. \CL relies on the standard \texttt{fftw3} library to perform Fourier transforms. In its current version, this library does allow to parallelize multi-dimensional Fourier transforms, but only along a single direction. As a result, if one is not willing to use any extra library besides \texttt{fftw3}, the parallelization of a lattice simulations can only be done along one spatial direction. This leads to the decomposition presented in the left-hand side of Fig.~\ref{fig:parageo}. Note that in the current implementation of \CLns, the linear size $N$ of the lattice must be an integer multiple of the number of cores you want to use. For instance, in Fig.~\ref{fig:parageo}, as we want to use three cores, $N$ must be a multiple of three.

It is very easy to activate this parallelization procedure in \CLns. Assuming you have installed \texttt{MPI} and a properly compiled version of \texttt{fftw3} (see Appendix~\ref{app:Installation} for more information, installation instructions for these libraries and guidance to use them on HPC clusters), you simply need to pass an extra flag \texttt{-DMPI=ON} to \texttt{CMake} before compiling your model:
\begin{shell-sessioncode}
 cmake -DMPI=ON -DMODEL=lphi4 ../
 make cosmolattice
\end{shell-sessioncode}
Of course, if you want to compile any other model (including the ones with gauge fields), you simply need to replace \texttt{lphi4} by the name of your model, as explained in Sections~\ref{sec:MyFirstModelScalars} and \ref{sec:MyFirstModelGauge}.

After having successfully compiled \CLns, you can run it with \texttt{nc} cores with \texttt{nc}$\geq 1$. Of course you need to have access to such number of CPU's; a typical laptop will have between one and four, whereas you can use even thousands of cores on a HPC cluster. This is done as follows,
\begin{shell-sessioncode}
mpirun -n nc lphi4 input=...
\end{shell-sessioncode}
Note that if you are using a high-performance-computation (HPC) cluster, you will typically have to use another command to run your parallel jobs.

\subsubsection{Parallelization in two directions: \texttt{MPI} and \texttt{PFFT}}
\label{subsubsec:para2D}

If we are willing to use some extra external libraries to compute Fourier transforms, we can actually overcome the limitation of \texttt{fftw3} and use a parallelization across multiple spatial directions. In the current implementation of \CLns, we use the \texttt{PFFT} library~\cite{Pi13}, see again Appendix~\ref{app:Installation} for installation instructions. This in principle allows us to parallelize the simulation in all directions. In practice, because of the overload due to the boundary exchanges, it is often a good compromise to parallelize in all dimensions except one, which involves less cores, but also less boundaries. We depict the resulting parallelization strategy for the case of three spatial dimensions in the right-hand side of Fig.~\ref{fig:parageo}. In this case, the number of sites/dimension $N$ of the lattice needs to be divisible by the number of cores used in each parallelized direction. In practice, 
when all directions have the same number of points, \texttt{N} needs to be an integer multiple of the number of cores.

To switch to this parallelization setting, again assuming you have a working installation of \texttt{MPI}, \texttt{fftw3} and now \texttt{PFFT} (see Apendix~\ref{app:Installation}), you simply need to pass the extra flag \texttt{-DPFFT=ON} to \texttt{CMake}, before compiling your favorite model
\begin{shell-sessioncode}
 cmake -DMPI=ON -DPFFT=ON -DMODEL=lphi4 ../
 make cosmolattice
\end{shell-sessioncode}
Note that this flag must be used together with the \texttt{-DMPI=ON} flag.

Nothing changes in this case to execute a run, as you can send a job again using the command:
\begin{shell-sessioncode}
mpirun -n nproc lphi4 input=...
\end{shell-sessioncode}
(or whichever is the equivalent command needed in your HPC cluster).

\subsubsection{Performances}

Before explaining some of the other features of the code, we want to show how good \CL can do as a parallel code. As an example, we will study how the execution time of the \texttt{lphi4SU2U1} model scales as a function of the used number of cores. Be aware that this kind of study has to be considered with care, as the quantitative results may depend on the type of hardware used, the actual state of the cluster when performed, the compiler, or the \texttt{MPI} implementation. Having noted this, we will show that the \CL parallelization performs very well, and that the possibility of having a Fourier transform in more than one dimension provides a significant advantage when a large number of cores are required, let it be because of memory or execution time requirements.

\begin{figure}
  \centering
  \includegraphics{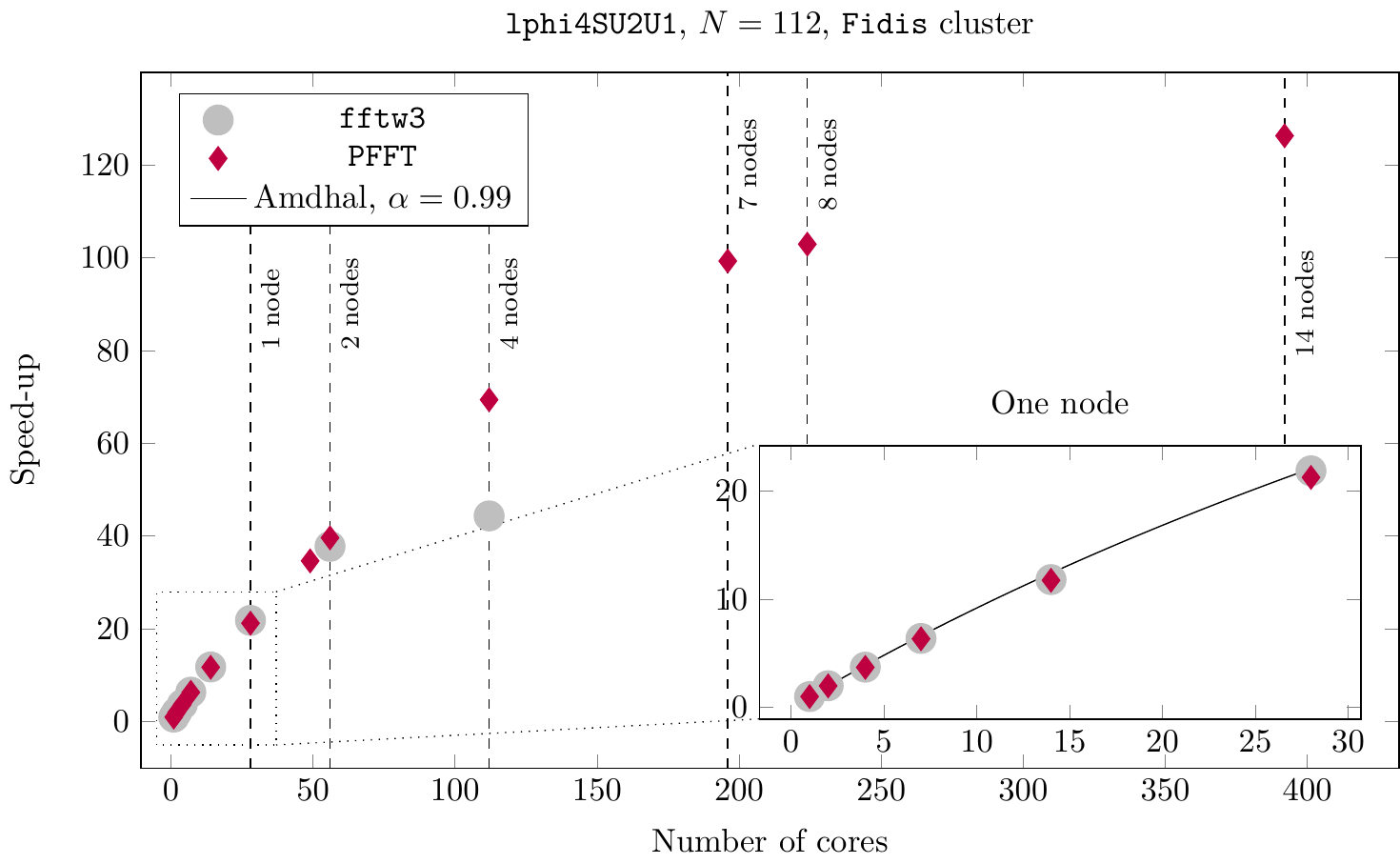}
  \caption{Speed up factor in parallelized simulations as a number of cores, as test on the \texttt{Gacrux} cluster from the EPFL HPC center SCITAS, Switzerland.}
  \label{fig:speedup}
\end{figure}

For simplicity, we choose a relatively small lattice with $N=112$ points/dimension, which we ran for $250$ time iterations. We perform $25$ ``frequent" measurements (mean values) and $6$ ``infrequent" measurements (spectra). We performed this benchmark on the \texttt{Gacrux} cluster\footnote{One node is made of two Intel Broadwell processors running at $2.6$ GHz, with $14$ cores each (hence $28$ codes/node). As node connectivity, it uses Infiniband EDR.} from the
\'Ecole polytechnique f\'ed\'erale de Lausanne (EPFL) HPC center SCITAS. We show the results in Fig.~\ref{fig:speedup}, where we plot the speed-up of the program as a function of the number of cores. In particular, we show the speed-up factor $S$, which is defined as the execution time in one core $T_1$ divided by the execution time in $n$ cores $T_n$, i.e.~$S \equiv T_1/T_n$. It is important to remark that our test case gives too much importance to the initialization of the fields relative to their evolution ($250$ time steps is orders of magnitude smaller than in a realistic simulation). The initialization functions are dominated by Fourier transforms, which are not expected to scale as good as the evolution routines. In any case, we obtain very satisfying speed-ups. Perhaps the most interesting feature of this figure is the comparison between the one-direction parallelization strategy via \texttt{fftw3} and the two-directions parallelization strategy via \texttt{PFFT}. It appears that up $\frac{N}{\# cores}\gtrsim 2$, both strategies perform equally well. The maximum number of cores we can have with the first strategy is $\# cores = N$. In this case, we already see it being outperformed by the second strategy. But more importantly, it increases the maximum number of cores we can use. For instance, in Fig.~\ref{fig:speedup} we show our benchmark running on up to $392$ cores with good performances.

To conclude this benchmark, we can attempt to make a more quantitative description of the goodness of the performance of \CLns, restricting our attention to the results obtained on one node; the one obtained on more nodes is harder to analyze as they can be relatively sensitive to hardware-dependent performance fluctuations. They are also sensitive to the efficiency of inter-nodes communications, which require some modeling beyond the scope of this section.

If we want to quantify how much of our code is actually parallelized, we can use a relation referred to as Amdhal's law \cite{conf/afips/Amdahl67}. Assume $\alpha\%$ of your code is parallelized. The execution time $T_1$ on a single core can then be written as $T_1=\alpha T_1 + (1-\alpha) T_1$. On $n_{cores}$ cores, it becomes $T_{n}=\left(\frac{\alpha}{n_{cores}} +1-\alpha\right)T_1$. Amdhal's law is the prediction of the speed-up you get from this relation,
\begin{equation}
  S=\frac{1}{\frac{\alpha}{n_{cores}} +1-\alpha} \ .
\end{equation}
By fitting our data on one node, we obtain $\alpha\approx 0.99$, which means that effectively $99\%$ of \CL is parallelized. Note that in actual simulations, we expect this number to be even better, as the invested amount of time in the fields' initialization will be even more subdominant with respect to their evolution. Note also that we performed this benchmark with the full matter content available to \CL lattice. We expect similar results for the scalar sector alone.

\subsection{Saving three-dimensional field distributions, backups and other options: \texttt{HDF5}}
\label{subsec:hdf5spec}

When running long simulations, it may come very handy to be able to stop them and restart them later on, or to have some kind of automatic backup in case some problem happens to the hardware you are using. In order to implement this type of features, we need to be able to save the field distributions to a file. For the sake of portability, the current version of \CL uses the \texttt{HDF5} library to perform this task in a binary format. This means that, if you want to use one of the features that involve saving a three-dimensional distribution of some fields to a file, you will need to have a working \texttt{HDF5} library installed (see Appendix~\ref{app:Installation} on how to do this). Assuming you have such installation, activating these features in \CL is as simple as using another \texttt{CMake} flag:
\begin{shell-sessioncode}
 cmake -DHDF5=ON -DMODEL=lphi4 ../
 make cosmolattice
\end{shell-sessioncode}
We will now survey what features this unlocks.

\subsubsection{Saving a simulation to disk}
After having activated \texttt{HDF5}, we can now save runs to disk. This is simply done via the argument \texttt{save\_dir}, which you can add to your input file, or simply pass it through the command line. For instance,
\begin{shell-sessioncode}
./lphi4 input=input.in save_dir=./
\end{shell-sessioncode}
will save this \texttt{lphi4} run at the end in the current folder. It is going to create a file named \texttt{lphi4\_DATE\_d**\_m**\_y**\_TIME\_h**\_m**\_s**.h5}, where \texttt{lphi4} is the model name, and the $**$ symbols will be replaced by the actual date and time.

You do not need to know anything about the actual content of the file in order to restart your simulation from there. However, thanks to the standardized \texttt{HDF5} format, you can easily go and explore the content of the file with your favorite data visualization tool, be it {\tt gnuplot}, {\tt Mathematica}, {\tt Matlab}, {\tt python}, {\tt Julia} or other, as long as it supports \texttt{HDF5}. To simplify, a \texttt{HDF5} file mimics a folder/file structure, folders being designated as ``groups" and files as ``datasets". In this case, every field is stored in a separate dataset. For simplicity, we also store the values of the scale factor, its time derivative, and the final time as separate datasets.

\subsubsection{Restarting a saved simulations}

Once a saved simulation file has been created, it is straightforward to restart the simulation from the same time when you stopped it and saved it. To do so, you only need to call your executable with the ``load\_dir" parameter set to the simulation file you want to restart from. It will also read the parameters of the previous simulation and use them. Except for the lattice size $N$, the length side $L$, and the infrarred and ultraviolet cutoffs $k_{\rm IR}$ and $k_{\rm UV}$, you can override the other parameters by either specifying them through the console line or in an output file. Not that if you try to override a parameter that you are not allowed to, the program will not crash, but simply ignore your changes.

To be concrete, let us assume that the above simulation was saved at the time \texttt{tMax=200}. If we want to continue the simulation, we can simply relaunch it with a different parameter (say  \texttt{tMax=500}) as follows,
\begin{shell-sessioncode}
./lphi4 load_dir=lphi4_DATE_d**_m**_y****_TIME_h**_m**_s**.h5 tMax=500
\end{shell-sessioncode}
As mentioned above, you can also use an input file as usual.

Note that when you run in restart mode, assuming you have not moved the previous output files, the new results will be overwrite the previous files. You can change this behavior by setting explicitly the parameters \texttt{appendToFiles} to \texttt{true}.

\subsubsection{Automatic backup}

With the ``start and stop" mechanism presented in the section above, it is natural to implement an automatic way of backing up the simulation to disk, in order to be able to recover from some hardware failure. This option is turned on by specifying the parameter \texttt{tBackupFreq}. Then, every \texttt{tBackupFreq} amount of program time, the simulation will write itself to disk in a file name \texttt{ModelName.backup} (\texttt{lphi4.backup} for instance). If a backup file is already present, it will first rename it to \texttt{ModelName.backup\~} before creating the backup. This extra amount of precaution allows you not to loose the whole simulation in case your hardware crashes while you are backing-up. By default, the backup file is saved in the same folder than the measurements. You can change this behaviour by specifying the \texttt{backup\_dir} parameter.

\textit{Note:} In the current implementation of \CLns, the saving of three-dimensional field configuration has not particularly been optimized for performances. As such, it is a good idea not to use a backing up frequency that is too high. You can determine what ``too high" means by trial and error, seeing how much the backing up affects performance on your hardware.

\subsubsection{Saving three dimensional energy distributions}

\CL is also capable of saving three-dimensional distributions of arbitrary observables in a file. At present, the user can save three-dimensional distributions of the various energy components of the system by adding different flags to the \texttt{energy\_snapshot} parameter in the parameter file. The different flags are indicated in Appendix \ref{App:TableParameters}. Let us show an example: suppose we are running the \texttt{lphi4SU2U1} gauge model and we want to save to file the scalars kinetic energies and the $SU(2)$ electric energy. We would then run
\begin{shell-sessioncode}
./lphi4SU2U1 input=input.in energy_densities="E_S_K E_B_K"
\end{shell-sessioncode}
Again, as usual, this parameter can go in the input file (in which case you do not need the quotes surrounding the arguments).

\subsubsection{A more user-friendly format for the spectra}
\label{subsubsec:hdf5spectra}

If you want to compute the field spectra with a very fine resolution binning on large lattices, the corresponding text files storing them may occupy a significant amount of disk memory. However, if you compile \CL with \texttt{HDF5}, you have access to a new way of storing the spectra. 
In particular, this problem can be mitigated if the spectra are saved in \texttt{HDF5} format, which are binary files. Furthermore, thanks to the internal structure of \texttt{HDF5} files, spectra at different times can easily be retrieved. Our \texttt{HDF5} spectra files are structured as follows. First, every time is its own group (``folder"). Inside this group, there will be a dataset called \texttt{momBinAverage}, which contains the average momentum in a bin, as well as another one called \texttt{momBinMultiplicity} which tells you how many values where binned in this bin. Then, there is a dataset for each of the $n$ spectra saved in the given file, named \texttt{spectAverage\_i} with \texttt{i}$=0,1,\dots,n-1$. This information is always printed for the default spectra verbosity, but if you choose a higher verbosity, you will also get a dataset containing the variance, minimum and maximum values of the momenta and spectra bins.

\section*{Acknowledgments}

We are very grateful to J.~Baeza, J.~Cline, C.~Cosme, D.~Croon, N.~Loayza, K.~Marschall and J.~S.~Roux for having helped us to test \CL in its developing phase. We also thank J.~Penedones for providing us with some computer time on the EPFL Fidis and Garcrux clusters.  DGF (ORCID 0000-0002-4005-8915) is supported by a
Ram\'on y Cajal contract with Ref.~RYC-2017-23493,
and by the grant “SOM: Sabor y Origen de la Materia” under no. FPA2017-85985-P. A.F. (0000-0002-7276-4515) is supported by the U.S.  Department  of  Energy,  Office  of  Science,  Office  of  Nuclear  Physics,  grants  Nos.DE-FG-02-08ER4145. F.T.~is supported by the Research Fund for Junior Researchers of Basel University (project number 4598519) and by the Swiss National Science Foundation (grant number 200020/175502).

\appendix

\section{Installation} \label{app:Installation}

In Section \ref{sec:MyFirstRun} we explain how to compile and run \CL assuming everything is setup correctly. In this Appendix, we will explain how to achieve such a set-up. We first present in detail how to install the relevant tools and libraries, including both the required and optional ones, on a UNIX (macOS and Linux)\footnote{\CL is written in \texttt{C++} and uses only standard packages, so it should work with minimum trouble on Windows. However, no one has verified this statement for now.} personal computer. These are summarized in Table \ref{tab:requirements}. We then make general comments on to make it work on High Performance Clusters (HPC).

In the following we assume that you have already a working \texttt{C++} compiler compatible with \CLns. The main requirement is for it to fully support the \texttt{C++14} standard, which is the case of all relatively modern compilers (see Table \ref{tab:requirements} for which ones are). We also assume that you have a working installation of \texttt{make}, \texttt{CMake} as well as a compatible \texttt{MPI} installation. These tools and libraries are very standard and easy to install. For completeness, we explain a way of installing these in Box \ref{Box:basicinstall}.

\subsection{\texttt{fftw3}}
\subsubsection{Serial Version: All you Need for \CL without Parallelization}

The first installation we will go through is the one of \texttt{fftw3}. It is very likely that you have some version of it already installed on your laptop, as many applications rely on it. However, it is also probable that such installation is not complete enough to work with \CL. For example, the installation required by \CL needs to have been compiled both for \texttt{double} and \texttt{float} precision. The easiest way of achieving this is to install \texttt{fftw3} from the source. To simplify the user's life, we provide a script which does that (almost) automatically; we will explain now how to use it. If you want to do it by yourself, you can skip to Box~\ref{Box:libsinstall}, where all the appropriate compilation flags for the external libraries are summarized. If you do this and choose a ``local" installation, do not forget to specify the path where you install \texttt{fftw3} to the \CL \texttt{CMake}, as explained in Section \ref{sec:MyFirstRun}.

In the root folder of \CLns, there is a folder called \texttt{dependencies}, which contains different scripts that help with the installation of external libraries. The script \texttt{fftw3.sh} deals with the installation of \texttt{fftw3}. We will install the library locally inside \texttt{dependencies}, in a new subfolder which we will call \texttt{MyFFTW3}. To do this, we simply need to run \texttt{fftw3.sh} with \texttt{MyFFTW3} as an argument.

\begin{shell-sessioncode}
  cd dependencies
  bash fftw3.sh MyFFTW3
\end{shell-sessioncode}

The lines above will download, unpack, compile and install \texttt{fftw3}. The compilation may take up to several minutes to complete. An advantage of using this script is that it automatically provides \texttt{CMake} with the correct path to the newly installed library, by writing it in \texttt{src/cmake/auto\_install\_paths.txt}. If the installation has gone well, everything is ready to run \CL in serial mode, as explained in Section \ref{sec:MyFirstRun}.

However, it is possible that the script does not manage to download the source. In this case, you can open the script and correct the url, by checking what the current one is at \url{http://www.fftw.org/download.html}. You can also create the directory \texttt{MyFFTW3} folder, download the archive yourself, put it there, and then run the \texttt{fftw3.sh} script.

\vspace{0.5cm}\noindent\texttt{dependencies/fftw3.sh:}~\inputminted[ firstline=8, lastline=10,linenos, frame=single]{bash}{code_files/fftw3.sh} 
The variable \mintinline{bash}{CURURL} contains the current URL. The variable \mintinline{bash}{CURNAME} should match the name of the folder extracted from the archive, as should the variable \mintinline{bash}{CURFOLDER} (but without double quotes). It is important not to add spaces around the equal signs.

If you cannot make this script work, you can proceed to Box~\ref{Box:libsinstall}, where we give instruction on how to compile all the libraries without the automated scripts.

\subsubsection{Parallel Version: All you Need for \CL with Parallelization in One Direction}

Assuming you have a working installation of \texttt{MPI}, the installation of the parallel version of \texttt{fftw3} proceeds exactly in the same way by using the same script. You simply need to pass an extra argument that specifies that you want the parallel installation:

\begin{shell-sessioncode}
  cd dependencies
  bash fftw3.sh MyFFTW3 --parallel
\end{shell-sessioncode}
We also provide relevant information for the installation of the parallel version of \texttt{fftw3} in  Box~\ref{Box:libsinstall}.

\subsection{\texttt{PFFT}: Required for Parallelization in $(N-1)$ Directions}

If you want to perform simulations parallelized in $N-1$ directions (e.g.~two for three dimensional simulations), you need to install the extra library \texttt{PFFT} \cite{Pi13}. You need to do this after installing the parallel \texttt{fftw3}, as \texttt{PFFT} is based on it. To do so, we also provide an automated script \texttt{pfft.sh} that installs it, which is also located in the \texttt{dependencies} folder. It takes two arguments. First, the path where you want it to be installed; which in this example will be in a subfolder called \texttt{MyPFFT} inside \texttt{dependencies}. Second, the folder where the parallel \texttt{fftw3} is located; \texttt{MyFFTW3} in this example.

\begin{shell-sessioncode}
  cd dependencies
  bash pfft.sh MyPFFT MyFFTW3
\end{shell-sessioncode}
 For the sake of compatibility and stability, we use a specific version of \texttt{PFFT}: version \texttt{1.0.8.alpha}, which is hosted at \url{https://www-user.tu-chemnitz.de/~potts/workgroup/pippig/software.php.en#pfft}. Also in this case, the script writes the path of this local installation  in the file \texttt{src/cmake/auto\_install\_paths.txt}. Again, instructions about how to manually install it are provided in Box~\ref{Box:libsinstall}.

\subsection{\texttt{HDF5}: To Backup your Simulations and Save $N$ Dimensional Field Distributions}

In order to save $N$ dimensional field distributions, we use the external library \texttt{HDF5}, which provides a standardized binary format which can be read by most data analysis tools/languages. To install it, you can use the script \texttt{hdf5.sh} provided in \texttt{dependencies}. This library needs to be compiled separately for serial and parallel usage. Contrary to \texttt{fftw3}, the serial and parallel version of \texttt{HDF5} are mutually incompatible, so if we want to use both of them, we will need to install them in separate folders. This is done using the \texttt{hdf5.sh} script as follows. The script first takes as an argument   the location where we want it to be installed, and second, an optional \texttt{--parallel} flag to specify we want the parallel version.

\begin{shell-sessioncode}
  cd dependencies
  bash hdf5.sh MyHDF5 # Only if you also want the serial version
  bash hdf5.sh MyHDF5 --parallel # Only if you also want the parallel version
\end{shell-sessioncode}
 As for the other scripts, the relevant path is directly set in the \CL \texttt{CMake}, in the file \texttt{src/cmake/auto\_install\_paths.txt}. In this case, if you install both the serial and parallel versions, it is important you comment out in this file the version you do not need when compiling \CLns. Namely, if you want to compile \CL in serial mode, you should comment out the link to the parallel \texttt{HDF5} version and vice versa. The same is true if you install the library following the instructions given in Box~\ref{Box:libsinstall}; to avoid conflict, \texttt{CMake} should be aware only of one \texttt{HDF5} version. Also, as in the case of \texttt{fftw3}, if the script has trouble downloading the archive, you can open the script and correct the url, by checking what the current one is at \url{https://www.hdfgroup.org/downloads/hdf5/}. You can also create the directory \texttt{MyHDF5} (or MyHDF5Parallel), download the archive yourself, put it there and then run the \texttt{hdf5.sh} script.

\vspace{0.5cm}\noindent\texttt{dependencies/hdf5.sh:}~\inputminted[ firstline=7, lastline=9,linenos, frame=single,breaklines=true]{bash}{code_files/hdf5.sh} 
Again, the variable \mintinline{bash}{CURURL} contains the current URL. The variable \mintinline{bash}{CURNAME} should match the name of the folder extracted from the archive, as should the variable \mintinline{bash}{CURFOLDER}, but without double quotes. Its important not to add spaces around the equal signs.

\subsection{All of It}
In case you want all of the libraries installed automatically, you can also run the \texttt{fetchall.sh} script in dependencies. It takes a single argument, the folder where you want the libraries to be installed, and an optional argument equal to \texttt{--no-pfft} or \texttt{--no-hdf5} if you want the script not to install \texttt{PFFT} or \texttt{HDF5}. It detects whether or not you have \texttt{MPI} installed. If so, it install the parallel version of the libraries. Otherwise, it asks you whether you want to proceed with the serial installation or not. The following execution would install all the libraries in a \texttt{MyLibs} folder inside \texttt{dependencies}.

\begin{shell-sessioncode}
  cd dependencies
  bash fetchall.sh MyLibs
\end{shell-sessioncode}

\subsection{Installing \CL on a HPC Cluster}

By its nature, \CL is intended to be used to run parallel simulations on many cores, so it is perfectly suited to be used on High-Performance Computing (HPC) Clusters. As every single HPC cluster is different, it is impossible to write a generic explanation of how to install \CL on a cluster. We still provide the user with some general guidelines and information about commonly encountered features. If you encounter trouble using a specific cluster, or you are missing some libraries/tools, you should directly contact the IT team maintaining it.

Often, when you connect to a cluster, you do not have access to any libraries by default; you need to first specify which ones you want to use from a predefined list of installed libraries. A common way of implementing this is through the use of Environment Modules. If this is the case of the cluster you are using, you will have access to the \texttt{module} command. In that case, you can see what libraries/tools are available by calling it with the \texttt{list} argument:

\begin{shell-sessioncode}
  module list
\end{shell-sessioncode}

This will show you a list of available packages, possibly with different versions. In order to run \CLns, you need appropriate version for the compiler, \texttt{CMake} and \texttt{MPI} distribution (see Table ~\ref{tab:requirements}). If this is not the case, you should contact your IT team. You can also expect to have \texttt{fftw3} available, both serial and parallel, and similarly for \texttt{HDF5}. If this is not the case, you can either ask for them to be installed (easiest option), or install them locally as explained in the previous sections (likely fastest option). You will need the parallel \texttt{fftw3} anyhow, but you will want to have \texttt{HDF5} installed only if you use such feature.

In any case, \texttt{PFFT} will not be installed by default and you will have to install it locally as explained above. If you do not have direct access to the internet from your cluster, you can download the archive somewhere else and upload it to the cluster, in the folder where you want to install the library. As explained above, you need to know where the appropriate version of \texttt{fftw3} is located in order to perform the installation of \texttt{PFFT}, which may not be clear if you use the version installed on the cluster. A useful trick to find its path is to run the \CL \texttt{CMake} with the \texttt{MPI=ON} and \texttt{PFFT=OFF} options. The \texttt{CMake} will automatically locate \texttt{fftw3}, and you will be able to recover the appropriate path by opening the generated \texttt{CMakeCache.txt} file. By looking for instance at \texttt{FFTW\_LIB}, you will see where the \texttt{fftw3} libraries are located. The path you want is the full path before the \texttt{lib} folder: for instance, if you see \texttt{/user/home/johndoe/fftw/lib/}, you must only give \texttt{/user/home/johndoe/fftw/}) to the \texttt{pfft.sh} script.

If your cluster is actually using Environment Modules, you can load the appropriate libraries/tools by using the \texttt{load} option of the \texttt{module} command. In that case, you will typically need to type something along the following lines before compiling \CLns,
\begin{shell-sessioncode}
  module list
  module load CMakeName
  module load CompilerName
  module load MPIName
  module load FFTW3Name
  module load HDF5Name # Only if you need it
\end{shell-sessioncode}
where \texttt{CMakeName, CompilerName, MPIName, FFTW3Name} and \texttt{HDF5Name} refer to the appropriate name returned by \texttt{module list}. Of course, you should not load \texttt{fftw3} or \texttt{hdf5} if you have loaded them locally.

At this point, you should be able to compile and run \CL with as many cores as you want.

  \begin{framed}
  \textit{Installing a} \texttt{C++} \textit{compiler,} \texttt{make}, \texttt{CMake}, \texttt{MPI} and \texttt{git} on your PC
  \bigskip

    \textbf{Ubuntu:} Ubuntu comes by default with the \texttt{apt-get} package manager, which makes the installation of all basic utilities easy. The following lines should be enough to install what you need
    \begin{shell-sessioncode}
      sudo apt-get install make
      sudo apt-get install g++
      sudo apt-get install cmake
      sudo apt-get install openmpi
      sudo apt-get install git
    \end{shell-sessioncode}
    The \texttt{sudo} command is necessary to give you the admin rights, which you need to have to install software ``globally" on your PC.

    \medskip

    \textbf{Fedora:} On Fedora, you can use \texttt{dnf} as a default package manager. Then the command are the same than on Ubuntu.

    \begin{shell-sessioncode}
      sudo dnf install make
      sudo dnf install g++
      sudo dnf install cmake
      sudo dnf install openmpi-devel
      sudo dnf install git
    \end{shell-sessioncode}

    By default, \texttt{dnf} does not install \texttt{openMPI} somewhere which is globally accessible. To fix that, go to your \texttt{home} folder and edit or create the \texttt{.bashrc}

    \begin{shell-sessioncode}
      cd ~
      res=`find /usr -name "mpirun"` #finds where openMPI was installed
      echo 'export PATH=$PATH:'"${res%/*}/" >> .bashrc #add it to your path
      source ~/.bashrc # reload it
    \end{shell-sessioncode}

    \textbf{macOS:} The first time you want to do something related to coding on your mac, you need to start by opening a terminal and run:
    \begin{shell-sessioncode}
      xcode-select --install
    \end{shell-sessioncode}
    This will enable your command line to be used to code and install some basic utilities. Then, to install the remaining missing software, it will be convenient to first install a package manager, which does not come by default on mac. Here we will use \texttt{Homebrew}, which can be installed as:
    \begin{shell-sessioncode}
/bin/bash -c "$(curl -fsSL https://raw.githubusercontent.com/Homebrew/install/HEAD/install.sh)"
    \end{shell-sessioncode}
  Once this is done, remaining packages can simply be installed as
  \begin{shell-sessioncode}
    brew install gcc
    brew install open-mpi
    brew install cmake
    brew install git
  \end{shell-sessioncode}
  \end{framed}
\captionof{boxfloat}{Summary of how to install basic utilities on a PC to code in \texttt{C++}.}
\label{Box:basicinstall}

\medskip

\begin{framed}
\textit{Installing the external libraries without \CL installation scripts}
  \bigskip

    \noindent\texttt{fftw3}:

    First, download the the source code (\url{http://www.fftw.org/download.html}) and extract the archive. Inside the extracted folder, which we will refer to as \texttt{fftw-3}, do the following:

    \begin{shell-sessioncode}
      cd fftw-3

      ./configure --prefix=/path/where/to/install/ --enable-threads --enable-sse2 --enable-avx --disable-shared --enable-static
      make -j
      make install

    \end{shell-sessioncode}

    \medskip

    \noindent\texttt{fftw3}, parallel:

    First, download the source code and extract the archive (it is the same than the ``normal" \texttt{fftw3}, so maybe you already did it). Inside the extracted folder, which we will refer to as \texttt{fftw-3}, do the following:
    \begin{shell-sessioncode}
      cd fftw-3

      ./configure --prefix=/path/where/to/install/ --enable-threads --enable-sse2 --enable-avx --disable-shared --enable-static --enable-mpi
      make -j
      make install

    \end{shell-sessioncode}
    With this, you should have a functional \texttt{fftw3} installation to work with \CLns.

    \medskip

    \noindent\texttt{PFFT}:

    Download version \texttt{1.0.8-alpha} from \url{https://www-user.tu-chemnitz.de/~potts/workgroup/pippig/software.php.en#pfft}.

    \begin{shell-sessioncode}
      cd pfft-1.0.8-alpha

      FFTWPATH=/path/where/fftw3/is/installed/

      export LDFLAGS="-L${FFTWPATH}/lib"
      export DYLD_LIBRARY_PATH="${FFTWPATH}/lib"
      export LIBS="-lfftw3_mpi -lfftw3"
      export CXX=mpic++
      export CC=mpicc
      export CFLAGS="-g3"

      ./configure --prefix=/path/where/to/install/ --with-fftw3=${FFTWPATH} --disable-fortran --disable-shared --enable-static
      make -j
      make install

    \end{shell-sessioncode}

\medskip

    \noindent\texttt{hdf5}:

    First, download the source code (\url{https://www.hdfgroup.org/downloads/hdf5/}) and extract the archive. Inside the extracted folder, which we will refer to as \texttt{hdf5}, do the following:

    \begin{shell-sessioncode}
      cd hdf5

      export CC=gcc
      ./configure --prefix=/path/where/to/install/
      make -j
      make install
    \end{shell-sessioncode}

    \medskip

    \noindent\texttt{hdf5}, parallel:

     First, download the the source code and extract the archive (it is the same than the ``normal" \texttt{HDF5}, so maybe you already did it). Inside the extracted folder, which we will refer to as \texttt{hdf5Parallel}, do the following:

    \begin{shell-sessioncode}
      cd hdf5

      export CC=mpicc
      ./configure --prefix=/path/where/to/install/ --enable-parallel
      make -j
      make install
    \end{shell-sessioncode}
    If you want to have both serial and parallel \texttt{HDF5}, it is important you do not install them in the same folder.

  \end{framed}

\captionof{boxfloat}{Installation of the external libraries without using the automated scripts provided with \CLns. As explained in the main text, while \texttt{fftw3} and its parallel version can perfectly be installed in the same place, one has to be careful to install \texttt{hdf5} and its parallel version in separate folders. Also, we are assuming here that you want to compile the libraries with \texttt{gcc}. If you want to use another compiler, replace the \texttt{CC} and \texttt{CXX} flags appropriately.}
\label{Box:libsinstall}


  \begin{tabularx}{\textwidth}{|>{\tt\small}c|>{\small}c|>{\small}X|>{\small}X|}
    \hline
    \multicolumn{4}{|>{\tt\small\bf}l|}{Required Tools} \\
    \hline
    \normalfont Name & \multicolumn{2}{>{\small}c|}{Minimal Version} & Notes \\
    \hline
     make & \multicolumn{2}{>{\small}c|}{-} & \\
     \hline
     CMake & \multicolumn{2}{>{\small}c|}{3.0} & \\
     \hline
     \multicolumn{4}{|>{\tt\small\bf}l|}{Required Compiler (one of the following)} \\
     \hline
     \normalfont Name & \multicolumn{2}{>{\small}c|}{Minimal Version} & Notes \\
     \hline
      g++ & \multicolumn{2}{>{\small}c|}{5.0} & Minimal version tested: 5.5\\
     \hline
      clang++ & \multicolumn{2}{>{\small}c|}{3.4}  & Minimal version tested: 3.9 \\
      \hline
      \multicolumn{4}{|>{\tt\small\bf}l|}{Required Libraries} \\
      \hline
      \normalfont Name & \multicolumn{2}{>{\small}c|}{Minimal Version} & Notes \\
      \hline
      fftw3 & \multicolumn{2}{>{\small}c|}{3} & Minimal version tested: 3.3.6. \\
      \hline
      \multicolumn{4}{|>{\tt\small\it}l|}{Optional Libraries} \\
      \hline
      \normalfont Name & Minimal Version & Extra Features & Notes \\
      \hline
      MPI & - & Parallelization & Needs an implementation of \texttt{MPI} compatible with the compiler you chose. The most common open-source ones are \texttt{OpenMPI} and \texttt{MVAPICH} (for \texttt{g++} and \texttt{clang++}).  Works only with one of the  parallel Fourier transforms libraries, see below and main text. \\
      \hline
      fftw3\normalfont, parallel version & 3 & Parallelization in one dimension &\texttt{fftw3} compiled for parallel use. Allows only for parallelization in one dimension. \\
      \hline
      PFFT& - &  Parallelization in $n-1$ dimensions. &External library based on the parallel \texttt{fftw3} library. Needs the parallel \texttt{fftw3}.\\
      \hline
      HDF5 & 5 & Saving of 3D distributions to file. Saving of whole simulations. Restarting simulations and automatic backup. & Needs to be separately compiled to work in parallel, see bulk text.  \\
      \hline
      \multicolumn{4}{|>{\tt\small\it}l|}{Optional Tools} \\
      \hline
      \normalfont Name & Minimal Version & Extra Features & Notes \\
      \hline
      git & - & Easy access to the code and easy way to update your code version. &\\
      \hline
  \end{tabularx}
  \captionof{table}{Summary of required and optional tools and libraries related to \CLns. See main text for more information. Note that the code should in theory work with a recent (at least above version \texttt{17.0}) \texttt{intel} compiler as well, but this has not been extensively tested. }
\label{tab:requirements}

\section{ Appendix: Parameters}\label{App:TableParameters}

In this appendix we list all the different parameters that can be specified when carrying out a simulation. Most of these parameters must be specified in program units, defined by the field and spacetime transformations of Eq.~(\ref{eq:FieldSpaceTimeNaturalVariables}), so that variables are dimensionless. For example, one must introduce the length side of the box as $\tilde{L} \equiv L \omega_*$, the infrared cutoff the lattice as $\tilde{k}_{\rm IR} \equiv k_{\rm IR}/\omega_*$, etc.

{
\begin{center}
  \textbf{ -- Run parameters -- }
\end{center}

\medskip

All these parameters are declared in \texttt{src/include/CosmoInterface/runparameters.}
\begin{center} \small
\begin{tabular}{ | m{3.7cm} | m{12.8cm}| } \hline
{\bf Parameters} & {\bf Explanation} \\ \hline
{\tt N} & Number of lattice points per dimension.\\ \hline
\tt kIR  & Infrared cutoff of the lattice \textbf{in program units}, i.e.~$\tilde{k}_{\rm IR} \equiv k_{\rm IR} / \omega_*$.  \\ \hline
\tt lSide  & Length of the box \textbf{in program units}, i.e.~$\tilde{L} \equiv L \omega_*$. \\ \hline
\tt dt & Time step of the evolution algorithm \textbf{in program units}, i.e.~$\delta \tilde{\eta}$.  \\ \hline
\tt expansion & Expanding universe or not.  If {\tt false}, the scale factor is fixed to unity and field dynamics occur in Minkowski. If {\tt true} (default value), the scale factor evolves self-consistently according to the Friedmann equations. A fixed background expansion rate can be further specified by the parameter \texttt{fixedBackground}.\\ \hline
\tt evolver & Type of evolution algorithm. Options `VV2', `VV4', `VV6', `VV8', and `VV10' solve the field equations with the velocity-verlet algorithm of the corresponding order, while `LF' solves them with the staggered-leapfrog method. Check \href{https://www.cosmolattice.net/technicalnotes}{\color{blue} https://www.cosmolattice.net/technicalnotes} for addition of new evolvers. \\ \hline
\tt t0 & Initial time of the simulation \textbf{in program units} (set to 0 by default).  \\ \hline
\tt tMax & Final time of the simulation \textbf{in program units}.  \\ \hline
\tt fixedBackground & If set to {\tt true}, turns off the self consistent expansion and replace it by a fixed background expansion. \\ \hline
\tt omegaEoS & Barotropic equation of state parameter $\omega\equiv p/\rho$ required for a fixed background expansion. Note that {\bf fractions are not allowed}, so one must write e.g.~for a RD universe, `{\tt omegaEoS=0.333}' instead of `{\tt omegaEoS=1/3}'. \\ \hline
\tt H0 & Initial Hubble rate ({\bf in GeV}) used for the fixed background expansion. \\ \hline
\end{tabular}
\end{center}
}

\underline{Note}: $\tilde{k}_{\rm IR}$ and $\tilde{L}$ obey $\tilde{k}_{\rm IR} = 2\pi /\tilde L$ , so only one of them must be specified for the simulation: the other one will be automatically computed by the code.

\medskip

{
\begin{center}\textbf{ -- Initial conditions -- }\end{center}

All these parameters are declared in \texttt{src/include/CosmoInterface/runparameters.}.

\begin{center} \small
\begin{tabular}{ | m{3.7cm} | m{12.8cm}| } \hline
{\bf Parameters} & {\bf Explanation} \\ \hline
\tt kCutOff & If specified, the given cutoff (\textbf{in program units}) is imposed in the spectrum of initial fluctuations for all scalar fields: the amplitude of the field modes at larger momenta is set to zero up to machine precision. Not specifying {\tt kCutOff} implies not having an initial cut-off, whereas {\tt kCutOff = 0} implies initially vanishing fluctuations. \\ \hline
\tt baseSeed & Seed for the random generator of initial field fluctuations. If not specified, the seed will be generated randomly in each simulation.   \\ \hline
\end{tabular}
\end{center}
}

\underline{Note}: Typically, you will also need to add some parameters to get the initial homogeneous components of your fields in the user-defined model file, see Section \ref{sec:MyFirstModelScalars} for more details.

\vspace{0.4cm}{
\begin{center}\textbf{ -- Gauge couplings and charges -- }\end{center}

The following parameters are defined in the constructor of the \texttt{src/include/CosmoInterface/abstractmodel.h}
\begin{center} \small
\begin{tabular}{ | m{3.7cm} | m{12.8cm}| } \hline
{\bf Parameters} & {\bf Explanation} \\ \hline
\tt CSU1Charges & U(1) charges $\{Q_A^{(\varphi)}\}$ of all complex scalars in vector form.\\ \hline
\tt SU2DoubletU1Charges & U(1) charges $\{Q_A^{(\Phi)}\}$ of all SU(2) doublets in vector form.  \\ \hline
\tt SU2DoubletSU2Charges & SU(2) charges $\{Q_B\}$ of all SU(2) doublets in vector form.  \\ \hline
\tt gU1s & U(1) gauge couplings $\{g_A\}$ in vector form.  \\ \hline
\tt gSU2s & SU(2) gauge couplings $\{g_B\}$ in vector form.  \\ \hline
\end{tabular}
\end{center}
}

\vspace{0.4cm}{\centering{{\bf -- Output format parameters -- }}
\begin{center} \small
\begin{longtable}{ | m{3.7cm} | m{12.8cm}| } \hline
{\bf Parameters} & {\bf Explanation} \\ \hline
\tt tOutputFreq & Time interval between the printing of \textit{frequent output} in program units.  \\ \hline
\tt tOutputInfreq & Time interval between the printing of \textit{infrequent output} in program units.  \\ \hline
\tt tOutputRareFreq & Time interval between the printing of \textit{very infrequent (rare) output} in program units.  \\ \hline
\tt tOutputVerb & Time interval between updates in the terminal in program units.  \\ \hline
\tt outputfile & Folder where output is saved. If unspecified, output will be printed in the compilation folder.  \\  \hline
\tt deltaKBin & Width of the bins in the field spectra, $\Delta k_{\rm bin}$ ($= 1$ by default). The total number of bins in the spectra is $N_{\rm bins} \simeq \sqrt{3} N / (2 \Delta k_{\rm bin})$. \\ \hline
\tt hdf5Spectra & If {\tt true}, field spectra are printed in HDF5 format instead of text format.   \\ \hline
\tt spectraVerbosity & If {\tt true}, additional information is printed in the spectra files.  \\   \hline
\tt print\_headers &  If {\tt true}, a header is printed in the first line of each text output file, with information of its contents.  \\  \hline
\tt energy\_snapshot & Vector which indicates for which energy contributions we print snapshots (in hdf5 format) The different options are:  \vspace*{0.1cm}
\begin{itemize}
    \item  {\tt E\_S\_K}: scalar singlet, kinetic energy. \vspace*{-0.2cm}
    \item {\tt E\_S\_G}: scalar singlet, gradient energy. \vspace*{-0.2cm}
    \item {\tt E\_CS\_K}: complex scalar, kinetic energy. \vspace*{-0.2cm}
    \item {\tt E\_CS\_G}: complex scalar, gradient energy. \vspace*{-0.2cm}
    \item {\tt E\_SU2D\_K}: SU(2) doublet, kinetic energy. \vspace*{-0.2cm}
    \item {\tt E\_SU2D\_G}: SU(2) doublet, gradient energy. \vspace*{-0.2cm}
    \item {\tt E\_A\_K}: U(1) gauge sector, electric energy. \vspace*{-0.2cm}
    \item {\tt E\_A\_G}: U(1) gauge sector, magnetic energy. \vspace*{-0.2cm}
    \item {\tt E\_B\_K}: SU(2) gauge sector, electric energy. \vspace*{-0.2cm}
    \item {\tt E\_B\_G}: SU(2) gauge sector, magnetic energy. \vspace*{-0.2cm}
    \item {\tt E\_V}: potential energy. \vspace*{-0.2cm}
\end{itemize} \\ \hline
\end{longtable}
\end{center}
}

\vspace{0.4cm}{\centering{{\bf -- Saving parameters -- }}
\begin{center} \small
\begin{tabular}{ | m{3.7cm} | m{12.8cm}| } \hline
{\bf Parameters} & {\bf Explanation} \\ \hline
\tt save\_dir & A copy of the simulation is saved at the specified folder at the end of the simulation, which can be loaded as the starting point of a new one.   \\ \hline
\tt backup\_dir & A copy of the simulation is saved at the specified folder at certain times  \\  \hline
\tt tBackupFreq & Time interval between backups in program units (see above).  \\ \hline
\tt load\_dir & The code checks if there is a previously saved simulation in the specified folder, and in that case, it loads it as the starting point.    \\ \hline
\tt appendToFiles & If {\tt false}, new output files are created when a new simulation starts, overwritting previous files with the same name that could exist in the same location. If {\tt true}, the output of the new simulation will be appended to the previously existing files. \\ \hline
\end{tabular}
\end{center}
}

\section{ Appendix: Generic Model variables}

Below we present the variable used throughout the \texttt{CosmoInterface} which are declared in the \texttt{AbstractModel} class and thus shared by all models. See \texttt{src/include/TempLat/abstractmodel.h}.

\begin{center}
\begin{longtable}{ | m{6cm} | m{9.5cm}| } \hline
{\bf Variable} & {\bf Definition} \\ \hline
{\tt fldS} & $\tilde \phi$ \\ \hline
{\tt piS}  & $\tilde \pi_{\phi}$  \\ \hline
{\tt fldCS}  & $\tilde \varphi$  \\ \hline
{\tt piCS}  & $\tilde \pi_{\varphi}$  \\ \hline
{\tt fldSU2Doublet}  & $\widetilde\Phi$  \\ \hline
{\tt piSU2Doublet}  & $\widetilde \pi_{\Phi}$  \\ \hline
{\tt fldU1}  & $\widetilde{A}_{i}$  \\ \hline
{\tt piU1}  & $\piApar_i$  \\ \hline
{\tt fldSU2}  & $\widetilde{B}_{i}^a$  \\ \hline
{\tt piSU2}  & $\piBpar^{a}_i$  \\ \hline
{\tt aI} & $a$  \\ \hline
{\tt aDotI}  & $a'$  \\ \hline
{\tt pi2AvI}  & $\langle \tilde \pi_{\phi}^2 \rangle$  \\ \hline
{\tt grad2AvI}  & $\sum_i  \langle (\tilde \partial_i \tilde \phi)^2 \rangle$  \\ \hline
{\tt CSpi2AvI}  & $\langle \tilde \pi_{\varphi}^2 \rangle$   \\ \hline
{\tt CSgrad2AvI}  & $ \sum_i  \langle(\widetilde D_i^A \varphi)^*(\widetilde D_i^A \widetilde \varphi) \rangle$  \\ \hline
{\tt SU2DblPi2AvI}  & $\langle \widetilde \pi_{\Phi}^2 \rangle$  \\ \hline
{\tt SU2DblGrad2AvI}  & $ \sum_i \langle (\widetilde  D_i\widetilde  \Phi)^\dag(\widetilde D_i \widetilde \Phi) \rangle$  \\ \hline
{\tt U1El2AvI}  & $\sum_i \langle \mathcal{E}_i^2 \rangle$  \\ \hline
{\tt U1Mag2AvI}  & $\sum_i \langle \mathcal{B}_i^2 \rangle$    \\ \hline
{\tt SU2El2AvI}  & $\sum_{i,a} \langle (\mathcal{E}_i^a)^2 \rangle$  \\ \hline
{\tt SU2Mag2AvI}  & $\sum_{i,a} \langle (\mathcal{B}_i^a)^2 \rangle$  \\ \hline
{\tt potAvI}  & $\langle \widetilde V \rangle$   \\ \hline
{\tt fldS0}  & $\langle \tilde \phi_* \rangle$   \\ \hline
{\tt fldCS0}  & $\langle \tilde \varphi_* \rangle$ \\ \hline
{\tt fldSU2Doublet0}  & $\langle \tilde \Phi_* \rangle$ \\ \hline
{\tt piS0}  & $\langle \tilde \pi_{\phi,*} \rangle$  \\ \hline
{\tt piCS0}  & $\langle \tilde \pi_{\varphi,*} \rangle$  \\ \hline
{\tt piSU2Doublet0}  & $\langle \tilde \pi_{\Phi,*} \rangle$  \\ \hline
{\tt pot0}  & $\langle \widetilde V_{*} \rangle$  \\ \hline
{\tt masses2S}  & $\tilde{m}_{\phi}^2$  \\ \hline
{\tt masses2CS}  & $\tilde{m}_{\varphi}^2$  \\ \hline
{\tt masses2SU2Doublet}  & $\tilde{m}_{\Phi}^2$  \\ \hline
{\tt alpha}  & $\alpha$  \\ \hline
{\tt fStar}  & $f_*$  \\ \hline
{\tt omegaStar}  & $\omega_*$  \\ \hline
\end{longtable}
\end{center}

\section{ Appendix: CMake Flags}
\label{app:CMakeArgs}

We collect here, the different flags the \CL CMake flags the user can pass to influence the compilation process.

\begin{center}
\begin{longtable}{ | m{4cm} | m{11cm}| } \hline
{\bf Flag} & {\bf Explanation} \\ \hline
{\tt -DMODEL} & Takes \texttt{modelname} as an argument, where \texttt{modelname} is the name of the model you want to compile. \\ \hline
{\tt -DMPI}  & Can be \texttt{ON} or \texttt{OFF}. It switches \texttt{MPI} parallelization on or off.  \\ \hline
{\tt -DPFFT}  & Can be \texttt{ON} or \texttt{OFF}. It switches \texttt{PFFT} on or off.  \\ \hline
\texttt{-DHDF5}  & Can be \texttt{ON} or \texttt{OFF}. It switches \texttt{HDF5} on or off.   \\ \hline
{\tt -DMYPFFT\_PATH}  & Takes a string as argument. Can be used to provide the path to \texttt{PFFT} in case CMake cannot find it.  \\ \hline
{\tt -DMYFFTW3\_PATH}  & Takes a string as argument. Can be used to provide the path to \texttt{fftw3} in case CMake cannot find it. \\ \hline
{\tt -DMYHDF5\_PATH}  & Takes a string as argument. Can be used to provide the path to \texttt{hdf5} in case CMake cannot find it.  \\ \hline
{\tt -DG++OPT}  & Only useful if compiled with \texttt{g++}. Takes one of the following as argument: \texttt{G}, \texttt{O1} , \texttt{O2} , \texttt{O3} , \texttt{Ofast} and sets the optimisation level of \texttt{g++} correspondingly (default is \texttt{Ofast}).  \\ \hline
{\tt -DG++SSE}  & Only useful if compiled with \texttt{g++}. Turns on the \texttt{SSE} instructions. Default is \texttt{OFF}, but if the code compiles and run on your platform when it is \texttt{ON}, this may speed up the code.  \\ \hline
{\tt -DG++AVX}  & Only useful if compiled with \texttt{g++}. Can be set to \texttt{OFF}, \texttt{mavx}, \texttt{mavx2}, \texttt{mavx512f} and turns on the \texttt{AVX} instructions up to the specified level. Default is \texttt{OFF}, but the more \texttt{AVX} instructions your platforms support, the better (this may speed up the code).  \\ \hline
{\tt -DTESTING}  & Can be \texttt{ON} or \texttt{OFF}. Enable the test modules, which can then be individually compiled. \\ \hline

\end{longtable}
\end{center}

\section{ List of Implemented Functions} \label{app:ImplemFunc}

Below we present all the currently implemented functions in the different algebra. Note that if unfortunately the function/operation you want is not implemented, you can do it very easily yourself by opening a similar function from the library, copy it and use it as a template for your missing function/operation.

\subsection{Scalar Algebra}

The files are located in \texttt{src/include/TempLat/lattice/algebra/operators/}.

\begin{tabularx}{\textwidth}{|>{\tt\small}c|>{\small}X|>{\tt\small}c|>{\small}X|}
  \hline
  \normalfont Function name & Operation & \normalfont File & Notes \\
  \hline
  operator+($\phi, \chi$) & $\phi + \chi$ & add.h & Addition of \texttt{ZeroType} is simplified to no addition. \\
  \hline
  operator-($\phi, \chi$)  & $\phi-\chi$ &subtract.h &  Simplifies to no subtraction if one of the input is of \texttt{ZeroType}. Simplifies $\phi-(-\chi)$ to $\phi+\chi$. Handles \texttt{HalfType} and \texttt{OneType} appropriately.\\
  \hline
  operator-($\phi$)  & $-\phi$ &unaryminus.h &  \texttt{-ZeroType} returns \texttt{ZeroType}. Simplifies $--\phi$ to $\phi$.\\
  \hline
  operator*($\phi, \chi$) & $\phi  \chi$ &multiply.h & Multiplication by \texttt{ZeroType} returns \texttt{ZeroType} and multiplication by \texttt{OneType} is simplified away.\\
  \hline
  operator/($\phi, \chi$) & $\phi/\chi$ & divide.h & \texttt{ZeroType}$/\chi$ is simplified to \texttt{ZeroType} and $\chi/$\texttt{OneType} is simplified to \texttt{OneType}.  \\  \hline
  safeDivide($\phi,\chi$) & $\phi/\chi$ & divide.h & Same as \texttt{operator/} except that if $\chi$ is a field, it checks point by point that the divisor is not too small and discards the division if this is so. Useful when dividing by some fields which fluctuates around $0$.  \\  \hline
  abs($\phi$) &$ |\phi|$ & absolutevalue.h&\\
  \hline
  arg($\phi, \chi$) &$\mathrm{arg}(\phi+ i \chi)$ & arg.h&\\
  \hline
  asinh($\phi$) & $\mathrm{asinh}(\phi)$ & asinh.h &  \\
  \hline
  conj($\phi$) & $\phi^*$ & complexconjugate.h &  Useful for the Fourier modes, as they are complex even for a scalar field.\\
  \hline
  cosh($\phi$) & $\mathrm{cosh}(\phi)$ & cosh.h &  \texttt{cosh}(ZeroType) is simplified to OneType.\\
  \hline
  cos($\phi$) & $\mathrm{cos}(\phi)$ & cosine.h &  \texttt{cos}(ZeroType) is simplified to OneType.\\
  \hline
  DiracDelta($\phi$) & $\delta(\phi)$ & diracdeltafunction.h &  \\ \hline
  exp($\phi$) & $\mathrm{exp}(\phi)$ & divide.h & \texttt{exp(ZeroType)} is simplified to \texttt{OneType}.\\
  \hline
  heaviside($\phi$) & $\theta(\phi)$ &heavisidestepfunction.h & \texttt{heaviside(ZeroType)}  and \texttt{heaviside(OneType)} are simplified to \texttt{OneType}.\\
  \hline
  log($\phi$) & $\log(\phi)$ &log.h & \texttt{log(OneType)}  is simplified to \texttt{ZeroType}.\\
  \hline
   pow<N>($\phi$) & $\underbrace{\phi\cdot\dots\cdot \phi}_{N-times}$ &pow.h & Computes the $\mathtt{N}^{th}$ power of $\phi$ by expanding the multiplication, for performance reasons. \texttt{N} must be an integer known at compile time.\\
   \hline
    pow($\phi, \chi$) & $\phi^\chi$ &pow.h & Computes arbitrary power, $\chi$ can even be a field. \texttt{pow(}$\phi$\texttt{, ZeroType)} returns \texttt{OneType} and \texttt{pow(ZeroType, }$\phi$\texttt{)} returns \texttt{ZeroType}.\\
    \hline
     shift<I>($\phi$)  &\texttt{shift<I>(}$\phi$\texttt{)}$|_{\vec n}=\phi|_{\vec n + \hat I}$  &shift.h & Shift the object by a unit vector in the $I^{th}$ direction. \\
    \hline
    shift($\phi$,Tag<I>)  & \texttt{shift(}$\phi$\texttt{, Tag<I>)}$|_{\vec n}=\phi|_{\vec n + \hat I}$ &shift.h & Shift the object by a unit vector in the $I^{th}$ direction. Same as above, different notation.   \\
      \hline
      shift<I,J,K,...>($\phi$)  & \texttt{shift<I,J,...>(}$\phi$\texttt{)}$|_{\vec n}$ \ $=\phi|_{(n_1+I,n_2+J,...)}$& shift.h & Allows to define an object shifted by an arbitrary vector. \\
   \hline
     sin($\phi$)  & $\sin(\phi)$ &sine.h &  \texttt{sin(ZeroType)} returns  \texttt{ZeroType}.\\
    \hline
    sinh($\phi$)  & $\sinh(\phi)$ &sinh.h &  \texttt{sinh(ZeroType)} returns  \texttt{ZeroType}.\\
    \hline
    sqrt($\phi$)  & $\sqrt{\phi}$ &squareroot.h &  Returns \texttt{pow(}$\phi$\texttt{,0.5}. \texttt{sqrt} of \texttt{ZeroType} resp. \texttt{ OneType} returns \texttt{ZeroType} resp. \texttt{ OneType}. \\
    \hline
      safeSqrt($\phi$) & $\sqrt{\phi}$ & squareroot.h & Same as \texttt{sqrt} except that if $\phi$ is a field, it checks point by point that it is positive and discards results (returns $0$) if not. Useful when taking square roots of fields which fluctuates close to $0$.  \\  \hline
    tanh($\phi$)  & $\tanh(\phi)$ &tanh.h &   \texttt{tanh(ZeroType)} returns  \texttt{ZeroType}.\\
    \hline

\end{tabularx}

\subsection{Complex Scalar Algebra}

The files are located in \texttt{src/include/TempLat/lattice/algebra/complexalgebra/}.

\begin{tabularx}{\textwidth}{|>{\tt\small\centering}X|>{\small}X|>{\tt\small}c|>{\small}X|}
  \hline
  \normalfont Function name & Operation & \normalfont File & Notes \\
  \hline
  operator+($\varphi_1, \varphi_2$) & $\varphi_1 + \varphi_2$ & complexfieldadd.h & Adding real objects to complex objects is supported. \\
  \hline
  operator-($\varphi_1, \varphi_2$)  & $\varphi_1-\varphi_2$ &complexsubtract.h & Subtracting real objects to complex objects is supported.\\
  \hline
  operator*($\varphi_1, \varphi_2$) & $\varphi_1 \varphi_2$ &complexmultiply.h & Multiplication by \texttt{ZeroType} returns \texttt{ZeroType} and multiplication by \texttt{OneType} is simplified away.\\
  \hline
  operator*($\phi, \varphi$),  operator*($\varphi, \phi$)& $\phi  \varphi, \varphi\phi$ & scalarcomplexmultiply.h & Scalar multiplication. $\phi$ is a real expression.\\
  \hline
  operator/($\varphi, \phi$) & $\varphi/ \phi$ & scalarcomplexmultiply.h & Division by a scalar. Implemented through scalar multiplication.\\
  \hline
  asFourier($\varphi$) & Treats a complex object as being a real expression in Fourier space & asfourier.h & \\
  \hline
Complexify($\phi, \chi$) & $\phi+i\chi$ & complexwrapper.h &  Makes a complex expressionout of two real expressions.\\
  \hline
  conj($\varphi$) & $\varphi^*$ & complexconjugate.h &  \texttt{conj(OneType)} return \texttt{OneType} and \texttt{conj(ZeroType)} return \texttt{ZeroType} \\
  \hline
  dagger($\varphi$) & $\varphi^*$ & complexconjugate.h &  Exactly same as \texttt{conj}.\\
  \hline
  Imag($\varphi$) & $\Im\varphi$ & imag.h & \\
  \hline
  norm2($\varphi$) &$ \Re\varphi^2+\Im\varphi^2$ & complexmultiply.h&\\
  \hline
  Real($\varphi$) & $\Re\varphi$ & real.h & \\
  \hline
  shift<I>($\varphi$)  &\texttt{shift<I>(}$\varphi$\texttt{)}$|_{\vec n}=\varphi|_{\vec n + \hat I}$  &complexshift.h & Shift the object by a unit vector in the $I^{th}$ direction. \\
 \hline
 shift($\varphi$,Tag<I>)  & \texttt{shift(}$\varphi$\texttt{, Tag<I>)}$|_{\vec n}=\varphi|_{\vec n + \hat I}$ &complexshift.h & Shift the object by a unit vector in the $I^{th}$ direction. Same as above, different notation.   \\
   \hline
   shift<I,J,K,...>($\varphi$)  & \texttt{shift<I,J,...>(}$\varphi$\texttt{)}$|_{\vec n}$ \ $=\varphi|_{(n_1+I,n_2+J,...)}$& complexshift.h & Allows to define an object shifted by an arbitrary vector. \\
\hline
\end{tabularx}

\subsection{$SU(2)$ Doublet Algebra}

The files are located in \texttt{src/include/TempLat/lattice/algebra/su2algebra/}.

\begin{tabularx}{\textwidth}{|>{\tt\small\centering}X|>{\small}X|>{\tt\small}c|>{\small}X|}
  \hline
  \normalfont Function name & Operation & \normalfont File & Notes \\
  \hline
  operator+($\Phi_1, \Phi_2$) & $\Phi_1 + \Phi_2$ & su2doubletsum.h &  \\
  \hline
  operator-($\Phi_1, \Phi_2$)  & $\Phi_1-\Phi_2$ &su2doubletsubtract.h & \\
  \hline
  operator*($\Phi, \varphi$),  operator*($\varphi, \Phi$) & $\varphi \Phi$, $\Phi\varphi$ &complexfieldsu2doubletmultiply.h & Multiplication of a $SU(2)$-doublet expression by a complex expression. Also support scalar (\texttt{double} and \texttt{float}) multiplication. \\
  \hline
  MakeSU2Doublet(i, expr) &Make an $SU(2) $ doublet object out of an expression which depends on \texttt{i}$\in \{0,1,2,3\}$, which labels the entries.  & su2doubletwrapper.h&\texttt{MakeSU2Doublet} is a macro. See the section on macros for more information.\\
    \hline
  norm2($\Phi$) &$\Phi\cdot\Phi$ & su2doubletdotter.h&\\
  \hline
  scalar\_prod($\Phi_1,\Phi_2$) &$\Phi_1\cdot\Phi_2=\varphi_1^{(0)*}\varphi_2^{(0)}+\varphi_1^{(1)*}\varphi_2^{(1)}$ & su2doubletdotter.h&Simplifies scalar product with \texttt{ZeroType} to \texttt{ZeroType}.\\
  \hline
  shift<I>($\Phi$)  &\texttt{shift<I>(}$\Phi$\texttt{)}$|_{\vec n}=\Phi|_{\vec n + \hat I}$  &su2doubletshift.h & Shift the object by a unit vector in the $I^{th}$ direction. \\
 \hline
 shift($\Phi$,Tag<I>)  & \texttt{shift(}$\Phi$\texttt{, Tag<I>)}$|_{\vec n}=\Phi|_{\vec n + \hat I}$ &su2doubletshift.h & Shift the object by a unit vector in the $I^{th}$ direction. Same as above, different notation.   \\
   \hline
   shift<I,J,K,...>($\Phi$)  & \texttt{shift<I,J,...>(}$\Phi$\texttt{)}$|_{\vec n}$ \ $=\Phi|_{(n_1+I,n_2+J,...)}$& su2doubletshift.h & Allows to define an object shifted by an arbitrary vector. \\
\hline
\end{tabularx}

\subsection{$SU(2)$ Algebra}

The files are located in \texttt{src/include/TempLat/lattice/algebra/su2algebra/}.

\begin{tabularx}{\textwidth}{|>{\tt\small\centering}X|>{\small}X|>{\tt\small}c|>{\small}X|}
  \hline
  \normalfont Function name & Operation & \normalfont File & Notes \\
  \hline
  operator+($U_1, U_2$) & $U_1 + U_2$ & su2sum.h & Not well defined inside $SU(2)$ group, but can still be useful. \\
  \hline
  operator-($U_1, U_2$) & $U_1 - U_2$ & su2subtract.h & Not well defined inside $SU(2)$ group, but can still be useful. \\
    \hline
  operator*($U_1, U_2$) & $U_1U_2$ &su2multiply.h & Matrix $SU(2)$ multiplication. \\
  \hline
  operator*($U, \Phi$) & $U\Phi$ &su2su2doubletmultiply.h & $SU(2)$ fundamental group action on the doublet. \\
  \hline
  operator*($\alpha,U$) & $\alpha U$ &scalarsu2doubletmultiply.h & Works only with $\alpha$ a \texttt{double} or \texttt{float}. Not well defined within the group, but sometimes useful.\\
  \hline
dagger($U$) & $U^\dagger$  & su2dagger.h&\\
  \hline  MakeSU2(i, expr) &Make an $SU(2) $ object out of an expression which depends on \texttt{i}$\in \{0,1,2,3\}$, which labels the entries.  & su2wrapper.h&\texttt{MakeSU2} is a macro. See the section on macros for more information.\\
  \hline
  shift<I>($U$)  &\texttt{shift<I>(}$U$\texttt{)}$|_{\vec n}=U|_{\vec n + \hat I}$  &su2shift.h & Shift the object by a unit vector in the $I^{th}$ direction. \\
 \hline
 shift($U$,Tag<I>)  & \texttt{shift(}$U$\texttt{, Tag<I>)}$|_{\vec n}=U|_{\vec n + \hat I}$ &su2shift.h & Shift the object by a unit vector in the $I^{th}$ direction. Same as above, different notation.   \\
   \hline
   shift<I,J,K,...>($U$)  & \texttt{shift<I,J,...>(}$U$\texttt{)}$|_{\vec n}$ \ $=U|_{(n_1+I,n_2+J,...)}$& su2shift.h & Allows to define an object shifted by an arbitrary vector. \\
   \hline
   toSU2($U$)  & Project to group a ``fake" $SU(2)$-like object, like the sum of two $SU(2)$ objects.& su2groupwrapper.h &  \\
   \hline
   trace($U$)  & $\mathrm{Tr}(U)$& su2trace.h & \\
\hline
\end{tabularx}

\subsection{General Purposes Functions and Macros}

The files are located in \texttt{src/include/TempLat/util/}.

\begin{tabularx}{\textwidth}{|>{\tt\small\centering}X|>{\small}X|>{\tt\small}c|}
  \hline
  \normalfont Function name & Operation & \normalfont File  \\
  \hline
  ForLoop(i, imin, imax,  expr) & Compile-time for-loop over the \texttt{Tag<I>} variable \texttt{i}. Varies from \texttt{imin} to \texttt{imax} included. & rangeiteration/for\_in\_range.h  \\
  \hline
  MakeArray(i, imin, imax,  expr) & Make a list of \texttt{\{expr(imin), expr(imin+1), ..., expr(imax)\}} compatible with the \texttt{listoperators} algebra. The resulting list is labeled from \texttt{0} to \texttt{imax-imin-1}. Useful to make array of expressions made out of fields for instance. & rangeiteration/make\_list\_tag.h  \\
  \hline
  MakeVector(i, imin, imax,  expr) & Make a list of \texttt{\{expr(imin), expr(imin+1), ..., expr(imax)\}} compatible with the \texttt{listoperators} algebra. The resulting list is labeled from \texttt{1} to \texttt{imax-imin} and is used to make ``physical" vectors. & rangeiteration/make\_list\_tag.h  \\
  \hline
  Total(i, imin, imax,  expr) & Computes the sum \texttt{expr(imin)+expr(imin+1)+ ...+ expr(imax)}. & rangeiteration/sum\_in\_range.h  \\
  \hline
  Function(x, expr) & Returns a a function of \texttt{x} (returns a lambda function whose body is \texttt{expr} with variable \texttt{x}). & function.h  \\
  \hline
  IfElse(condition, ifExpr, elseExpr) & Depending on the known-at-compile-time boolean condition \texttt{condition}, returns \texttt{ifExpr} if \texttt{true} or \texttt{elseExpr} if \texttt{false}. & staticif.h  \\
  \hline
  If(condition, ifExpr) & Depending on the known-at-compile-time boolean condition \texttt{condition}, returns \texttt{ifExpr} if \texttt{true} or \texttt{ZeroType} if \texttt{false}. & staticif.h  \\
  \hline
  IsLess(i, j) & Returns a known-at-compile-time boolean, checking whether \texttt{i<j} with \texttt{i} and \texttt{j} \texttt{Tag} variables. & rangeiteration/tag.h  \\
  \hline
  IsLessOrEqual(i, j) & Returns a known-at-compile-time boolean, checking whether \texttt{i<=j} with \texttt{i} and \texttt{j} \texttt{Tag} variables. & rangeiteration/tag.h  \\
  \hline
  IsMore(i, j) & Returns a known-at-compile-time boolean, checking whether \texttt{i>j} with \texttt{i} and \texttt{j} \texttt{Tag} variables. & rangeiteration/tag.h  \\
  \hline
  IsMoreOrEqual(i, j) & Returns a known-at-compile-time boolean, checking whether \texttt{i>=j} with \texttt{i} and \texttt{j} \texttt{Tag} variables. & rangeiteration/tag.h  \\
  \hline
  IsEqual(i, j) & Returns a known-at-compile-time boolean, checking whether \texttt{i==j} with \texttt{i} and \texttt{j} \texttt{Tag} variables. & rangeiteration/tag.h  \\
  \hline
  TempLatVector & Overload of the \texttt{C++} \texttt{std::vector} compatible with our vector algebra. Also has \texttt{operator()} overloaded to access its component, to match the syntax of the other objects in the library.  & templatvector.h  \\
  \hline
  TempLatArray & Overload of the \texttt{C++} \texttt{std::array} compatible with our vector algebra. Also has \texttt{operator()} overloaded to access its component, to match the syntax of the other objects in the library.  & templatarray.h  \\
  \hline
\end{tabularx}

\subsection{Accessing into a given location}

\CL is designed to work at an abstract level; all iterations over the lattice are handled internally. However, in some specific circumstances, one may need to \textit{exceptionally} retrieve a given point on the lattice. This is possible by calling the function \texttt{operator()(bool\& test, vector<ptrdiff\_t> position)} defined in \texttt{src/include/TempLat/lattice/field/views/fieldviewconfig.h} (for coordinate space) and \texttt{src/include/TempLat/lattice/field/views/fieldviewfourier.h}  (for Fourier space). The first argument is a boolean, which is set to \texttt{false} by the function on all processors, except the one that is host to the requested coordinates (when using parallelization, not all processes have access to all coordinates, see Sec.~\ref{subsec:para} for more details). The second argument is the coordinate we want to access, passed as an array. The coordinates run from $n_i = -N/2+1$ to $n_i = N/2$ for all directions in coordinate space $i = 1,2,3$ and from $\tilde n_i = -N/2+1$ to $ \tilde n_i = N/2$ for $i = 1,2$ directions except for $i = 3$, for which it runs only between $\tilde n_3 = 0$ and $\tilde n_3 =  N/2-1$ in Fourier space.

Similarly, one can set the value of a field at a given location by calling a \texttt{set(value, position)} function, where value is the value one wants to set the field to at position \texttt{position}. Value is expected to be a number/complex number in configuration/Fourier space.\\

\textit{Example in configuration space:}
\begin{C++code}
phi.set(5.0, {-1,2,3}); //set the field phi to 5 at position (-1,2,3)

bool test;
double res = phi(test, {-1,2,3}) //reads the value of the field phi at position (-1,2,3)
if(test) std::cout << res << std::endl; //prints the above value of the field
\end{C++code}

Assuming \texttt{phi} is a real scalar field, this will first set the field \texttt{phi} to $5$ at position $(-1,2,3)$. It then retrieve this value and store it in {\tt res}. If run in parallel, only one processor knows about this value and this processor is asked to print it on the last line.\\

\textit{Example in Fourier space:}
\begin{C++code}
phi.inFourierSpace().set(std::complex<double>(5.0, 10.0), {1,-2,3}); //set the field phi to 5 + I10 at position (1,-2,3)

bool test;
std::complex<double> var = phi.inFourierSpace()(test, {1,-2,3}) //reads the value of the Fourier amplitude of phi at coordinate (1,-2,3) of the reciprocal lattice
if(test) std::cout << var.real() << " + I" << var.imag() << std::endl; //prints the above Fourier amplitude of the field at coordinate (1,-2,3) of the reciprocal lattice
\end{C++code}

This will set the Fourier amplitude of the field \texttt{phi} to $5+I10$ at coordinate $(1,-2,3)$ of the reciprocal lattice. It then retrieves this value and stores it in {\tt var}. If run in parallel, only one processor knows about this value and this processor is asked to print it on the last line.

\section{Under the Hood: Expression Templates and \CL}
\label{app:ExprTemp}
While it is beyond the scope of this user-manual to expose the whole mechanics behind the code and most specifically the \texttt{TempLat} library, we want to elaborate on the main concept behind the implementation of the fields and their related algebra, the one of ``expression templates". Templates, which appear throughout \CLns, are a \texttt{C++} mechanism which allows for ``generic" programming; types become themselves ``variables", to be specified at compile times. Templates can be used as their own programming language (this is often referred to as ``template metaprogramming", see
ref.~\cite{10.5555/3175809} for more information on this paradigm)\footnote{Which is even Turing-complete, meaning that any software whatsoever could in principle be written only out of templates.}; atop of allowing for generic (type independent) functions, they can be used to move computations from being performed at run time to be performed at compile time. As we will see shortly, this extra level of abstraction can be used to implement ``symbolic" computations in \texttt{C++}, which is what ``expression templates" are.

By ``symbolic computations", we mean some software which is capable as understanding expressions of the type ``$a + b - b$", simplifying them to ``$a $" and evaluate them once  ``$a$" and ``$b$" have been assigned a value. Instead of spending time on the various subtleties on expression templates, we will refer the reader to Ref.~\cite{Falcou15} and simply explain as an example how to create such a software using template metaprogramming. The challenge is to have an object which can at the same time represent the abstract expression ``$ a + b - b$", being able to manipulate it at the abstract level and then also being able to evaluate it. A key realisation is that this challenge can be addressed by using template metaprogramming, evaluated at compile time, to represent the abstract structure and use normal code, evaluated at run time, to take care of the evaluation. To achieve this, every member of an algebraic expression will be represented as a ``type".

Let us be specific. Let us first show how one use templates to implement abstract expressions. We first create two empty \texttt{Number} classes,
\begin{C++code}
  class A{};
  class B{};
\end{C++code}
and a class to represent the ``addition" operation. It can simply be implemented as follow
\begin{C++code}
  template<class X, class Y>
  class Addition{};
  template<class X, class Y>
  class Subtraction{};
\end{C++code}

Now, an object of the type \texttt{Addition<A, B>} can be used to represent ``$a+b$". \texttt{Addition<A,Subtraction<B,B>>} would then be interpreted as ``$a+b-b$". To actually be able to to manipulate these types, we add some operators
\begin{C++code}
  template<class X, class Y>
  Addition<X,Y> operator+(X, Y) // Not necessary to give names to function's arguments in C++.
  {
    return Addition<X,Y>();
  }

  template<class X, class Y>
  Subtraction<X,Y> operator-(X, Y)
  {
    return Subtraction<X,Y>();
  }
\end{C++code}
 Now, atop of this generic definition, we can add more specific ones to deal with special case. First, let us define a type to represent the number
$0$, which will be the special case we will be dealing with, as for instance $a+0=a$ and $a-a=0$.
\begin{C++code}
  class ZeroType{};
\end{C++code}

We can now specify our operators to behave differently when in presence of \texttt{ZeroType}
 \begin{C++code}
   template<class X>
   X operator+(X, ZeroType)
   {
      return X();
   }
   template<class X>
   X operator+(ZeroType, X)
   {
      return X();
   }
 \end{C++code}
 We can also use this to simplify expression of the type ``$a-a$" to "$0$"
 \begin{C++code}
   template<class X>
   ZeroType operator-(X, X)
   {
      return ZeroType();
   };
 \end{C++code}

 Altogether, this set of definitions would simplify the expression \texttt{A()+B()-B()} to simply \texttt{A()}.

 Note that at this point we have only taken care of the abstract expression; everything happens at compile-time and nothing happens at execution time. Note however that we have already achieved something interesting; we can now store and manipulate ``formulas", as you would do with a program such as Mathematica.

 We can easily take care of the evaluation. For that we modify our class \ \texttt{A} and \texttt{B} so that they represent some given number and give them a ``getter" to retrieve this number

 \begin{C++code}
   class A{
    double get()
    {
      return 9.0;
    }
   };
   class B{
    double get()
    {
     return 0.5;
    }
   };
 \end{C++code}
We also modify the operators so that they hold a copy of the object they operate on and provide them with a getter responsible for the evaluation.

\begin{C++code}
  template<class X, class Y>
  class Addition{
  public:
  Addition(X pX, Y pY):
  x(pX),
  y(pY)
  {
  }

  double get()
  {
    return a.get() + b.get();
  }

  private:
    X x;
    Y y;

  };
  template<class X, class Y>
  class Subtraction{
  public:
  Subtraction(X pX, Y pY):
  x(pX),
  y(pY)
  {
  }

  double get()
  {
    return a.get() - b.get();
  }

  private:
    X x;
    Y y;

  };
\end{C++code}

Now we can evaluate our expression. For instance, we can write
\begin{C++code}
  A a;
  A b;

  auto expr = a + b; // Only store the expression, nothing is computed here.

  std::cout << expr.get() << std::endl;  // When we call expr.get(), the addition is done.
\end{C++code}
This way, we achieved to have symbolic expression which can be stored, in a way which is completely unrelated to the evaluation.

While being simple, this example illustrate the most important features of expression templates. For a more complicated example, we invite the interested reader to explore the \texttt{Field} class of \CLns, located in the \texttt{src/include/TempLat/lattice/field/} (the equivalent of $A$ and $B$) and the operators defined in the \texttt{src/include/TempLat/lattice/algebra/operators/} folder. Fields have a getter which takes as an argument the index of a lattice point. Operators are implemented in the same way as presented above. Evaluation happens only in the \texttt{operator=} of the \texttt{Field} class.

In a sophisticated situation like this, this separation between abstracts expression and evaluation also allow to affect the evaluation depending on the actual expression we want to evaluate, something which would not be possible otherwise. In particular, this is precisely this mechanism which allows \CL to completely hide under the hood the parallelization of the program. By having at hand the expression on an abstract level, we can analyze them to see whether or not they contain operator involving interaction between neighboring sites. If this is the case, we know this means that boundary needs to be synchronized before evaluation (see Section \ref{subsec:para}) and we can trigger it automatically. In this way, it is completely hidden from the user.

\begin{multicols}{2}

\end{multicols}

\end{document}